\documentclass[showpacs,aps,prd,preprintnumbers,amsmath,amssymb,superscriptaddress,groupedaddress,nofootinbib,twocolumn]{revtex4}
\usepackage{dcolumn,lipsum}
\usepackage{bm}
\usepackage{natbib}
\usepackage[mathscr]{eucal}
\usepackage{doi}
\usepackage{soul}
\usepackage{dsfont}
\usepackage[toc,page]{appendix}

\usepackage{amsmath,amssymb}
\usepackage{float}
\usepackage{epsfig} 
\usepackage{graphicx,multirow}
\usepackage[caption = false,,justification=justified]{subfig}
\usepackage{slashed}
\usepackage{slashbox}
\usepackage{multirow}
\usepackage{shorthand}
\makeatletter

\usepackage{xcolor}

\begin{document}
\title{Precise probing of the inert Higgs-doublet model at the LHC}
\author{Anupam Ghosh}
\email{anupam@prl.res.in}
\affiliation{Physical Research Laboratory (PRL), Ahmedabad - 380009, Gujarat, India}
\affiliation{Indian Institute of Technology, Gandhinagar-382424, Gujarat, India}
\author{Partha Konar}
\email{konar@prl.res.in}
\affiliation{Physical Research Laboratory (PRL), Ahmedabad - 380009, Gujarat, India}
\author{Satyajit Seth}
\email{seth@prl.res.in}
\affiliation{Physical Research Laboratory (PRL), Ahmedabad - 380009, Gujarat, India}

\begin{abstract}
	The inert Higgs-doublet model provides a simple framework to accommodate a viable Higgs portal scalar dark matter candidate, together with other heavier scalars of mass 100 GeV or more. We study the effect of next-to-leading order (NLO) QCD corrections in this scenario in the context of the Large Hadron Collider. 
${\cal{O}}(\alpha_s)$ corrections to the gluon-gluon-Higgs effective coupling have been taken into account in this study wherever appropriate. We find such corrections have a significant impact on various kinematic distributions and reduce scale uncertainties substantially. Fixed order NLO results are matched to the {\sc Pythia8} parton shower (PS) and the di-fatjet signal associated with the missing transverse momentum is analyzed, as this channel has the ability to explore its entire parameter space during the next phase of the LHC run. A closer look at the NLO+PS computation indicates a sizable NLO effect together with a subdued contribution from associated production of the heavy scalar compared to the pair production, thereby leading to a refined analysis strategy during the multivariate analysis of this signal.
\end{abstract}


\maketitle

\section{Introduction}
\label{sec:into}

The Standard Model (SM) of particle physics elegantly describes three of the four fundamental interactions of the Universe. As a major triumph of this model, the last piece {\em i.e,} the only scalar in the SM, Higgs boson, was discovered by the ATLAS \cite{ATLAS:2012yve} and CMS \cite{CMS:2012qbp} Collaborations in 2012. Certain drawbacks of the SM, either in terms of theoretical consistency or due to the lack of explanation for different fundamental observations, such as non-zero masses of the neutrinos, existence of the dark matter, matter antimatter asymmetry, {\em etc.}, have prompted both theoretical as well as experimental communities to look further into the ideas and evidences beyond the Standard Model (BSM). Different cosmological observations, namely, galactic rotation curves, gravitational lensing, bullet cluster structure formation, {\em etc.}, already establish the existence of some exotic nonluminous, weakly (to feebly) interacting matter, known as dark matter (DM) and has shown its existence almost at all length scales of the universe.  Apart from the fact that DM has gravitational interaction, precise measurement of cosmic microwave background from the WMAP \cite{WMAP:2012nax} and PLANCK \cite{Planck:2013pxb, Planck:2015fie, Planck:2018vyg} data, has established the relic abundance corresponding to $26\%$ of the present energy budget of our universe. However, any microscopic nature of the DM is yet unknown. Despite of considerable amount of effort by the astroparticle and high energy physics communities in order to detect and explain the microscopic nature of the DM, no experiment has so far been able to make any detection.

In the present study, we consider the inert Higgs doublet model (IDM)~\cite{Barbieri:2006dq, Cirelli:2005uq} as a prospective BSM scenario that gives a viable DM candidate. This model is renormalizable and is constructed by a simple particle extension of the SM containing an extra $SU(2)_L$ scalar doublet, which is odd under the discrete $\mathbb{Z}_2$ symmetry \footnote{Promoting such $\mathbb{Z}_2$ parity into a global U(1) symmetry is possible that provides a heavily constrained DM model with fewer model parameters \cite{Jueid:2020rek}.}. In contrast, all SM particles are even under the $\mathbb{Z}_2 $ and therefore such a symmetry arrangement prevents any interaction between the SM fermions and BSM scalars -- this stabilizes the lightest neutral scalar, which acts as a suitable DM candidate. The parameter space of this model is constrained \cite{Ilnicka:2015jba, Belyaev:2016lok, Arhrib:2013ela, Dercks:2018wch} from the dark matter direct detection (DD) experiment, LHC data, and various other astronomical and cosmological observations. In its simplest form, the present model can satisfy the whole amount of observed relic density of the DM in some particular parameter space, the so-called {\it resonance region} and {\it degenerate region}. In the former case, the relic density of the DM is produced thermally through the resonant Higgs portal annihilation. Hence, the DM mass is required to be nearly half of the Higgs boson mass and other BSM scalars carry larger masses.  
This region is also known as the {\it hierarchical mass region} as DM is the lightest, while others are quite heavy. On the other hand, DM and all other BSM scalars are nearly of equal mass ($\sim 500 $ GeV or more) in the degenerate region \cite{Goudelis:2013uca, Blinov:2015qva, Diaz:2015pyv}. As expected, this region is harder to probe at the LHC because of the kinematic suppression due to heavy final state production, narrow mass gap and poor detection efficiency of the soft products coming from the decay of the BSM scalars. In spite of these complexities, this region is partially probed by the CMS Collaboration \cite{CMS:2020atg}. There are two additional regions: ({\em i}) heavy DM, $m_{DM}\sim 550$ GeV with the hierarchical mass spectrum where the mass difference, $\Delta M \sim \mathcal{O}(10-100)$ GeV between DM and other heavy BSM scalars, ({\em ii}) light DM, $m_{DM} < 80$ GeV with the degenerate mass spectrum, $\Delta M \sim \mathcal{O}(1-10)$ GeV. However, these two regions gather only a few percent of the observed relic density of the DM \cite{Bhardwaj:2019mts} and are hence less exciting. The light DM degenerate scenario is known to be satisfying only $\sim 10\%$ of the observed relic density \cite{Bhardwaj:2019mts} and that can be probed at the LHC with the monojet search \cite{Belyaev:2018ext}. Several other search strategies focused on different final states such as, multilepton production with missing transverse energy \cite{Miao:2010rg, Gustafsson:2012aj, Hashemi:2016wup, Datta:2016nfz}, dilepton with dijet \cite{Hashemi:2016wup}, and dijet with missing transverse momentum ($\slashed{E}_T$) \cite{Poulose:2016lvz}.

In this study, we consider the hierarchical mass region that satisfies the total observed relic density of the DM and all other constraints. The significant mass difference between BSM scalars and the DM leads to a very interesting signal topology due to the boosted vector boson created through heavy scalar decay. We focus on the associated production and pair production of heavy scalars. Among heavy scalar pair production channels, pseudoscalar pair production is only possible through the Higgs boson mediator. Another channel (charged-scalar pair production) also gets contribution from the Higgs mediated diagrams. The major contribution to such production comes from the gluon fusion, which contains a loop at the leading order (LO). We work in the heavy top mass limit and that reduces the one loop diagram into an effective vertex. We consider ${\cal{O}}(\alpha_s)$ corrections to that effective term which is known to be as large as the LO alone. 
Therefore, the total Lagrangian is the sum of the IDM Lagrangian and the gluon-Higgs effective Lagrangian, and we consider ${\cal O}(\alpha_s)$ corrections to the total Lagrangian. The $p p \rightarrow H H$ process does not contribute to our phenomenological analysis, while loop induced $p p \rightarrow A A$ process has a minimal contribution compared to the other processes and we take this contribution into account in the heavy top mass limit. So even in the phase space regions where the heavy top mass limit is not a reasonable assumption, we restrict ourselves to this approximation as the effect coming from the above-mentioned channels are considerably smaller.
This hierarchical mass spectrum of the IDM is hard to probe at the LHC because of its tiny cross section over an immense SM background. We recognize in this study that the $K$ factor due to NLO correction for the associated production of the heavy scalar ranges from 1.33 to 1.37. Vector boson mediated diproduction channels get corrections ranging from 1.35 to 1.56, while for the Higgs boson mediated ones it varies from 1.7 to 1.9. Correct understanding of signal distribution is important, where the signal is tiny compared to the background. These corrections give precise predictions improving the signal's statistical significance and lead us to change the analysis strategy based on modified signal output.

It was already demonstrated that the dijet+$\slashed{E}_T$ signal could barely give $2\sigma$ statistical significance of the signal at high luminosity LHC \cite{Poulose:2016lvz}. 
The mono-fatjet signal was also studied in \cite{CMS:2016xus, ATLAS:2018nda}, which is not sufficient to achieve discovery potential. Hybrid topology (admixture of the mono-fatjet and di-fatjet+$\slashed{E}_T$ signal) was analyzed in \cite{Bhardwaj:2019mts}. 
Fatjets originate from boosted $W^\pm/Z$ boson (denoted as $J_V$ later on) after the decay of heavy BSM scalars and therefore possess a two-prong structure. Naturally, the pruned fatjet mass ($M_J$) and subjettiness ($\tau_{21}$) become crucial variables to separate the tiny signal from the background. A sophisticated multivariate analysis (MVA) with jet-substructure variables is adopted in this analysis. It is demonstrated that di-fatjet detection from hybrid topology, together with the full potential of jet substructure variables, could effectively bring nearly the entire available parameter space well within reach of the 14 TeV HL-LHC. In MVA, we choose cuts on different variables optimally to increase the signal to background ratio and perform the analysis at NLO accuracy with jet substructure variables. Note that the di-fatjet signature relies upon a particular phase space region; hence the differential NLO $K$ factor due to NLO correction plays a vital role to get more accurate signal efficiency. 
In this analysis, we find that NLO computation reduces the contribution in the di-fatjet final state arising from the associated production of the heavy scalar processes compared to the pair production channels, leading to rectified analysis strategies during MVA and that helps to reach a higher discovery potential of the IDM.

We organize the paper as follows: \autoref{sec:model} briefly describes the IDM model and the Higgs-gluon effective Lagrangian that we adopt in this computation. \autoref{sec:BP}, points out various constraints on the IDM model and lists benchmark points accordingly in the hierarchical mass region. 
In \autoref{sec:result1}, we mainly discuss the computational setup and show numerical results including the differential NLO $K$ factor and scale uncertainties. \autoref{sec:result2} presents the distributions of different high-level kinematical variables involving jets at LO and NLO for the associated and pair production channels, demonstrating the importance of the QCD corrections.
\autoref{sec:collider} explains the reason to consider $2J_V+\slashed{E}_T$ as the signal while dealing with a tiny IDM signal over an immense background. 
We also discuss here the MVA, which uses a highly non-linear cut, and use the full potential of NLO computation and jet-substructure variables to separate this tiny signal from the large background. Finally, we conclude in \autoref{sec:conclude}.

\section{Theoretical framework}
\label{sec:model}  

IDM has a new $SU(2)_L$ doublet $\Phi_2$ in addition to the SM Higgs doublet, $\Phi_1$, and a discrete $\mathbb{Z}_2$ symmetry is being imposed on it. All the fields of the SM are even under $\mathbb{Z}_2$ transformations.  $\Phi_2$ is odd under $\mathbb{Z}_2$ transformation and therefore the inert doublet can not acquire vacuum expectation value (vev), as vev can not change sign under any internal symmetry. As $\Phi_2$ has no vev, we can write this doublet in terms of physical fields. $\mathbb{Z}_2$ symmetry also prevents the interaction between inert scalars and SM fermions at any order in the perturbation series, aiding the lightest inert neutral scalar to act as a dark matter. The doublet, $\Phi_2$, has hypercharge $Y=\frac{1}{2}$, which is equal to the hypercharge of $\Phi_1$. These two doublets can be written in the unitary gauge as 
\begin{equation}
\Phi_1= \begin{pmatrix}
G^+\\
\dfrac{1}{\sqrt{2}}(v+h+iG^0)
\end{pmatrix}
, 
\quad\Phi_2= \begin{pmatrix}
H^+\\
\dfrac{H + i \hspace{1mm} A}{\sqrt{2}}
\end{pmatrix},
\end{equation}
where $G^+$ and $G^0$ are the Goldstone bosons and the vev $v=246 \hspace{1mm} \text{GeV}$. $H^+$ is the charged BSM scalars.  $H$ and $A$ are both neutral scalars; one is CP even, and the other is CP odd. Note that CP properties of the neutral scalars are basis-dependent. The most general potential \cite{Belyaev:2016lok} can be written as,
\begin{equation} 
\begin{split}
V_{IDM}  = & \, \mu_1^2\Phi_1^\dagger\Phi_1+ \mu_2^2\Phi_2^\dagger\Phi_2+ \dfrac{\lambda_1}{2} (\Phi_1^\dagger\Phi_1)^2+ \dfrac{\lambda_2}{2} (\Phi_2^\dagger\Phi_2)^2  \\ & 
+ \lambda_3 (\Phi_1^\dagger\Phi_1)(\Phi_2^\dagger\Phi_2)  + \lambda_4 (\Phi_2^\dagger\Phi_1)(\Phi_1^\dagger\Phi_2) \\ &
+ \dfrac{\lambda_5}{2}\hspace{1mm} [(\Phi_1^\dagger\Phi_2)^2+(\Phi_2^\dagger\Phi_1)^2] \, . 
\end{split}
\end{equation}

After electroweak symmetry breaking through SM Higgs doublet, $\Phi_1$, the masses of the BSM scalars at the tree level can be expressed as, 
\begin{equation}
\begin{split}
& m_h^2= \lambda_1 v^2, \hspace{1mm} m^2_{H^\pm}=\mu_2^2+\dfrac{1}{2}\lambda_3 v^2, \hspace{1mm}  \\ &
m^2_{A} =\mu_2^2+\dfrac{1}{2}\lambda_c v^2, 
 \hspace{1mm} m^2_{H} =\mu_2^2+\dfrac{1}{2}\lambda_L v^2 .
\end{split}
\end{equation}
All free parameters are real, so the scalar sector does not contain any $CP$ violations and $\lambda_{L/c}=(\lambda_3+\lambda_4\pm \lambda_5)$. Higgs portal coupling $\lambda_{L}$, which can be positive or negative, plays an important role in the DM sector as it determines the annihilation rate of the DM in the hierarchical mass region. $m_h$ is the SM Higgs boson mass, and $m_{H^\pm,A,H}$ are the masses of the BSM scalars. The parameters $\lambda_1$ and $\mu_1$ can be written in terms of the mass of the Higgs boson and vev.  So, IDM has five parameters -- three masses of the inert scalars, self-coupling between inert scalars $\lambda_2$ and Higgs portal coupling $\lambda_L$. Self-coupling $\lambda_2$ does not affect the scalar masses and their phenomenology.
In our study, we choose the inert scalar $H$ as the dark matter candidate, but one can also choose the $A$ as the dark matter without changing any phenomenology, just by flipping the sign of $\lambda_5$ preserving the $CP$ properties of
the DM candidate. The full IDM Lagrangian can be written as 
\begin{equation}
\mathcal{L}_{IDM}= \mathcal{L}_{SM} +(\mathcal{D}_{\mu} \Phi_2)^\dagger (\mathcal{D}^\mu \Phi_2)+ V_{IDM}
\label{idmL}
\end{equation}
where the covariant derivative, $\mathcal{D}_{\mu} =(\partial_\mu -ig_{Y}  Y  B_\mu -ig \frac{\sigma^i}{2} W^i_\mu)$, and $\sigma^i$ are the Pauli matrices; $g$ and $g_{Y}$ are the coupling strength of the weak and hypercharge interactions, respectively.
In addition, we consider the following five-dimensional effective term to take into account Higgs interactions with gluons in the heavy top mass limit,  

\begin{equation}
\mathcal{L}_{HEFT}=-\dfrac{1}{4} C_{eff} \hspace{1mm} h \hspace{1mm} G^a_{\mu\nu} G^{a\mu\nu} .
\label{eftL}
\end{equation}
Here, $G^a_{\mu\nu}$ represents QCD field strength tensor and
$C_{eff}=\dfrac{\alpha_s}{3\pi v}\hspace{1mm}(1+\dfrac{11}{4}\dfrac{\alpha_s}{\pi}) = C_0 \hspace{1mm}(1+\dfrac{11}{4}\dfrac{\alpha_s}{\pi})$
contains terms up to ${\cal O}(\alpha_s^2)$, that basically takes part in the one loop corrected amplitude for Higgs boson mediated production channels. 
%

\section{Constraints and benchmark points}
\label{sec:BP}

The parameter space of the IDM is very constrained from theoretical calculations, various experimental data and cosmological observations. We briefly demonstrate all these constraints and then set few benchmark points that will cover almost the entire hierarchical region of the IDM. Further details are provided in \cite{Belyaev:2016lok},\cite{Bhardwaj:2019mts}.

The Potential must be bounded from below for any realistic model, and the vacuum should be neutral, which leads to the following constraint \cite{Bhardwaj:2019mts}:
\begin{equation}
\begin{split}
& \lambda_1 > 0, \hspace{1mm} \lambda_2 > 0, \hspace{1mm} \lambda_3 + 2\sqrt{\lambda_1 \lambda_2} > 0, \\
& \hspace{1mm} \lambda_3 + \lambda_4 + \lambda_5 +  2\sqrt{\lambda_1 \lambda_2} > 0.
\end{split}
\end{equation}

The condition $\lambda_4 + \lambda_5 < 0$ ensures the inert vacuum to be charge neutral. Generically, depending on the nature of additionally imposed symmerty, the electroweak symmetry breaking pattern has the following possibilities,
\begin{equation}
\begin{split}
& v_1=v,\hspace{1mm} v_2=0, \hspace{2mm} \text{inert vacuum}\\
& v_1=0,\hspace{1mm} v_2=v, \hspace{2mm} \text{pseudo-inert vacuum}\\
& v_1\neq 0,\hspace{1mm} v_2\neq 0, \hspace{2mm} \text{mixed vacuum}
\end{split}
\end{equation}
where $v_1$ denotes the vev of the doublet $\Phi_1$, and $v_2$ is the vev of the $\Phi_2$. $v$ is the electroweak scale, $(G_F\sqrt{2})^{-1/2}=246$ GeV. We want the inert vacuum as the global minima, which leads to the following constraint \cite{Ginzburg:2010wa},\cite{Swiezewska:2012ej}
\begin{equation}
\dfrac{\mu_1^2}{\sqrt{\lambda_1}} - \dfrac{\mu_2^2}{\sqrt{\lambda_2}} > 0 .
\end{equation}
The eigenvalues of the $2 \rightarrow 2$ scalar scattering processes \cite{Arhrib:2012ia} are given in Eq.~\ref{eq:condition_4}, and each eigenvalue ($|\Lambda_i|$) should be $\leq 8\pi$, coming from the perturbativity and unitarity constraints:

\begin{equation}
\begin{split}
& \Lambda_{1,2} = \lambda_3 \pm \lambda_4, \hspace{1mm} \Lambda_{3,4} = \lambda_3 \pm \lambda_5, \hspace{1mm} \Lambda_{5,6} = \lambda_3 + 2 \lambda_4 \pm 3\lambda_5, \\&
\Lambda_{7,8} = -\lambda_1 - \lambda_2 \pm \sqrt{(\lambda_1- \lambda_2)^2 + \lambda_4^2}, \hspace{1mm}\\& 
\Lambda_{9,10} = -3 \lambda_1 - 3 \lambda_2 \pm \sqrt{9(\lambda_1- \lambda_2)^2 + (2\lambda_3 + \lambda_4)^2}, \\&
\Lambda_{11,12} = -\lambda_1 - \lambda_2 \pm \sqrt{(\lambda_1- \lambda_2)^2 + \lambda_5^2} .
\end{split}
\label{eq:condition_4}
\end{equation}

The contribution that navigates from the BSM physics to the electroweak radiative correction is parametrized by the $S$, $T$, $U$ parameters \cite{Peskin:1991sw}, known as oblique parameters. The central values of the oblique parameters that we use in our analysis are \cite{Haller:2018nnx}
\begin{equation}
S = 0.04\pm 0.11, \hspace{2mm} T = 0.09 \pm 0.14, \hspace{2mm} U = -0.02 \pm 0.11 \, .
\end{equation}
The following parameter space of the IDM is ruled out from the neutralino search results at LEP-II \cite{Lundstrom:2008ai},\cite{Belanger:2015kga}:
\begin{equation}
m_{H}<80\hspace{1mm}\text{GeV},\hspace{1mm} m_A <100\hspace{1mm}\text{GeV} \hspace{1mm}, ~\mbox{and}~ \hspace{1mm} (m_A-m_H) >8\hspace{1mm}\text{GeV} \,.
\end{equation}

THe charged Higgs mass gets the following bound from the chargino search results at LEP-II \cite{Pierce:2007ut}:
\begin{equation}
m_{H^\pm}>70\hspace{1mm}\text{GeV} \, .
\end{equation}
More recently, analyzing a pair of boosted hadronically decaying bosons together with MET from 13 TeV LHC data, ATLAS gave constraints on the masses of the charginos and neutralinos of the minimal supersymmetric model~\cite{ATLAS:2021yqv}. Based on a similar production mechanism from IDM, Ref.~\cite{Banerjee:2021oxc}
carried out a recasting analysis to show that the Higgs portal DM scenario and hierarchical heavy scalars of mass 123 GeV or above are allowed from this exclusion limit. 

In the hierarchical region, the decay channels, $\Gamma(Z\rightarrow HA,H^+H^-)$ and $\Gamma(W^\pm\rightarrow H^\pm A,H^\pm H)$  are kinematically forbidden.
The signal strength of the Higgs boson decay into the diphoton final state relative to SM prediction is \cite{ParticleDataGroup:2018ovx, ATLAS:2015egz, CMS:2018piu}:
\begin{equation}
\mu_{\gamma \gamma}=\dfrac{\sigma(p p \rightarrow h \rightarrow \gamma \gamma)}{\sigma(p p \rightarrow h \rightarrow \gamma \gamma)_{SM}}=1.10^{+0.10}_{-0.09} \, .
\end{equation}

The Higgs boson production rate is the same in both the SM and IDM models, dominated by the gluon gluon fusion channel, and so the signal strength turns out to be

\begin{equation}
\mu_{\gamma \gamma } = \dfrac{BR(h \rightarrow \gamma \gamma)_{IDM}}{BR(h \rightarrow \gamma \gamma)_{SM}}.
\end{equation}

A sufficiently large value of the $\lambda_3$ coupling and lighter charged Higgs mass can lead to enhanced decay of $h \rightarrow \gamma \gamma$, thereby pushing the ratio beyond the experimental limit and hence excluded.
The upper limit of the Higgs invisible decay branching ratio measured by the ATLAS Collaboration \cite{ATLAS:2020kdi} is 0.11 at $95\%$ C.L. This measurement puts stringent constraints on the Higgs portal coupling ($\lambda_L$) and DM mass ($m_H$) in the region $m_{H} < \dfrac{m_h}{2}$. The Higgs invisible decay width in the IDM model is given by
\begin{equation}
\Gamma_{h\rightarrow H H}= \dfrac{\lambda_L^2 v^2}{64\pi m_h}\sqrt{1-\dfrac{4m_H^2}{m_h^2}} \, .
\end{equation}

$\Gamma_{h\rightarrow H H}/(\Gamma_{SM}+\Gamma_{h\rightarrow H H})\leq 0.11$ must be satisfied in the kinematically allowed region of the decay of the Higgs boson into pair of the DM.
Moreover, extremely precise measurements from WMAP \cite{WMAP:2012nax} and PLANCK \cite{Planck:2013pxb, Planck:2015fie, Planck:2018vyg} have established that the relic abundance of the DM is $\Omega_{DM}h^2=0.120 \pm 0.001$ \cite{Planck:2018vyg} with $h = \dfrac{\text{Hubble Parameter}}{(100 \mbox{km}\hspace{1mm} \mbox{s}^{-1} \mbox{Mpc}^ {-1})}$.
The dark matter annihilates into SM particles and the relic density of the DM is inversely proportional to this annihilation rate. The observed relic density of DM sets a rigid constraint on the parameter spaces of the IDM so as not to overproduce the relic in the IDM.
The spin-independent cross section of the DM-nucleon scattering processes at leading order mediated by the Higgs boson is given by \cite{Barbieri:2006dq}
\begin{equation}
\sigma = \dfrac{\lambda_L^2 \hspace{1mm} f^2}{4\pi}\hspace{1mm} \dfrac{\mu^2 \hspace{1mm} m_n^2}{m_h^4 \hspace{1mm} m_{DM}^2},
\end{equation}
\begin{table}[!t]
\begin{center}
 \begin{tabular}{|c|c|c|c|c|c|c|c|}
\hline
Input Parameters & BP1 & BP2 & BP3 & BP4 & BP5 & BP6 & BP7   \\
\hline
$m_{H^\pm}$(GeV) & 255.3 & 304.8 & 350.3 & 395.8 & 446.9 & 503.3 & 551.8 \\
\hline
$m_{A}$(GeV) & 253.9 & 302.9 & 347.4 & 395.1 & 442.4 & 500.7 & 549.63 \\
\hline
$\lambda_2$ & 1.27 & 1.07 & 0.135 & 0.106 & 3.10 & 0.693 & 0.285 \\
\hline
 \end{tabular} 
\caption{Input parameters, masses of the BSM scalars ($m_{H^\pm},m_A$), and the self-coupling constant ($\lambda_2$) between dark sector particles for several selected benchmark points that satisfy theoretical, DM relic density, DD data, and collider constraints listed in the text. Three other parameters are DM mass, $m_{H}=53.71$ GeV, Higgs portal coupling, $\lambda_L=5.4\times 10^{-3}$ and Higgs boson mass $m_h=125$ GeV.
}
\label{tab:parameters}
\end{center}
\end{table}
where $m_n$ is the mass of the nucleon and  $\mu=\dfrac{m_n \hspace{1mm} m_{DM}}{m_n+m_{DM}}$. $f$ is the Higgs-nucleon coupling strength and the allowed range of $f$ is 0.26 - 0.63 \cite{Mambrini:2011ik}. However, the recent study suggests the value of $f$ is 0.32 \cite{Giedt:2009mr}. The upper bound of the DM-nucleon scattering cross section from the DM DD experiments like LUX \cite{LUX:2016ggv} and Xenon1T \cite{XENON100:2012itz} poses a firm limit on the allowed values of $\lambda_L$.
As already stated, we can divide the entire parameter space of the IDM into four distinct regions depending on the mass of the DM and the mass splitting between DM and other scalars - among these four, only the following two regions satisfy the observed relic density of the DM entirely.

The {\em hierarchical mass region} consists of a Higgs portal mass region with $m_{DM} \equiv m_H < 80 \hspace{1mm}$ GeV, and the mass gap with other BSM scalars as, $\Delta M \equiv \Delta M_{charged} \simeq \Delta M_{neutral} \sim 100 \hspace{1mm} $ GeV or more, where $\Delta M_{charged}=(m_{H^\pm} - m_{DM})$ and $\Delta M_{neutral}=(m_{A} - m_{DM})$. In this region, no bound on the DM mass comes from the LEP $Z$-boson width measurements. Since the DM mass is less than 80 GeV, the annihilation of the DM into the pair of weak gauge bosons is significantly suppressed. In this region, relic density of DM is achieved only through the Higgs portal annihilation channel. Since the mass differences between DM and other BSM scalars are significant, the co-annihilation effects are absent. As the annihilation cross section is proportional to  $\lambda_L$, any small value of $\lambda_L$ leads to overproduction of relic density.  We get the total observed relic density of the DM  in the range where the DM mass varies between 53 and 70 GeV for substantial $\lambda_L$ values, constrained from DD of DM.

The {\em degenerate mass region} consists of  high mass region, $m_{DM}\geq 500 \hspace{1mm}$ GeV, with rather tiny mass gap $\Delta M \sim 1 \hspace{1mm} $ GeV.
In this regime, the following annihilation and co-annihilation processes open up:
\begin{equation}
\begin{split}
 \text{annihilation} & 
\begin{cases}
H \hspace{1mm} H \rightarrow W^+ \hspace{1mm} W^- \\
H \hspace{1mm} H \rightarrow Z \hspace{1mm} Z \\
\end{cases}  
\lambda_L \hspace{1mm} \text{sensitive} \\
 \text{co-annihilation} &
\begin{cases}
H^+ \hspace{1mm} H^- \rightarrow W^+ \hspace{1mm} W^-\\
A \hspace{1mm} A \rightarrow W^+ \hspace{1mm} W^-\\
A \hspace{1mm} A \rightarrow Z \hspace{1mm} Z \\
\end{cases}
\lambda_L \hspace{1mm} \text{sensitive} \\
 \text{co-annihilation} &
\begin{cases}
H^\pm \hspace{1mm} H \rightarrow W^\pm \hspace{1mm} \gamma & \text{gauge couplings} \\
H^\pm \hspace{1mm} A \rightarrow W^\pm \hspace{1mm} \gamma \\
\end{cases} 
\end{split}
\end{equation}

The quartic coupling between DM and the longitudinal gauge bosons in the annihilation processes $H \hspace{1mm} H \rightarrow W^+_L \hspace{1mm} W^-_L$ and $H \hspace{1mm} H \rightarrow Z_L \hspace{1mm} Z_L$ is ($4\hspace{1mm} m_{DM}\hspace{1mm}\Delta M/v^2+\lambda_L$). In this degenerate mass spectrum $\Delta M  \rightarrow 0$, and so this coupling remains sensitive to $\lambda_L$ mostly. The relic density of DM increases with the DM mass and decreases with the annihilation cross section. Those combined effects set the correct relic density of DM in this region for $m_{DM}\geq 500 $ GeV. Although this region is difficult to probe, with a charged long-lived Higgs boson, one can explore this region at LHC with the charged track signal \cite{CMS:2014gxa}.

The different benchmark points that we pick out for this study are given in \autoref{tab:parameters}, and all of them satisfy the constraints discussed above.

\section{Computational setup and numerical results}
\label{sec:result1}

\begin{figure}[!t]
\centering
  \subfloat[]{\label{Feyn:a0h2}\includegraphics[scale=0.261]{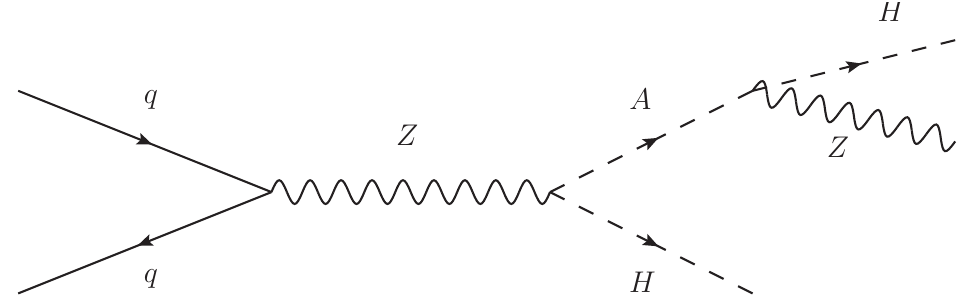}} \hspace{0.cm}
  \subfloat[]{\label{Feyn:hpmh2}\includegraphics[scale=0.261]{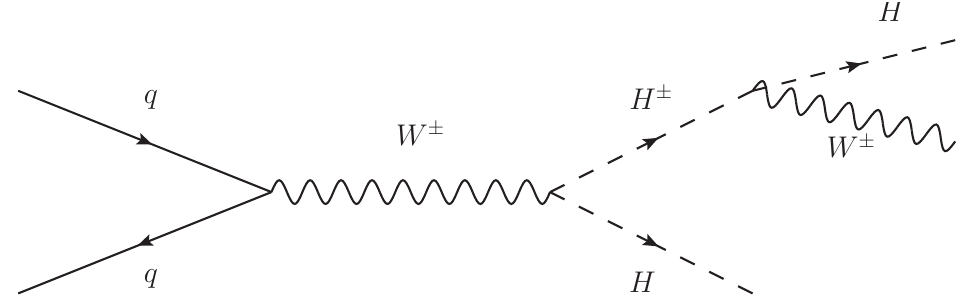}}\\
  \subfloat[]{\label{Feyn:hpma0}\includegraphics[scale=0.261]{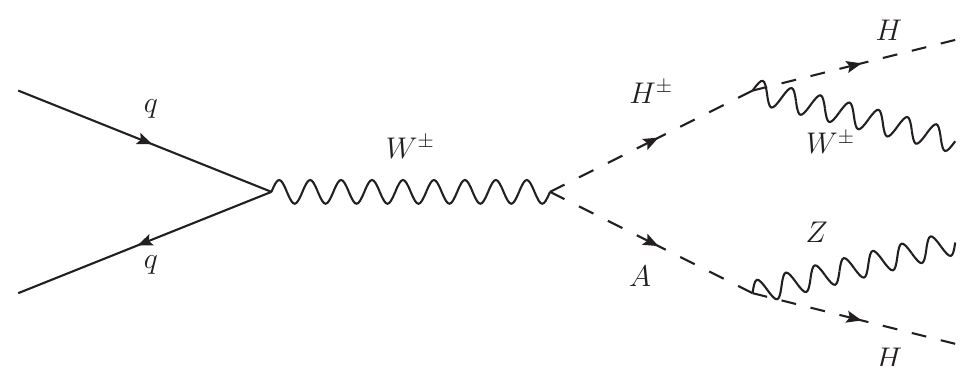}} \hspace{0.cm}
  \subfloat[]{\label{Feyn:hpm}\includegraphics[scale=0.261]{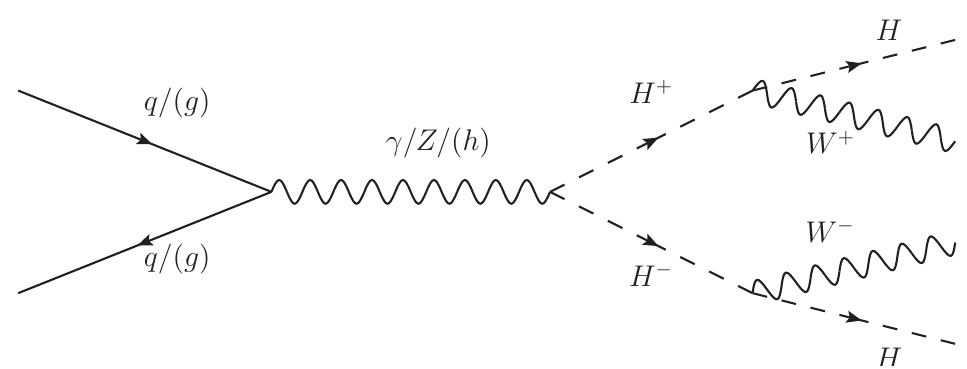}}\\
   \subfloat[]{\label{Feyn:a0a0}\includegraphics[scale=0.261]{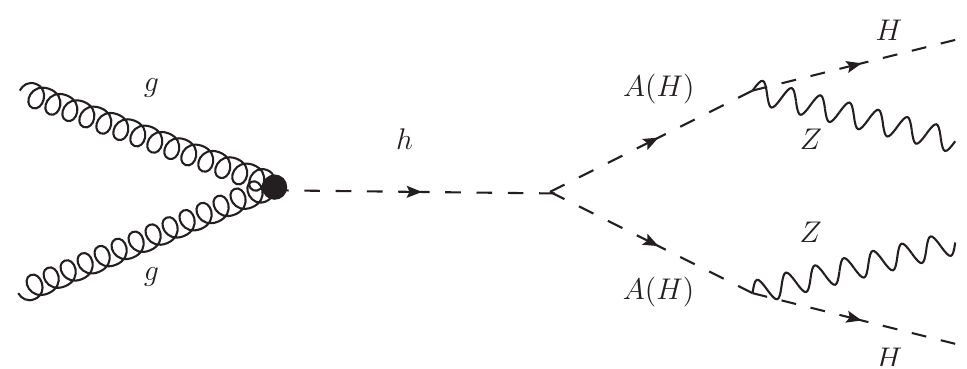}}\\
  \caption{Parton level representative diagrams at LO of (a), (b) associate production of heavy scalar, and (c), (d), (e) pair production of heavy scalars. In our study, we consider one loop correction in $\alpha_S$ of all these diagrams.}
\label{fig:Feyn1}
\end{figure}

We implement the Lagrangian given in \autoref{idmL} together with the leading term of \autoref{eftL} in {\sc FeynRules}~\cite{Alloul:2013bka} and employ {\sc NLOCT}~\cite{Degrande:2014vpa} to generate $UV$ and $R_2$ counterterms of the SM Lagrangian in order to have a NLO UFO model that we use under the {\sc MG5\_aMC@NLO} environment~\cite{Alwall:2014hca}. Inside this environment, real corrections are performed following the FKS subtraction method~\cite{Frixione:1995ms}, whereas OPP technique~\cite{Ossola:2006us} is the one that is being used to take care of the virtual contributions. Nevertheless, for $AA$, $HH$ and $H^+H^-$ pair production processes, gluon-gluon initiated processes mediated by Higgs propagator play a significant role and we insert the corresponding analytic form of the one loop amplitude in {\sc MadGraph5} virtual routine and that in $d=(4-2\epsilon)$ dimension reads as,
\begin{eqnarray} \nonumber
\overline{2\mathcal{R}(M_0 M_v^\dagger)}
&=&\Big(\frac{\alpha_s}{2\pi}\Big)\, \frac{(4\pi)^\epsilon}{\Gamma(1-\epsilon)} \left(\frac{\mu^2}{s_{12}}\right)^{\epsilon} |\overline{M_0}|^2\, \\
&& \Bigg[-\frac{6}{\epsilon^2} - \frac{2\,b_0}{\epsilon} + 11 + 3\pi^2  \Bigg],
\end{eqnarray}
while setting the renormalization scale $\mu^2=s_{12}$, partonic center-of-mass (CM) energy. $\mathcal M_0$ and $\mathcal M_v$ represent tree-level and one-loop amplitudes respectively. The leading term of the QCD $\beta$-function $b_0=\frac{11}{6} C_A - \frac{2}{3} n_f T_F$, where $n_f$ represents the number of active quark flavors and $C_A=3, T_F=1/2$. Note that the strong coupling is renormalized following the $\overline{\rm MS}$ scheme and the ${\cal{O}}(\alpha_s^2)$ term of the Lagrangian given is \autoref{eftL} is taken into account in the above expression. The colour and spin averaged tree level squared amplitude in $d=(4-2\epsilon)$ dimension can be written as,
\begin{eqnarray}
|\overline{M_0}|^2=\frac{1}{128} (1+\epsilon+\epsilon^2) \frac{C_{0}^2\, \Lambda^2\, v^2\, s_{12}^2}{(s_{12}-m_h^2)^2+\Gamma_h^2}.
\end{eqnarray}
Here $C_0=\frac{\alpha_s}{3\pi v}$, $\Gamma_h$ is the Higgs boson width, and $\Lambda$ corresponds to $\Lambda_{L/c/3}$ as given in the Feynman rules furnished in \autoref{appendixFR}. Final state heavy scalar particles are decayed via {\sc MadSpin}~\cite{Artoisenet:2012st} which retains spin information at the tree level accuracy. NLO events thus obtained are then matched to {\sc Pythia8}~\cite{Sjostrand:2001yu, Sjostrand:2014zea} parton shower following the {\sc MC@NLO} formalism~\cite{Frixione:2002ik} to avoid any double counting. For the signal, we use in-built NN23LO1 and NN23NLO PDF sets \cite{Ball:2012cx} for LO and NLO respectively. 
We use {\sc Delphes3}~\cite{deFavereau:2013fsa} to include the detector effects in our simulation, where we use the default card of the CMS. Jets are formed by clustering the particle-flow tower objects and particle-flow tracks. We employ $\mbox{anti-k}_T$ \cite{Cacciari:2008gp} clustering algorithm to form jets, where we have set radius parameter R=0.5. Using {\sc Fastjet 3.2.2}~\cite{Cacciari:2011ma} package, we reconstruct fatjets, utilizing Delphes tower objects as input for clustering. Cambridge-Achen (CA) \cite{Dokshitzer:1997in} algorithm is hired for fatjets clustering where radius parameter is set to R=0.8. Fatjets are characterized by the radius parameter, $R \sim 2m_V/P_T $ $(V\equiv \{W^\pm,Z\})$, where $P_T$ being the transverse momentum and $m_V$ is the mass of the weak boson. We apply minimum $P_T=180$ GeV for each fatjet formation. MVA analysis is done in the {\sc TMVA} framework \cite{Hocker:2007ht}. We implement the Boosted Decision Tree (BDT) algorithm in our MVA analysis. A decision tree splits the high-level input data recursively depending on a set of input features. The method that combines many trees (weak learners) into a strong classifier is called boosting.
\autoref{fig:Feyn1} displays representative LO Feynman diagrams of the associated production of the heavy scalar and pair production of the heavy scalars, which we ultimately decay hadronically. 
%
\begin{table*}[!tb]
\begin{center}
 \scriptsize
 \begin{tabular}{|c|c|c|c|c|c|c|}
\hline
\multirow{2}{1em}{BP} & \multicolumn{3}{| c |}{$\sigma(p p \rightarrow A H)$ (fb)} & \multicolumn{3}{| c |}{ $\sigma(p p \rightarrow H^\pm H)$ (fb)}  \\
\cline{2-7}
& NLO & LO & K-fac & NLO & LO & K-fac  \\
\hline
BP1 & $46.55^{+0.79(1.7\%)}_{-0.65(1.4\%)} \pm 8.0\times 10^{-2} $ & $35.12^{+0.04(0.1\%)}_{-0.14(0.4\%)} \pm 1.3\times 10^{-1}$ & 1.33 & $82.42^{+1.26(1.5\%)}_{-0.98(1.2\%)} \pm 2.0\times 10^{-2}$ & $61.67^{+0.12(0.2\%)}_{-0.37(0.6\%)} \pm 2.2\times 10^{-1}$ & 1.34  \\
\hline
BP2 &$25.34^{+0.40(1.6\%)}_{-0.29(1.1\%)} \pm 4.0\times 10^{-2}$ & $19.01^{+0.17(0.9\%)}_{-0.23(1.2\%)} \pm 7.0\times 10^{-2}$ & 1.33 & $45.27^{+0.87(1.9\%)}_{-0.40(0.9\%)} \pm 8.0\times 10^{-2}$ & $33.77^{+0.37(1.1\%)}_{-0.47(1.4\%)} \pm 1.0\times 10^{-2}$ & 1.34  \\
\hline
BP3 & $15.50^{+0.21(1.4\%)}_{-0.21(1.4\%)} \pm 3.0\times 10^{-2}$ & $11.60^{+0.17(1.5\%)}_{-0.21(1.8\%)} \pm 4.0\times 10^{-2}$ & 1.34 & $27.78^{+0.44(1.6\%)}_{-0.3(1.1\%)} \pm 5.0\times 10^{-2}$ & $20.69^{+0.37(1.8\%)}_{-0.41(2\%)} \pm 8.0\times 10^{-2}$ & 1.34   \\
\hline
BP4 & $9.68^{+0.15(1.6\%)}_{-0.14(1.4\%)}\pm 2.0\times 10^{-2}$ & $7.19^{+0.16(2.2\%)}_{-0.17(2.3\%)}\pm 3.0\times 10^{-2}$ & 1.35 & $17.86^{+0.27(1.5\%)}_{-0.29(1.6\%)} \pm 3.0\times 10^{-2}$ & $13.23^{+0.32(2.4\%)}_{-0.33(2.5\%)} \pm 4.0\times 10^{-2}$ &1.35  \\
\hline
BP5 & $6.32^{+0.11(1.7\%)}_{-0.10(1.6\%)} \pm 1.0\times 10^{-2}$ & $4.67^{+0.13(2.8\%)}_{-0.13(2.8\%)} \pm 2.0\times 10^{-2}$ & 1.35 & $11.34^{+0.2(1.8\%)}_{-0.19(1.7\%)} \pm 2.0\times 10^{-2}$ & $8.39^{+0.25(3\%)}_{-0.25(3\%)}\pm 3.0\times 10^{-2}$ & 1.35   \\
\hline
BP6 & $3.90^{+0.07(1.8\%)}_{-0.07(1.8\%)} \pm 6.5\times10^{-3}$ & $2.87^{+0.10(3.5\%)}_{-0.10(3.5\%)} \pm 1.0\times 10^{-2}$ & 1.36 & $7.19^{+0.14(1.9\%)}_{-0.13(1.8\%)}\pm 1.0\times 10^{-2}$ & $5.33^{+0.19(3.6\%)}_{-0.19(3.6\%)} \pm 2.0\times 10^{-2}$ & 1.35  \\
\hline
BP7 & $2.69^{+0.05(1.9\%)}_{-0.05(1.9\%)} \pm 4.6\times 10^{-3}$ & $1.97^{+0.08(4\%)}_{-0.07(3.6\%)} \pm 7.3\times 10^{-3}$ & 1.37 & $5.01^{+0.1(2\%)}_{-0.1(2\%)} \pm 8.9 \times 10^{-3}$ & $3.68^{+0.15(4\%)}_{-0.15(4\%)}\pm 1.0\times 10^{-2}$ & 1.36 \\
\hline
 \end{tabular}   
\caption{Cross-sections for the associated production of heavy scalar at LO and NLO with integrated K-factor are given in this table at 14 TeV LHC  before the decay of heavy scalars into DM and SM particles. The superscript and subscript denote the scale uncertainties in the total cross-section (the percentages are given in bracket), while the last entry is the Monte Carlo uncertainty. $AH$ and $H^\pm H$ channels are produced in five and four massless quark flavors, respectively.}
\label{tab:crosssection_2}
\end{center}
\end{table*}

\begin{table*}[!tb]
\begin{center}
 \scriptsize
 \begin{tabular}{|c|c|c|c|c|c|c|}
\hline
 \multirow{2}{1em}{BP} & \multicolumn{3}{| c |}{$\sigma(p p \rightarrow H^\pm A)$ (fb)} & \multicolumn{3}{| c |}{ $\sigma(p p \rightarrow H^+ H^-)$ (fb)} \\
\cline{2-7}
& NLO & LO & K-fac & NLO & LO & K-fac \\
\hline
BP1  & $16.93^{+0.28(1.7\%)}_{-0.25(1.5\%)}\pm 3.0\times 10^{-2}$ & $12.53^{+0.34(2.7\%)}_{-0.34(2.7\%)} \pm 5.0\times 10^{-2}$ & 1.35 & $11.01^{+0.43(3.9\%)}_{-0.42(3.8\%)}\pm 2.0\times 10^{-2}$ & $7.98^{+0.48(6\%)}_{-0.4(5\%)} \pm 2.0\times 10^{-2}$ & 1.38 \\
\hline
BP2  &  $8.41^{+0.14(1.7\%)}_{-0.18(2.1\%)} \pm 2.0\times 10^{-2}$ & $6.21^{+0.23(3.7\%)}_{-0.22(3.5\%)}\pm 3.0\times 10^{-2}$ & 1.35  & $6.01^{+0.37(6.2\%)}_{-0.32(5.3\%)}\pm 8.5\times 10^{-3}$ & $4.17^{+0.37(8.9\%)}_{-0.29(7\%)}\pm 1.0\times 10^{-2}$ & 1.44 \\
\hline
BP3  & $4.78^{+0.1(2.1\%)}_{-0.1(2.1\%)}\pm8.3\times10^{-3}$ & $3.48^{+0.15(4.3\%)}_{-0.14(4\%)} \pm 1.0\times 10^{-2}$ & 1.37  & $3.76^{+0.29(7.7\%)}_{-0.24(6.4\%)}\pm 5.8 \times 10^{-3}$ & $2.57^{+0.29(11.3\%)}_{-0.23(8.9\%)}\pm 7.6\times 10^{-3}$ & 1.46 \\
\hline
BP4 &  $2.81^{+0.06(2.1\%)}_{-0.06(2.1\%)} \pm 5.0\times 10^{-3}$ & $2.04^{+0.1(4.9\%)}_{-0.1(4.9\%)} \pm 7.8 \times 10^{-3}$ & 1.38  & $2.50^{+0.23(9.2\%)}_{-0.19(7.6\%)} \pm 4.1\times10^{-3}$ & $1.68^{+0.24(14.3\%)}_{-0.18(10.7\%)} \pm 5.3\times 10^{-3}$ & 1.49\\
\hline
BP5  & $1.69^{+0.04(2.4\%)}_{-0.04(2.4\%)}\pm 3.0\times 10^{-3}$ & $1.22^{+0.07(5.7\%)}_{-0.06(4.9\%)}\pm 4.7\times 10^{-3}$ & 1.38  & $1.68^{+0.18(10.7\%)}_{-0.15(8.9\%)} \pm 3.0\times10^{-3}$ & $1.10^{+0.19(17.3\%)}_{-0.14(12.7\%)} \pm 3.1\times 10^{-3}$  & 1.52\\
\hline
BP6  & $0.97^{+0.02(2.1\%)}_{-0.02(2.1\%)}\pm 1.7\times 10^{-3}$ & $0.70^{+0.05(7.1\%)}_{-0.04(5.7\%)} \pm 2.4\times 10^{-3}$ & 1.38 & $1.14^{+0.14(12.3\%)}_{-0.12(10.5\%)} \pm 2.4\times 10^{-3}$ & $0.74^{+0.15(20.3\%)}_{-0.11(14.9\%)} \pm 2.2\times 10^{-3}$ & 1.54\\
\hline
BP7  & $0.63^{+0.018(2.8\%)}_{-0.018(2.8\%)} \pm 1.1\times 10^{-3}$ & $0.45^{+0.03(6.7\%)}_{-0.03(6.7\%)} \pm 1.6 \times 10^{-3}$ & 1.39 & $0.85^{+0.11(12.9\%)}_{-0.09(10.6\%)} \pm 1.9\times 10^{-3}$ & $0.55^{+0.13(23.6\%)}_{-0.09(16.4\%)} \pm 1.7 \times 10^{-3}$ & 1.56\\
\hline
 \end{tabular}  
 \begin{tabular}{|c|c|c|c|}
\hline
\multirow{2}{1em}{BP} & \multicolumn{3}{| c |}{ $\sigma(p p \rightarrow A A)$ (fb)} \\
\cline{2-4}
&  NLO & LO & K-fac  \\
\hline
BP1  & $0.88^{+0.18(20.4\%)}_{-0.14(15.9\%)} \pm 3.3\times 10^{-3}$ & $0.46^{+0.15(32.6\%)}_{-0.11(23.9\%)} \pm 1.6\times 10^{-3}$ & 1.92 \\
\hline
BP2  & $0.72^{+0.14(19.4\%)}_{-0.12(16.7\%)} \pm 2.6\times10^{-3}$ & $0.38^{+0.13(34.2\%)}_{-0.09(23.7\%)} \pm 1.4\times 10^{-3}$ & 1.87 \\
\hline
BP3  & $0.59^{+0.12(20.3\%)}_{-0.10(16.9\%)} \pm 2.2\times10^{-3}$ & $0.32^{+0.11(34.3\%)}_{-0.08(25\%)} \pm 1.1\times10^{-3}$ & 1.86\\
\hline
BP4  & $0.48^{+0.1(20.8\%)}_{-0.08(16.7\%)} \pm 1.7\times10^{-3}$ & $0.27^{+0.09(33.3\%)}_{-0.07(25.9\%)} \pm 9.6\times10^{-4}$ & 1.80 \\
\hline
BP5  & $0.40^{+0.08(20\%)}_{-0.07(17.5)} \pm 1.7\times10^{-3}$ & $0.22^{+0.08(36.4\%)}_{-0.05(22.7\%)} \pm 8\times10^{-4}$ & 1.78\\
\hline
BP6  & $0.31^{+0.06(19.4\%)}_{-0.05(16.1\%)} \pm 1.1\times10^{-3}$ & $0.18^{+0.06(33.3\%)}_{-0.05(27.8\%)} \pm 6.5\times10^{-4}$ & 1.75 \\
\hline
BP7  & $0.26^{+0.05(19.2\%)}_{-0.05(19.2\%)} \pm 9.3\times 10^{-4} $ & $0.15^{+0.05(33.3\%)}_{-0.04(26.7\%)} \pm 5.4\times10^{-4}$ & 1.70\\
\hline
 \end{tabular} 
\caption{Same as \autoref{tab:crosssection_2} but for the pair production of the heavy scalars. $H^\pm A$ is produced in four massless quark flavors, while $H^+ H^-$ and $AA$ channels are produced in five massless quark flavors.}
\label{tab:crosssection_3}
\end{center}
\end{table*}

\begin{figure}[t]
\centering
\subfloat{\includegraphics[scale=0.56]{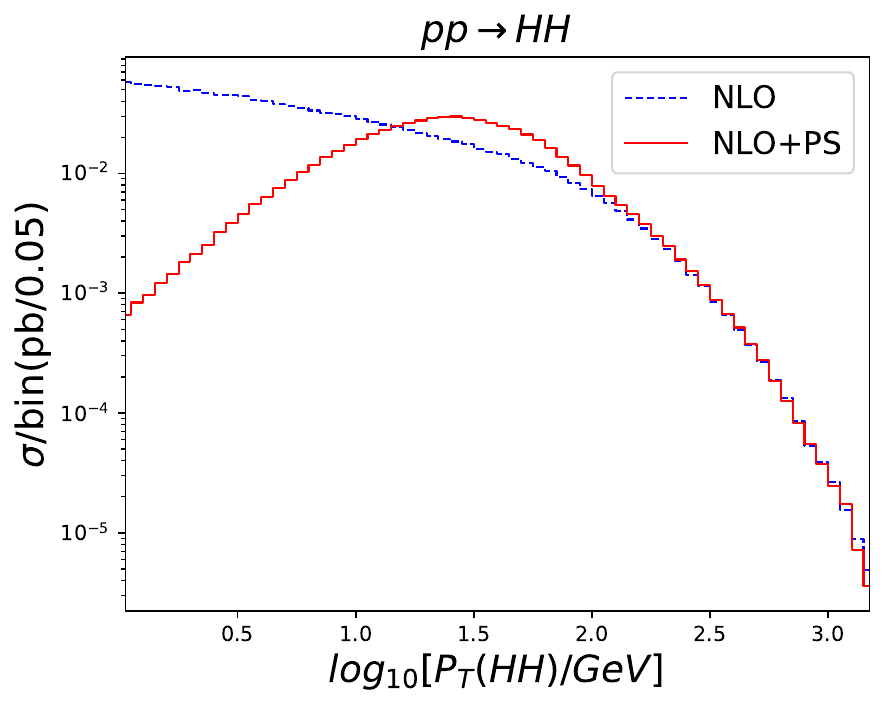}}\\
  \caption{Differential distribution of the total transverse momentum of the DM pair for the channel $pp\rightarrow HH$ at fixed order NLO (dashed blue) and NLO+PS (solid red) accuracy.}
  \label{fig:sud_h2h2}
\end{figure}

\begin{figure*}[t]
\centering
  \subfloat[]{\label{fig:inv_a0h2}\includegraphics[scale=0.50]{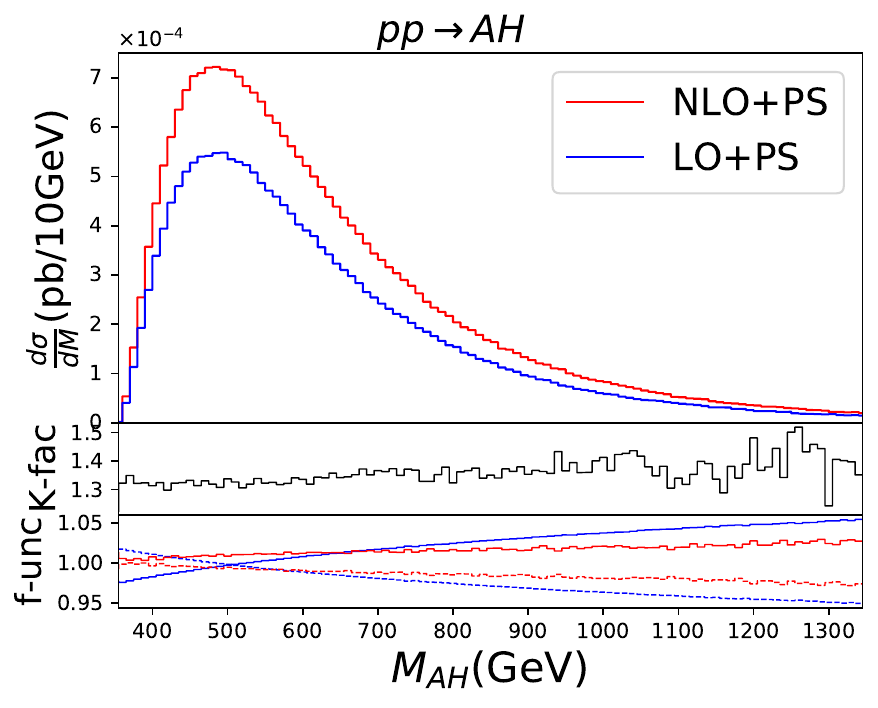}} \hspace{0.4cm}
  \subfloat[]{\label{fig:sud_a0h2}\includegraphics[scale=0.50]{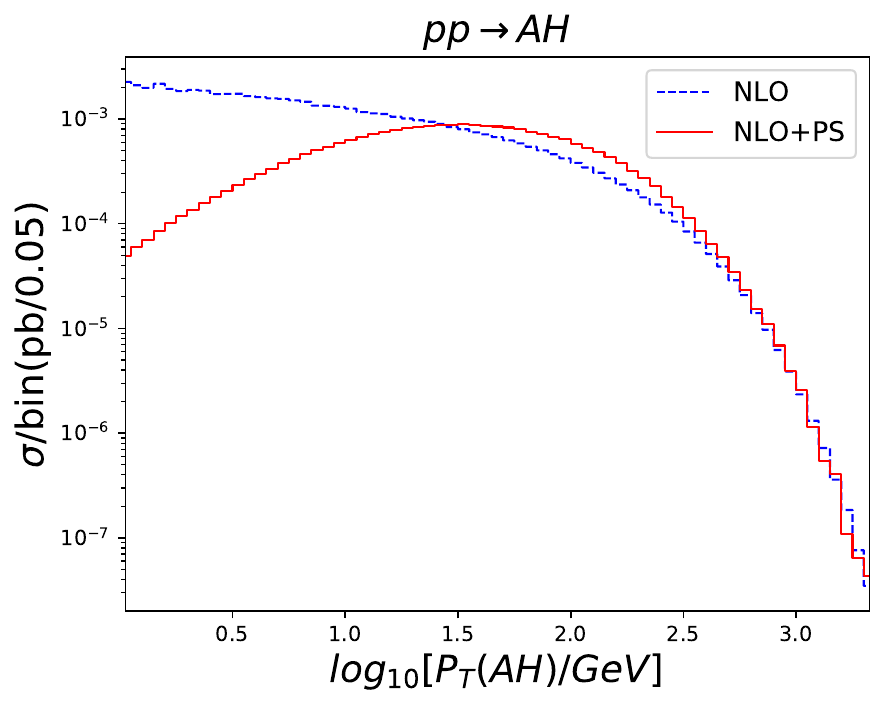}}\\
    \subfloat[]{\label{fig:inv_hpmh2}\includegraphics[scale=0.50]{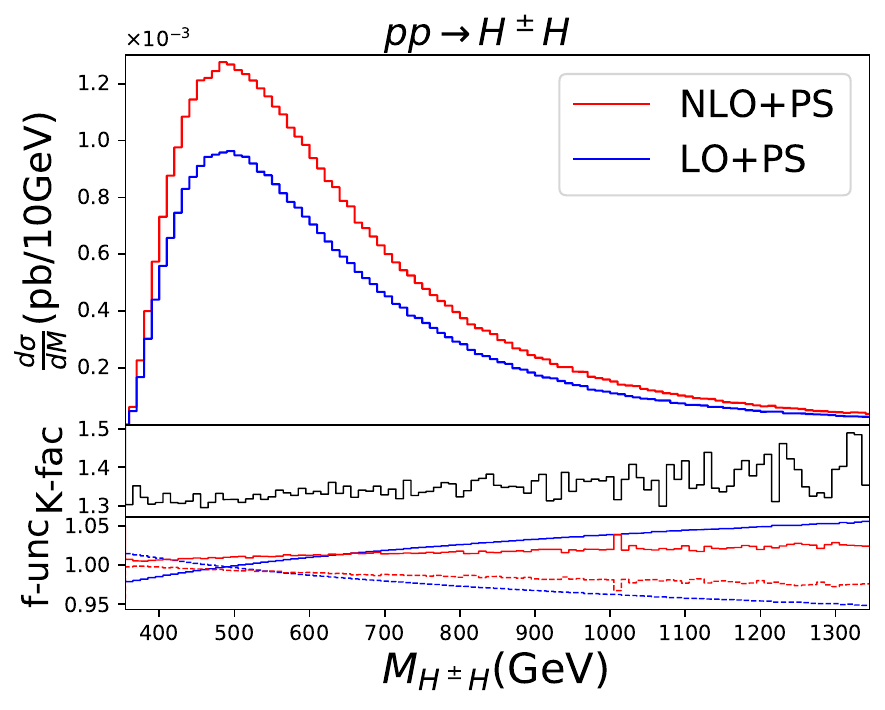}} \hspace{0.4cm}
  \subfloat[]{\label{fig:sud_hpmh2}\includegraphics[scale=0.50]{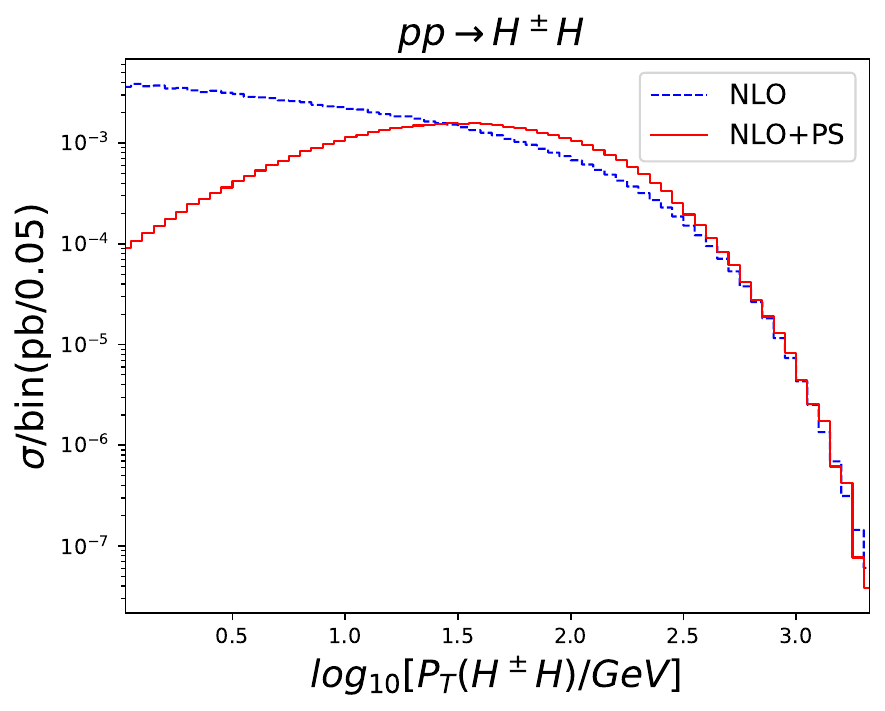}}\\ 
  \caption{NLO effects on the associated production of heavy scalar channels, such as, $pp\rightarrow AH$ (subfig.[a], [b]), $pp\rightarrow H^\pm H$(subfig.[c], [d]). 
In each plot of the left panel, the top subplot shows the invariant mass distribution of the heavy scalar and DM pair at $\text{LO}+\text{PS}$ (dashed blue) and $\text{NLO}+\text{PS}$ (solid red) accuracy. The middle subplot displays the differential NLO K factor, the ratio of the $\text{NLO}+\text{PS}$ cross-section to the $\text{LO}+\text{PS}$ one in each bin, while the bottom subplot presents the scale uncertainties for $\text{LO}+\text{PS}$ (blue) and $\text{NLO}+\text{PS}$ (red).  
The right panel shows the differential distribution of the total transverse momentum of the heavy scalar and DM pair for the respective channel at fixed order NLO (dashed blue) and NLO+PS (solid red) accuracy. All distributions are given for sample benchmark point BP2.}
\label{fig:3_1}
\end{figure*}

Production cross-sections for these channels before hadronic decay of the heavy scalars are given in \autoref{tab:crosssection_2}, and \autoref{tab:crosssection_3} at 14 TeV LHC. We choose the renormalization scale and the factorization scale as $\mu_R=\zeta_R \sqrt{s_{12}}$ and $\mu_F=\zeta_F \sqrt{s_{12}}$ respectively, where $\zeta_R=\zeta_F=1$ represents the central scale choice. We vary $\zeta_R,\zeta_F=\{1/2,1,2\}$, which has a total of nine datasets. All the cross-sections are given corresponding to the central scale where superscripts and subscripts denote the envelope of those nine scale choices. The Monte Carlo uncertainties are also given in those tables. We get reduced scale uncertainty in the total cross-section at NLO than LO for both the associated and pair production processes, except in a few benchmark points for the associated production processes and the reason could be the cross-over of the envelopes around the maximum differential LO cross-section, unlike NLO (see Fig.\autoref{fig:inv_a0h2}: bottom, \autoref{fig:inv_hpmh2}: bottom). Fractional scale uncertainty is defined as the envelope of the ratios of the differential cross-sections at eight additional ($\zeta_R,\zeta_F$) choices to the central one. Dashed and solid lines in the fractional scale uncertainty subplot correspond to the lower and upper envelope respectively. Our study includes one order in $\alpha_S$ corrections to all these channels. Cross-section of $p p \rightarrow HH$ channel at LO is 0.332 pb and at NLO it is 0.617 pb ({\em i.e,} K factor = 1.858) at 14 TeV LHC, independent of benchmark points since cross-section depends only on $m_{H}$  and $\lambda_L$, and both remain same for chosen benchmark points. This channel has a larger cross-section than any other pair or associated production channels because of being less s-channel suppressed due to the presence of an on-shell Higgs boson mediator. Total transverse momentum distribution of the DM pair for BP2 of the channel, $pp\rightarrow HH$ is shown in \autoref{fig:sud_h2h2} for fixed order NLO (dashed blue) and NLO matched with parton-shower (solid red). It is clear from this figure that NLO+PS describes the low $P_T$ region more vividly compared to a fixed order estimation. Note that, although such calculation is essential for traditional mono-jet search, possible contributions of $pp \rightarrow HH$ can only come in our di-fatjet study at the NNLO level. Characteristically, this process is background like and we find that much of the events will not pass the event selection criteria even while starting from a reasonably significant contribution. This channel is shown here for completeness, but we would not add such a contribution to our conservative estimate. 

\begin{figure*}[t]
\centering
  \subfloat[]{\label{fig:inv_hpma0}\includegraphics[scale=0.48]{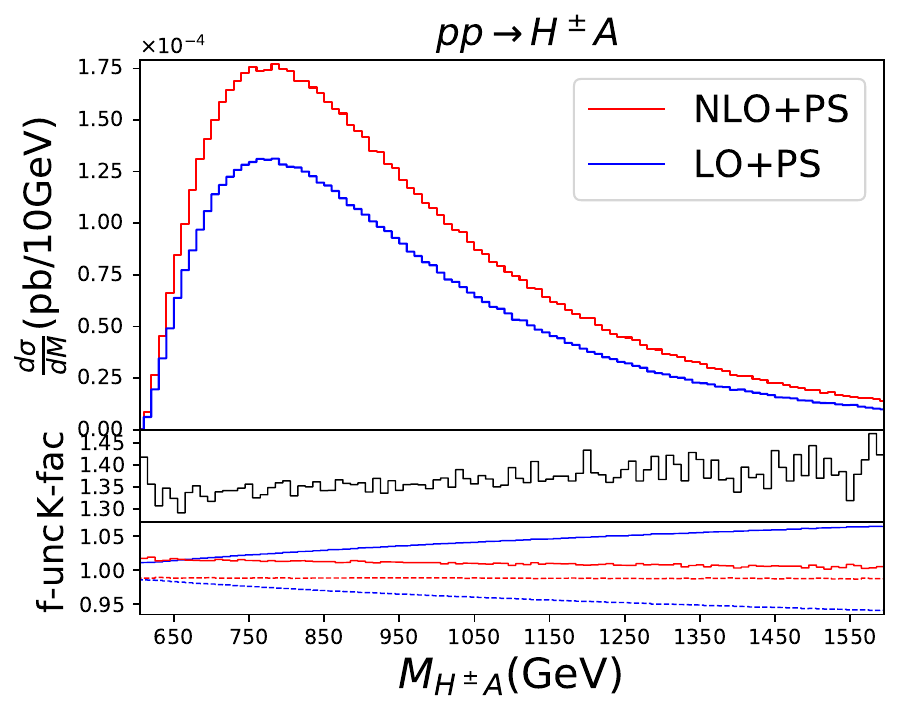}} \hspace{0.4cm}
  \subfloat[]{\label{fig:sud_hpma0}\includegraphics[scale=0.48]{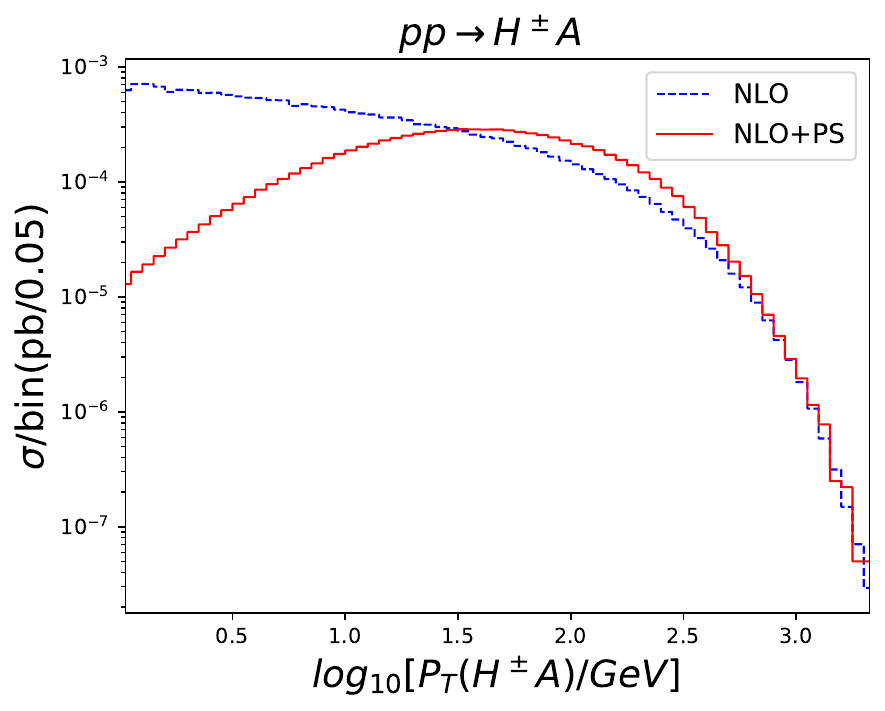}} \\
  \subfloat[]{\label{fig:inv_hpm}\includegraphics[scale=0.48]{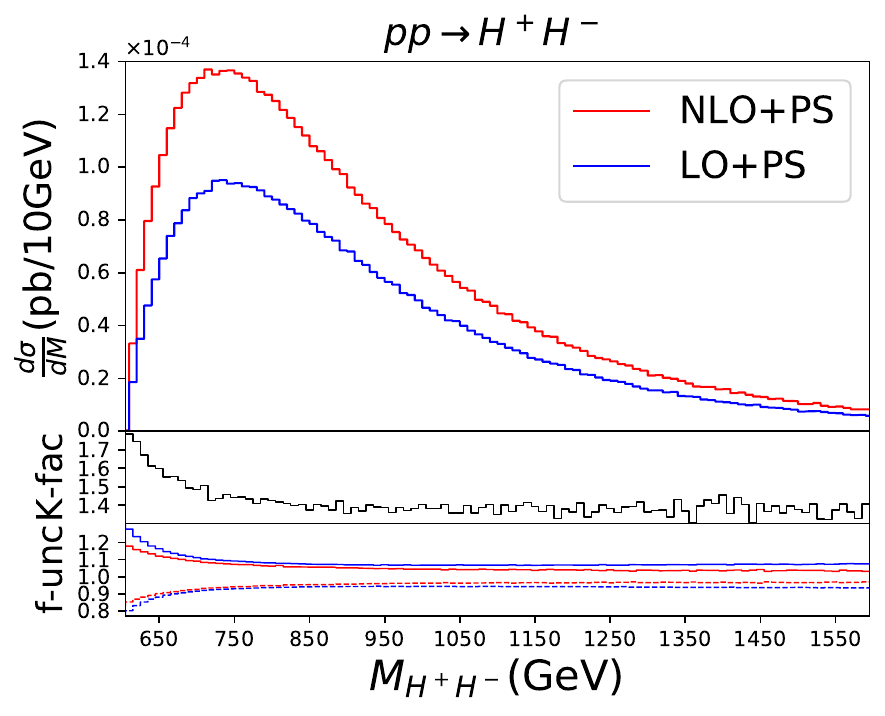}} \hspace{0.4cm}
  \subfloat[]{\label{fig:sud_hpm}\includegraphics[scale=0.48]{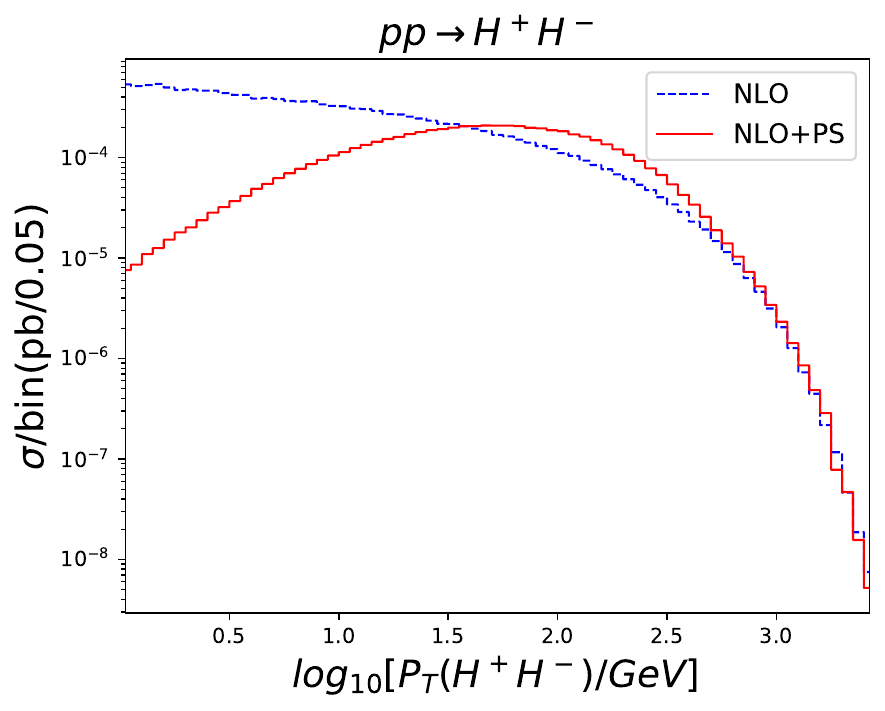}} \\
   \subfloat[]{\label{fig:inv_a0a0}\includegraphics[scale=0.48]{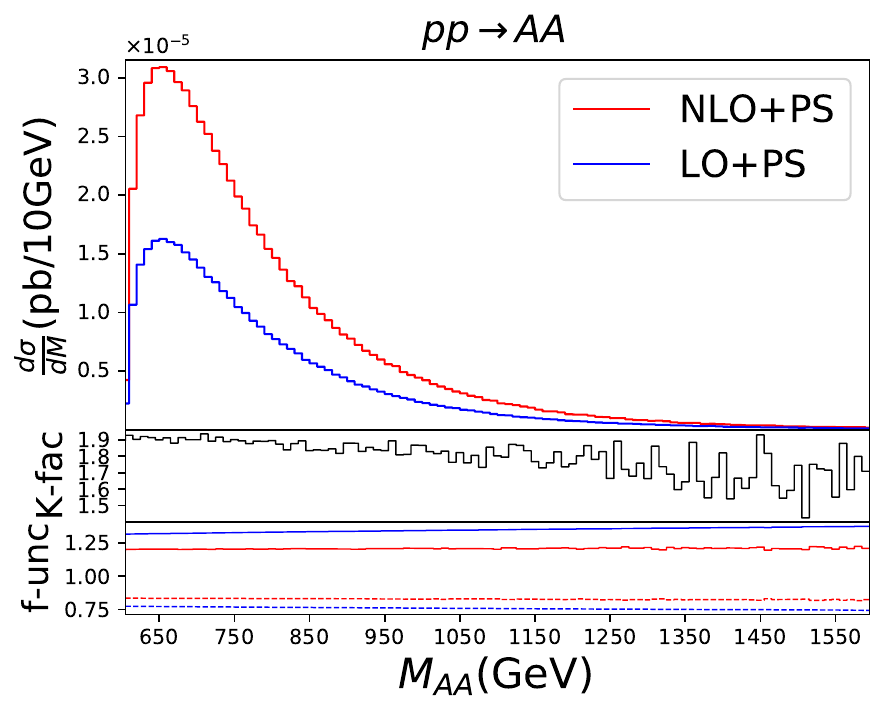}} \hspace{0.4cm}
  \subfloat[]{\label{fig:sud_a0a0}\includegraphics[scale=0.48]{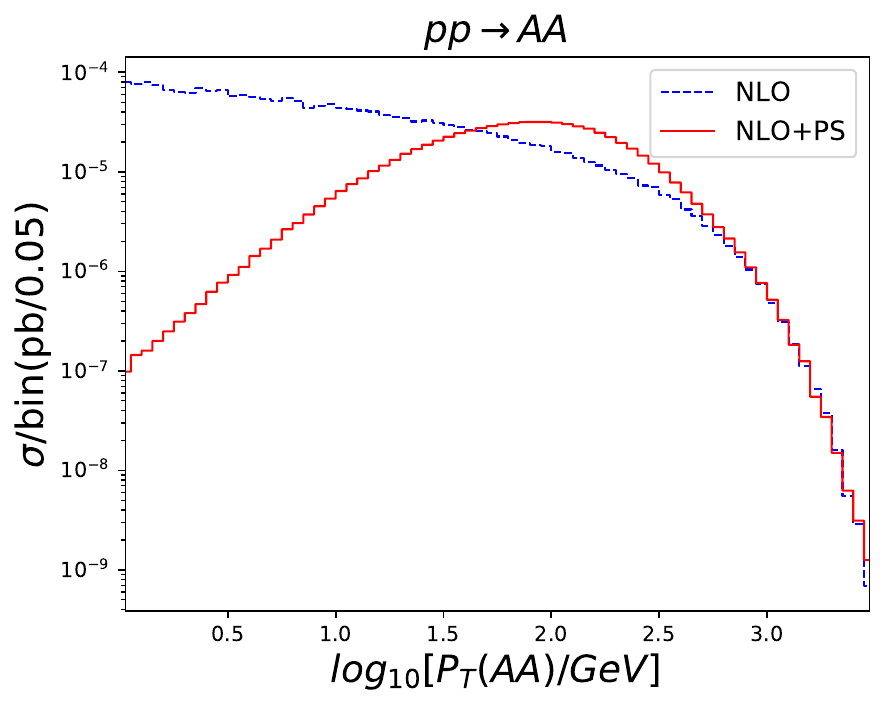}}\\
  \caption{Kinematic variables in the left and right panels are same as in \autoref{fig:3_1}, but these are for the pair production of the heavy scalar channels, such as, $pp\rightarrow H^\pm A$ (subfig.[a], [b]), $pp\rightarrow H^+ H^-$ (subfig.[c], [d]), and $pp\rightarrow A A$ (subfig.[e], [f]).
  }
\label{fig:3_3}
\end{figure*}

In the subsequent figures, on the left panel, we show the improvement in NLO+PS results over the LO+PS ones on the invariant mass distribution (top) along with differential K-factor (middle) and fractional scale uncertainties (bottom) for all remaining production channels. Differential K-factor is vital in extracting correct signal efficiency, as most collider analyses usually do not cover the entire phase space and apply various kinematical cuts to distinguish signal from the background. Fractional scale uncertainty denotes how stable the NLO result is as compared to the LO under scale variation. On the right panel, Sudakov suppression due to NLO+PS computation is explicitly shown for each corresponding channel and that ensures resummation of large logarithm terms in the low $P_T$ region because of incorporating parton shower effect on top of the fixed order calculation. Note that, in these sets of representative figures, hadronic decays of final state heavy scalars are not considered for the time being. \autoref{fig:3_1} collects all the associated production channels of heavy scalars, whereas \autoref{fig:3_3} contains various pair production channels of heavy scalars. In all these figures, BP2 is considered as the representative benchmark point. 
The invariant mass distributions for the associated production channels peak around the same region, close to 485 GeV for both $pp\to AH$ (\autoref{fig:inv_a0h2}: top) and $pp\to H^\pm H$ (\autoref{fig:inv_hpmh2}: top). 
However, among the pair production channels, vector boson mediated processes {\em viz.} $pp\to H^\pm A$ (\autoref{fig:inv_hpma0}: top) and $pp\to H^+ H^-$ (\autoref{fig:inv_hpm}: top) peak around 785 GeV and 730 GeV respectively, but the peak for the other one {\em i.e,} $pp\to AA$ (\autoref{fig:inv_a0a0}: top) occurs near to 650 GeV which is solely scalar mediated. This indicates that the final state particles coming from the associated production processes would be softer compared to the pair production processes. K-factor varies substantially, and in some kinematic regions, it indicates correction up to 90\% .  
Nature of scale uncertainties for associated production processes are quite similar. Among pair production processes, fractional scale uncertainties for $pp\to H^+H^-$ (\autoref{fig:inv_hpm}: bottom) and $pp\to AA$ (\autoref{fig:inv_a0a0}: bottom) are mostly stable in the high invariant mass region, whereas for $pp\to H^\pm A$ (\autoref{fig:inv_hpma0}: bottom) such uncertainties are monotonically increasing. Although these results are metaphorical as hadronic decay of the final state heavy scalars are not being considered here, they show the importance of doing ${\cal O}(\alpha_s)$ corrections to all the production channels to have better estimation of production rate and reduced scale uncertainty. 

\section{QCD jets from heavy scalar decay}
\label{sec:result2}

\begin{figure*}[t]
\centering
  \subfloat[]{\label{fig:m_jo_a0h2}       \includegraphics[scale=0.22]{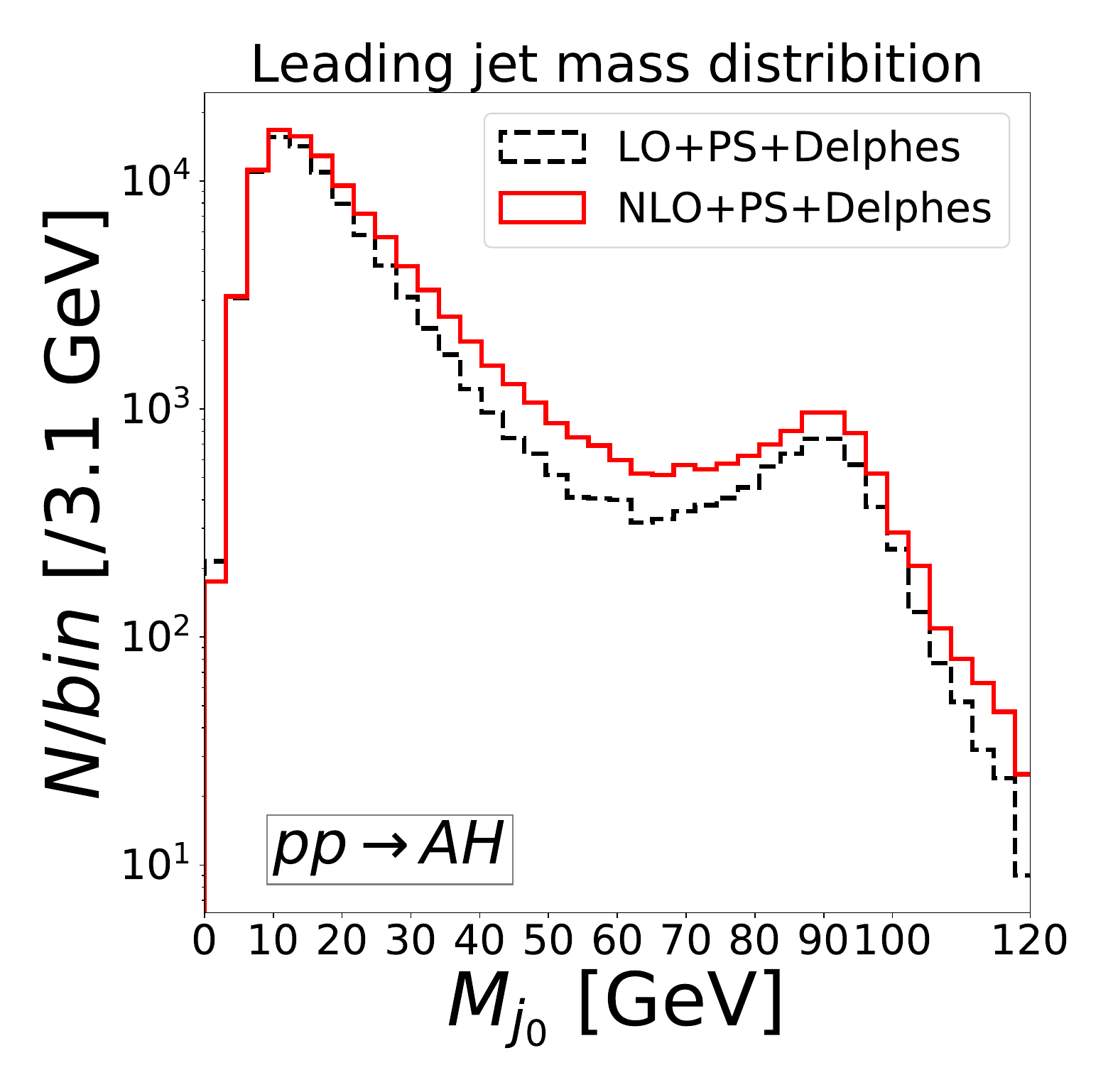}} \hspace{0.1cm}
  \subfloat[]{\label{fig:m_j1_a0h2}       \includegraphics[scale=0.22]{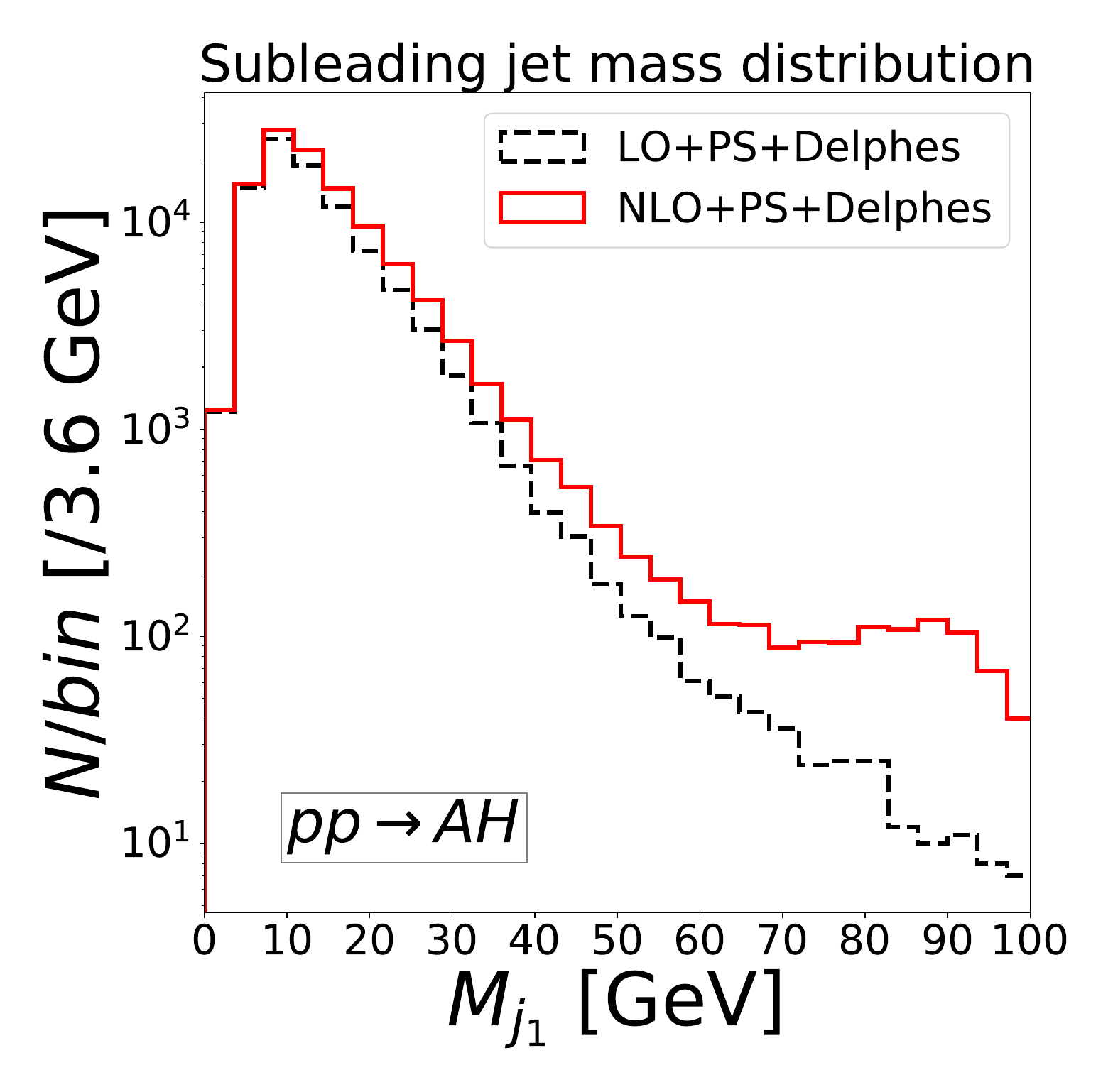}} \hspace{0.1cm}
  \subfloat[]{\label{fig:delR_joj1_a0h2}\includegraphics[scale=0.22]{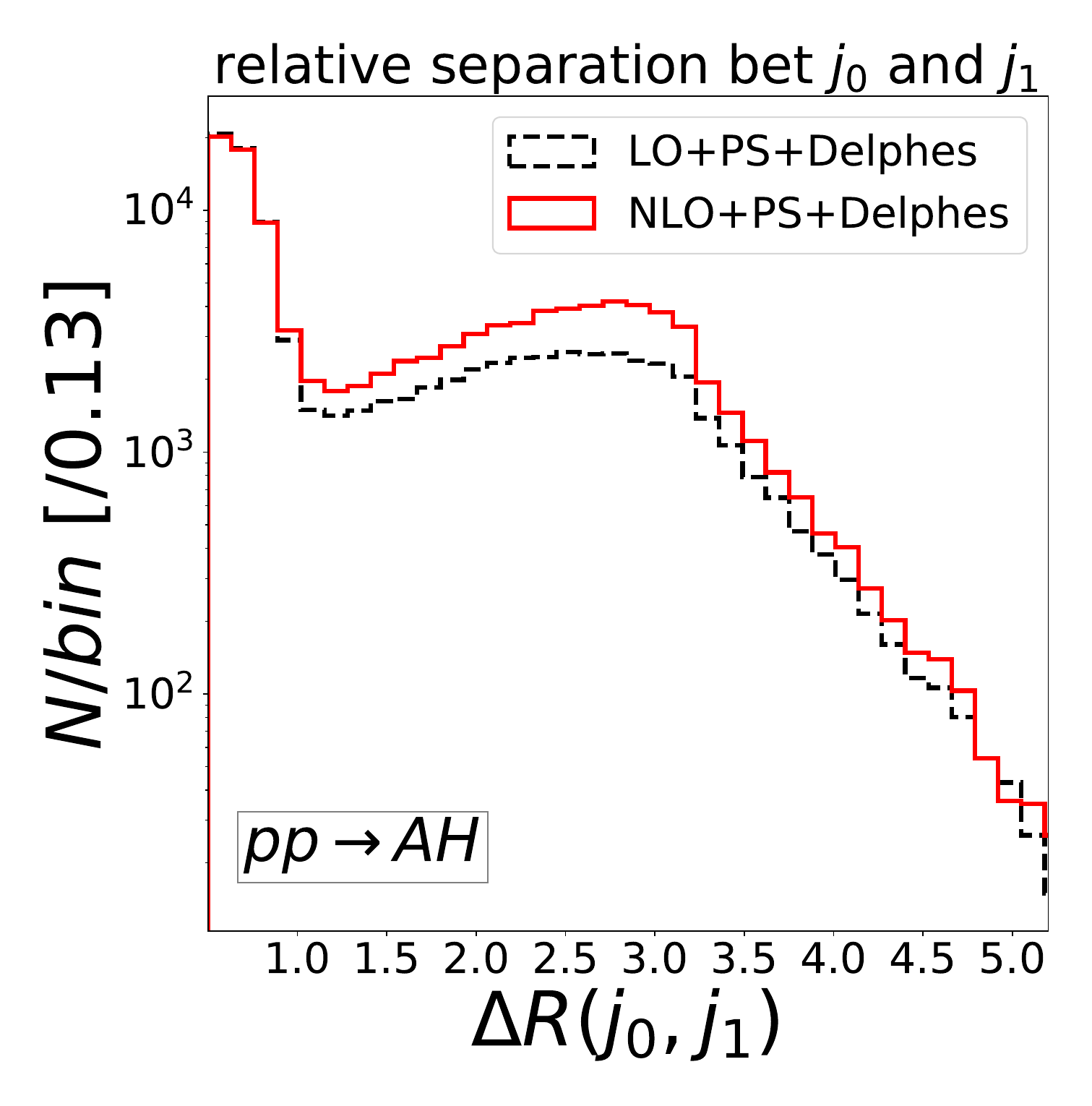}} \\
  \subfloat[]{\label{fig:pt_j0_a0h2}       \includegraphics[scale=0.215]{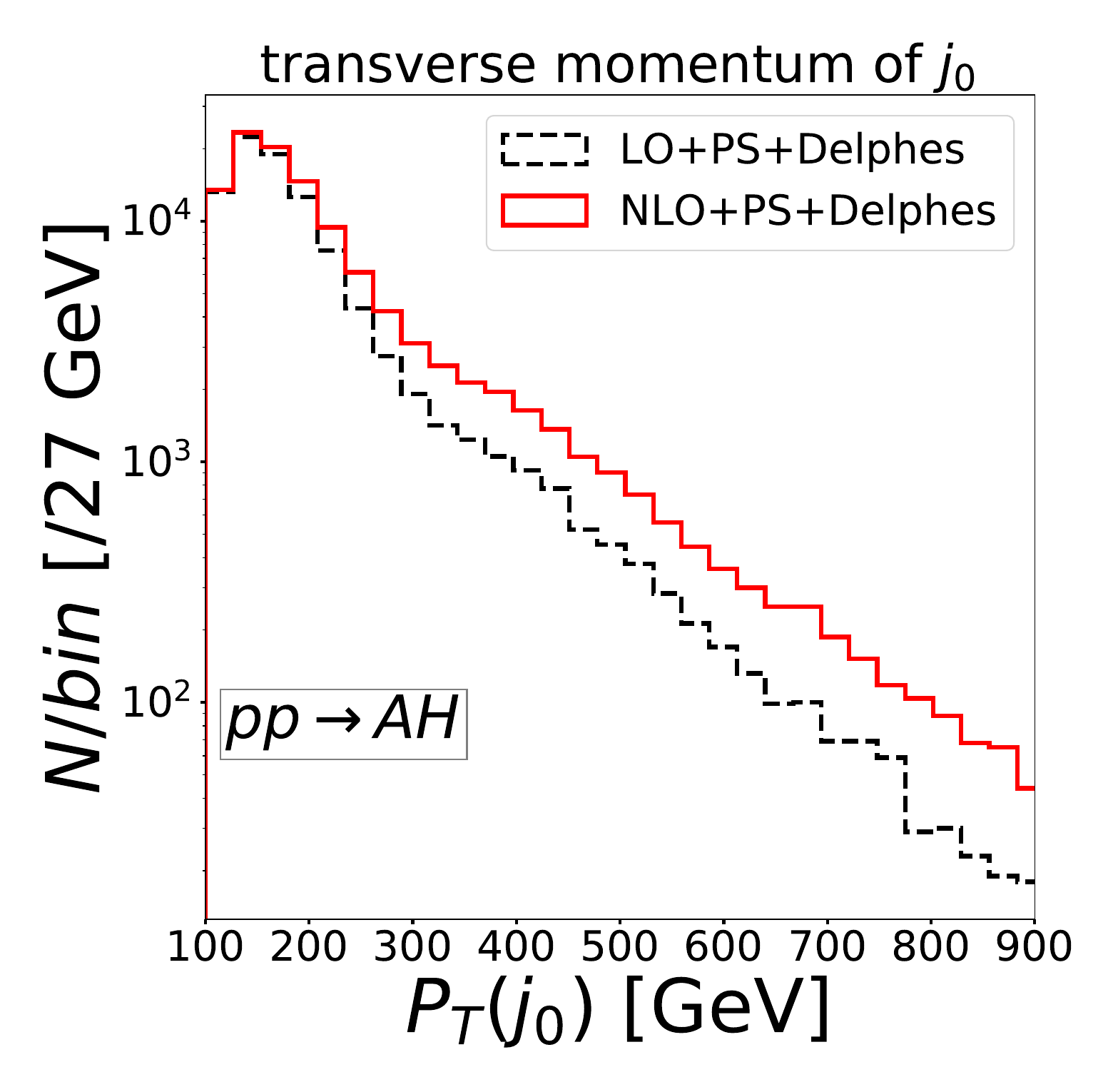}} \hspace{0.1cm}
   \subfloat[]{\label{fig:pt_j1_a0h2}      \includegraphics[scale=0.215]{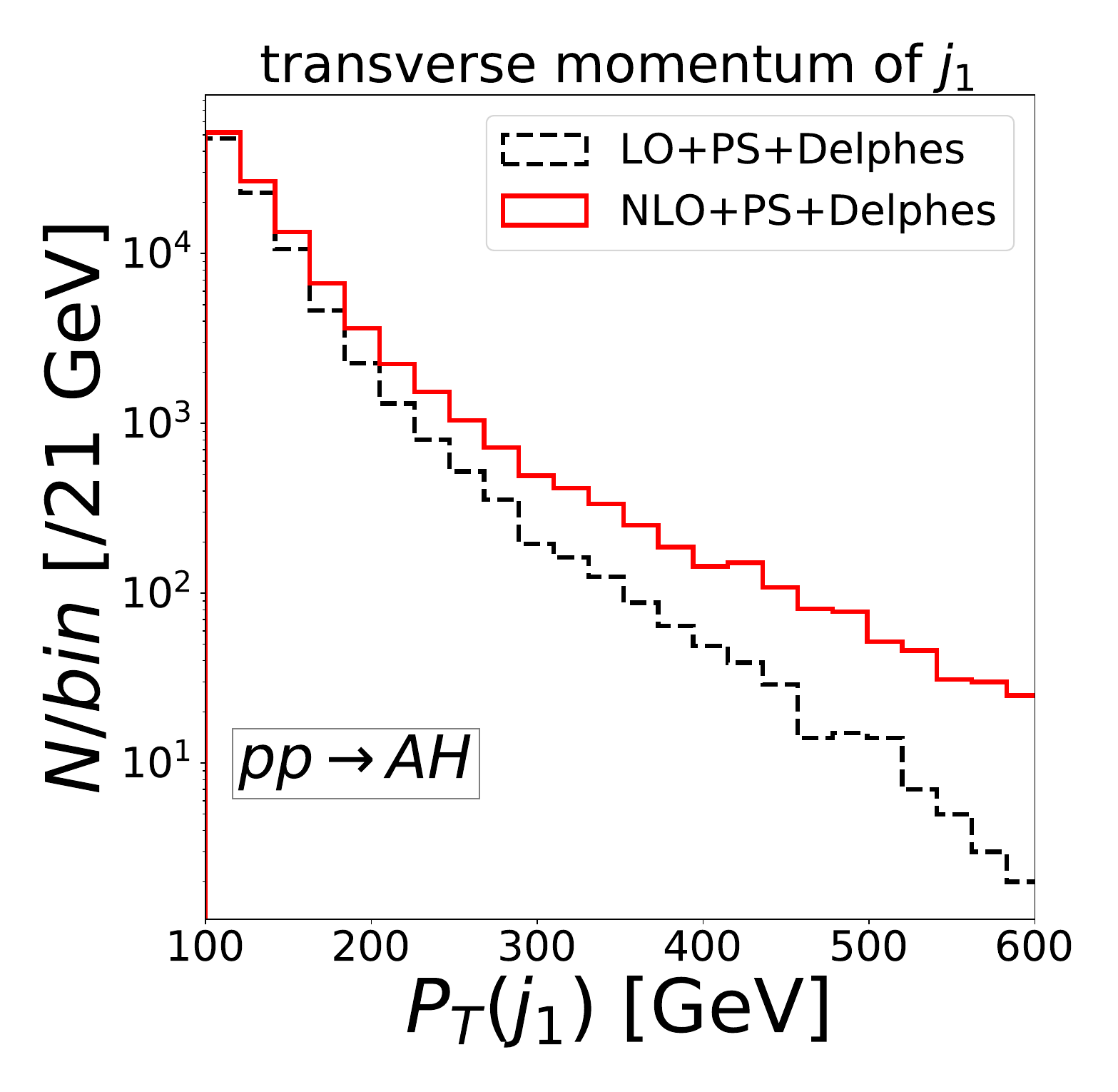}} \hspace{0.1cm}
   \subfloat[]{\label{fig:MET_a0h2}      \includegraphics[scale=0.215]{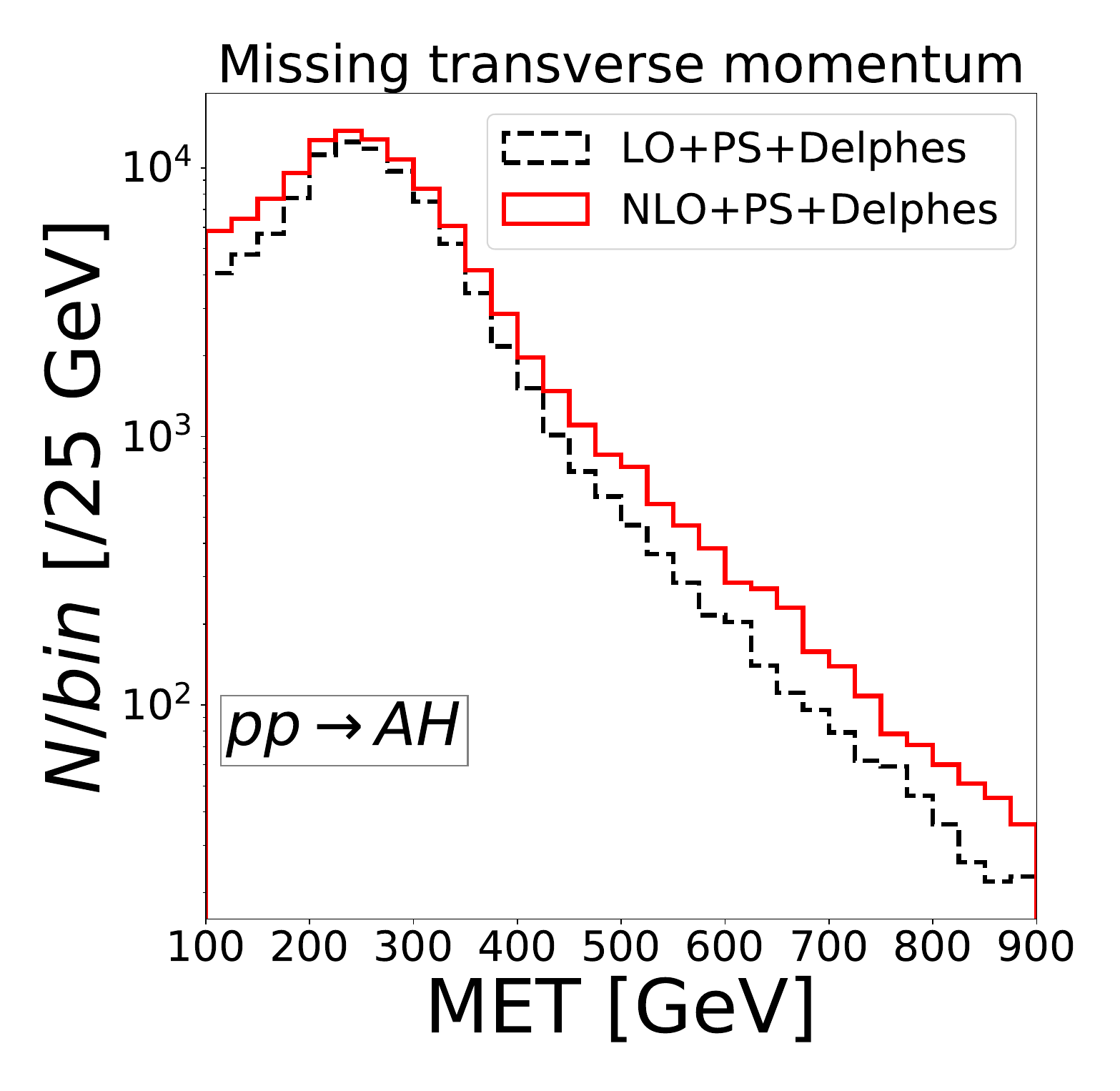}}\\
  \caption{Distributions of the various kinematic observables at LO (dashed black) and NLO (solid red) for the selected events with $\slashed{E}_T,\, P_T(j_0), P_T(j_1)>100 \, GeV $ from the channel $pp\rightarrow AH$, where $A$ decay hadronically. This demonstration is for the benchmark point BP2. Plots (a) and (b) show distributions of the leading ($j_0$) and subleading ($j_1$) jet mass ($M_{j_0}, \, M_{j_1}$) respectively, (c) is the distribution of the relative separation between these two leading jets $\Delta R(j_0,j_1)$, while (d) and (e) are transverse momentum distribution of $j_0$ and $j_1$ respectively. Plot (f) shows the distribution of the total missing transverse energy; here the label MET represents $\slashed{E}_T$.}
\label{fig:4_1}
\end{figure*}

\begin{figure*}[t]
\centering
  \subfloat[]{\label{fig:m_jo_hpma0}       \includegraphics[scale=0.215]{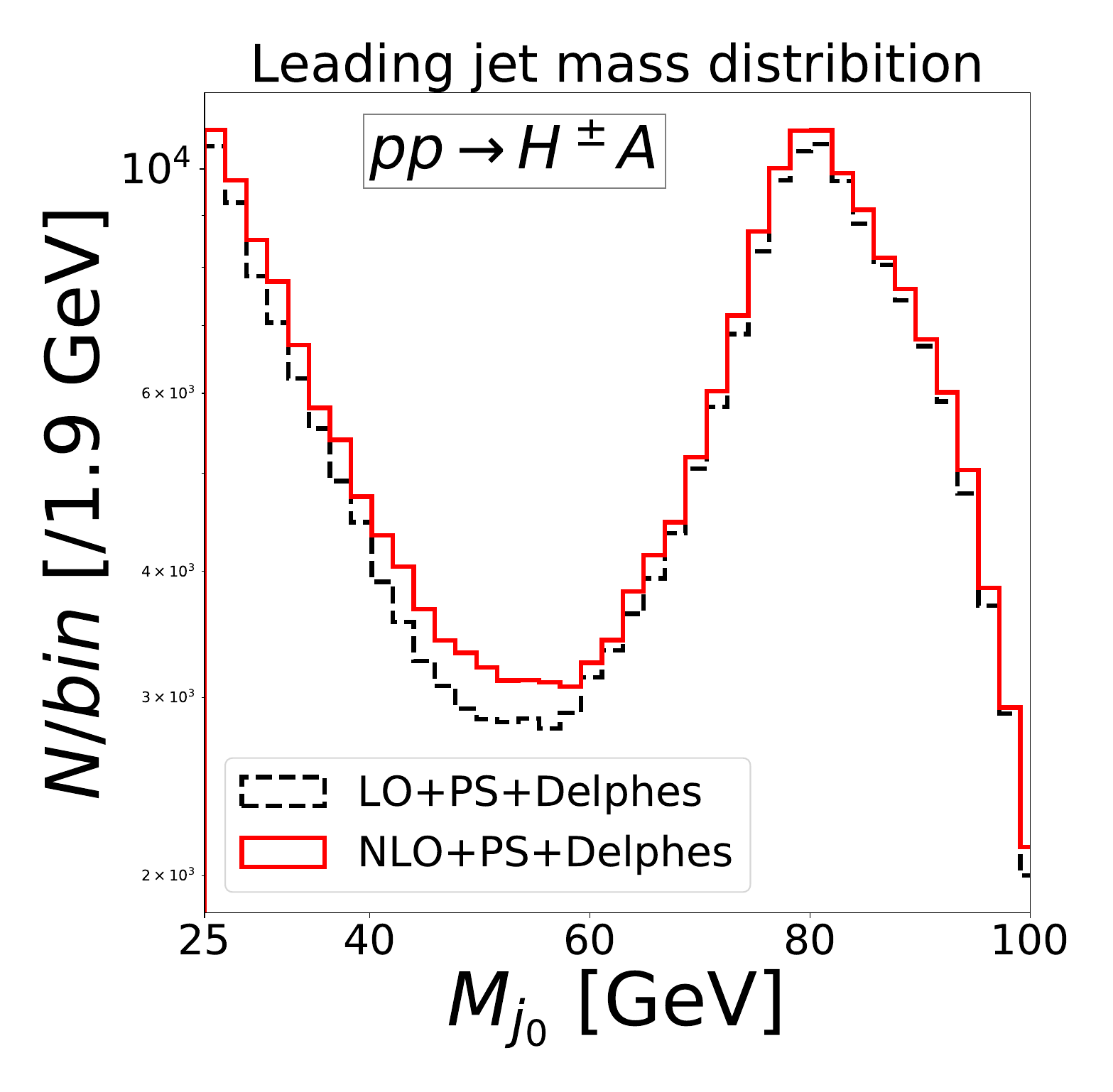}} \hspace{0.1cm}
  \subfloat[]{\label{fig:m_j1_hpma0}       \includegraphics[scale=0.215]{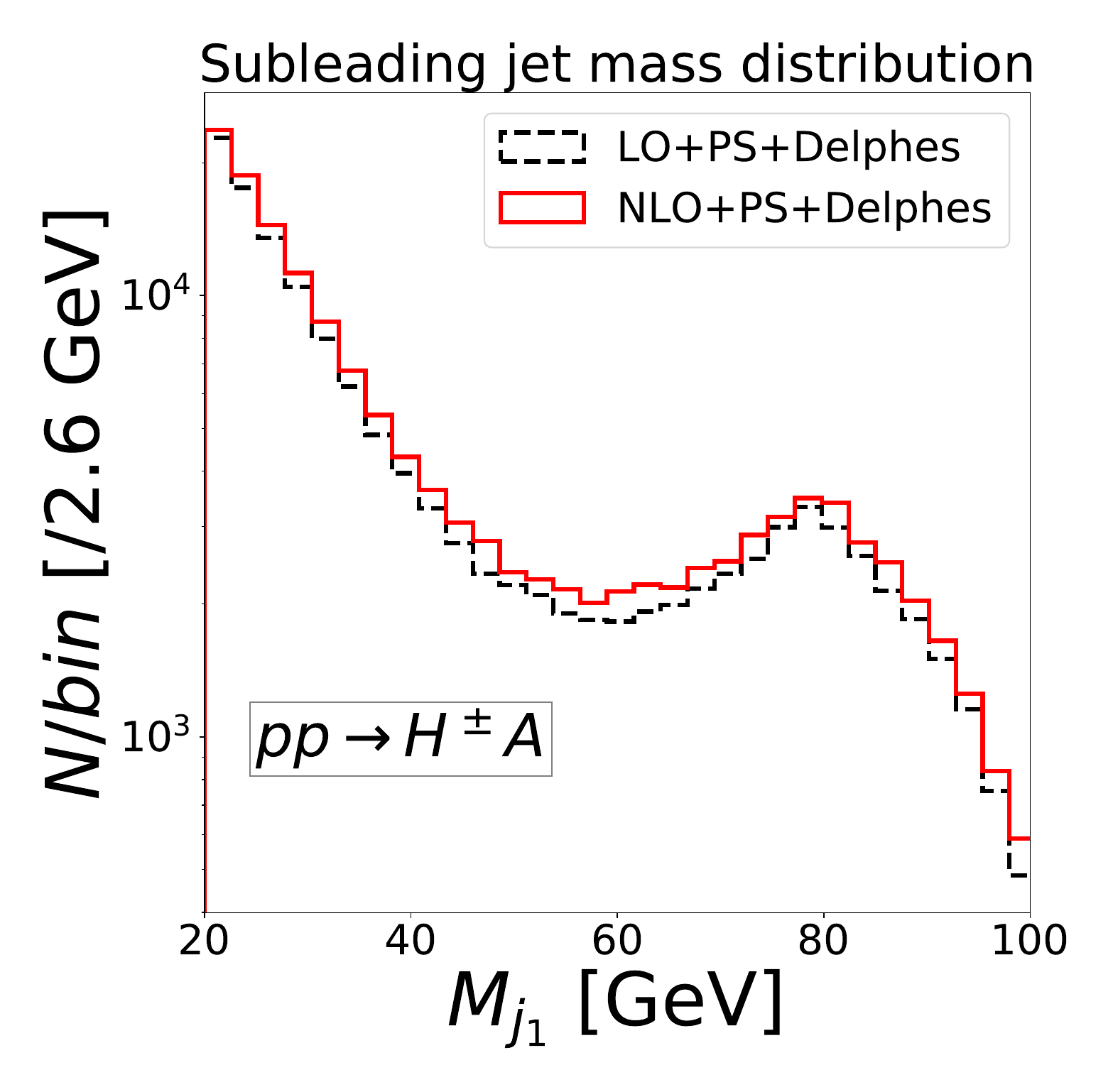}} \hspace{0.1cm}
  \subfloat[]{\label{fig:delR_joj1_hpma0}\includegraphics[scale=0.215]{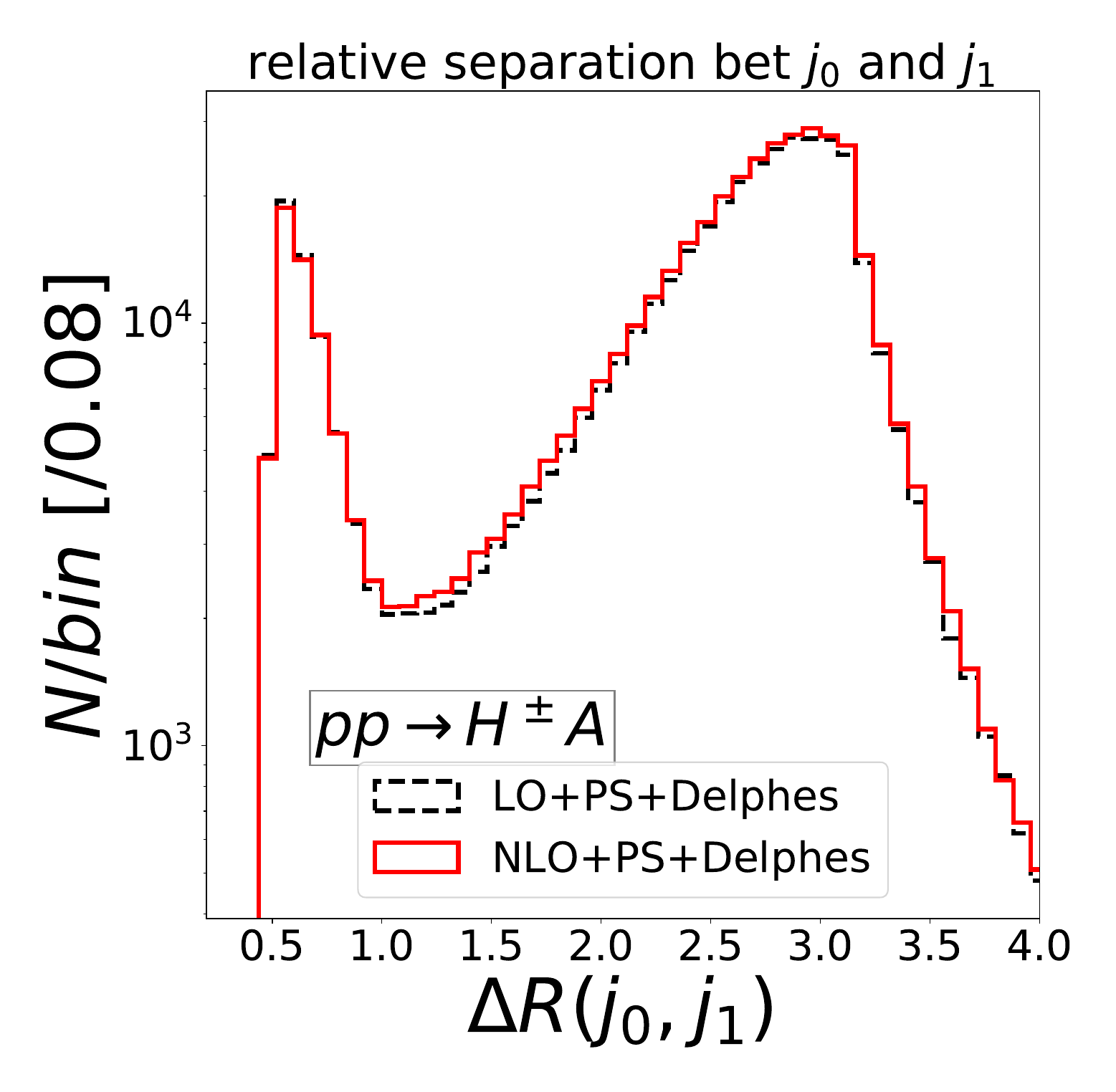}}\\
  \subfloat[]{\label{fig:pt_j0_hpma0}       \includegraphics[scale=0.215]{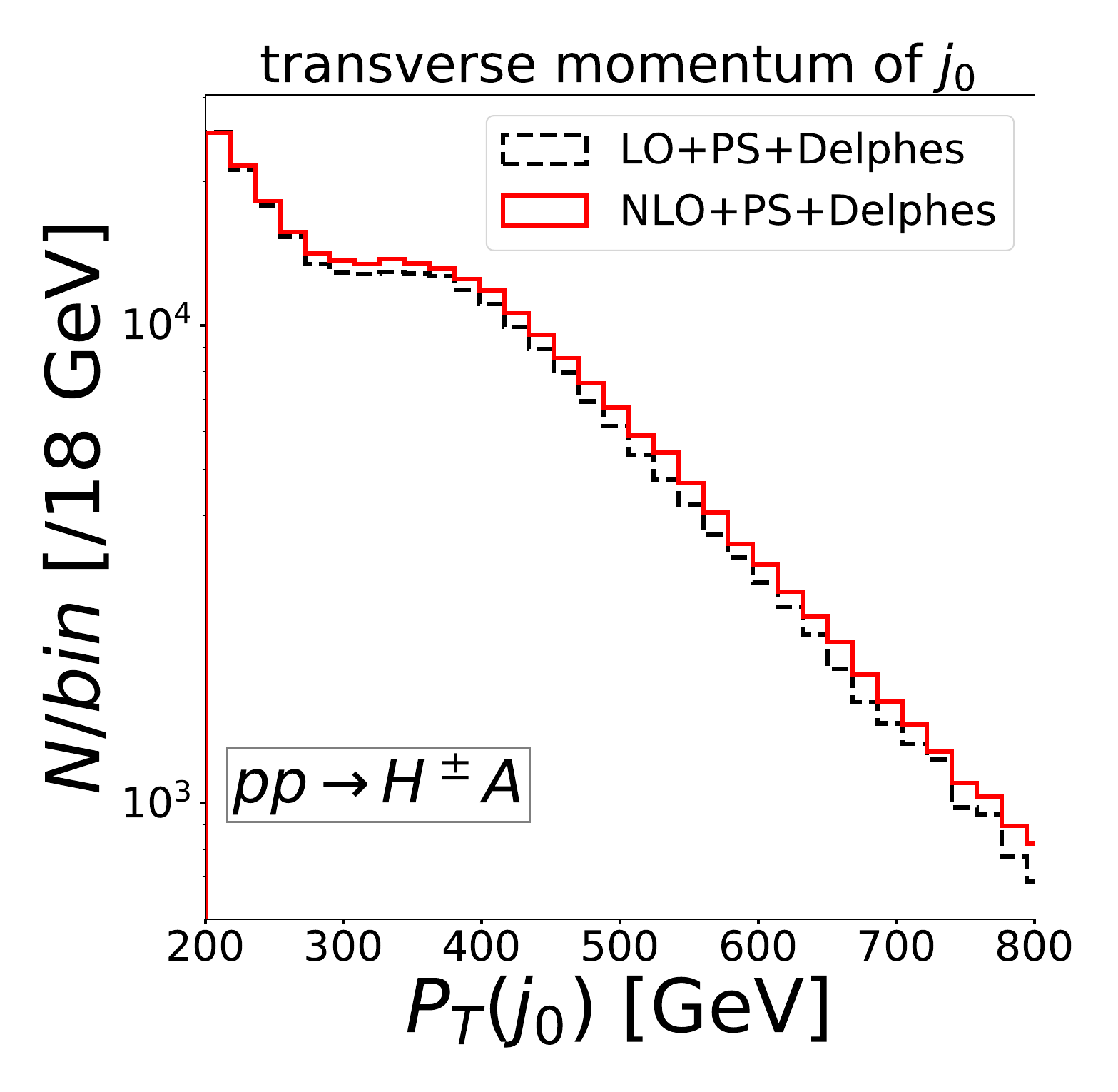}} \hspace{0.1cm}
   \subfloat[]{\label{fig:pt_j1_hpma0}      \includegraphics[scale=0.215]{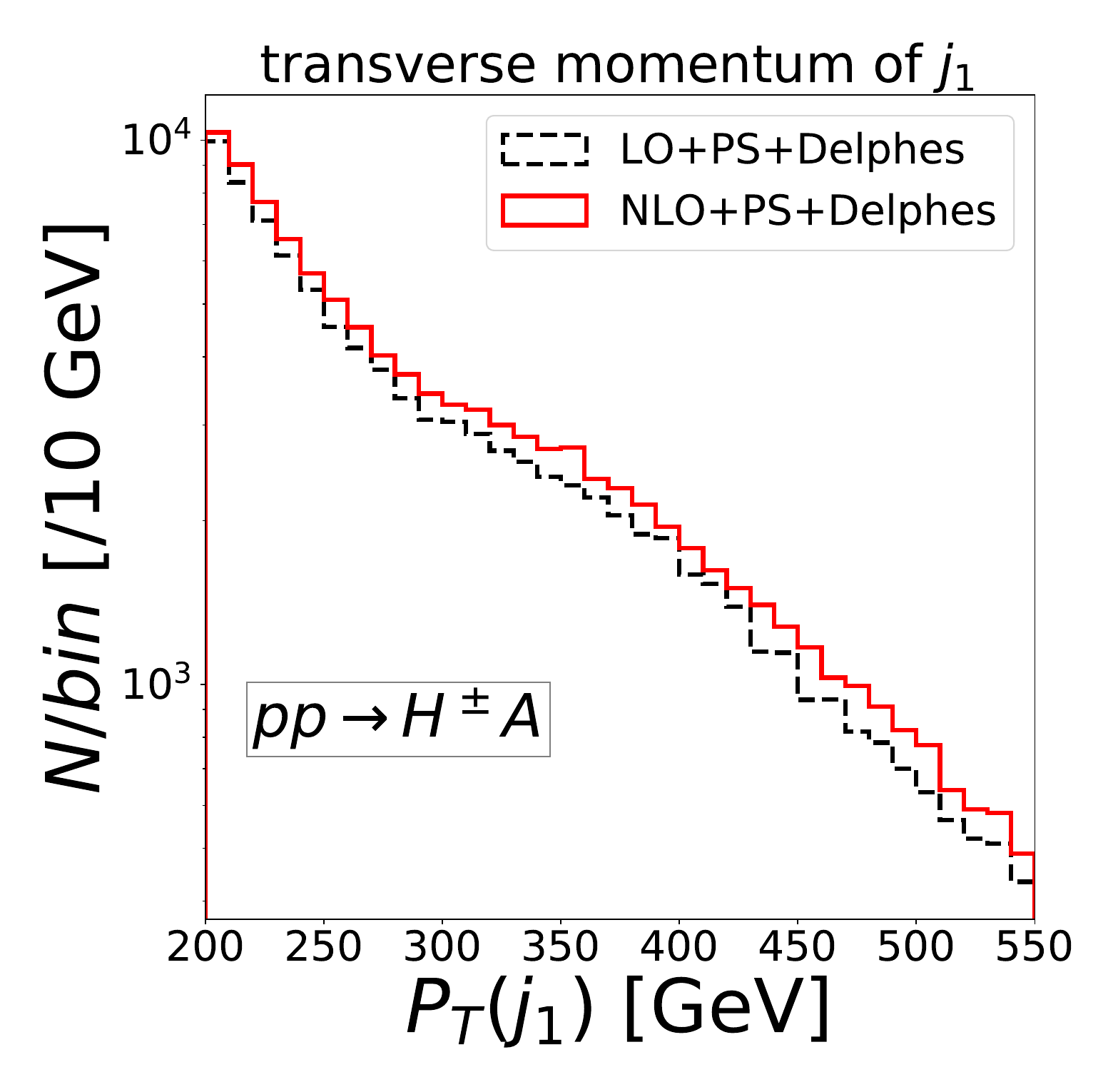}} \hspace{0.1cm}
   \subfloat[]{\label{fig:MET_hpma0}      \includegraphics[scale=0.215]{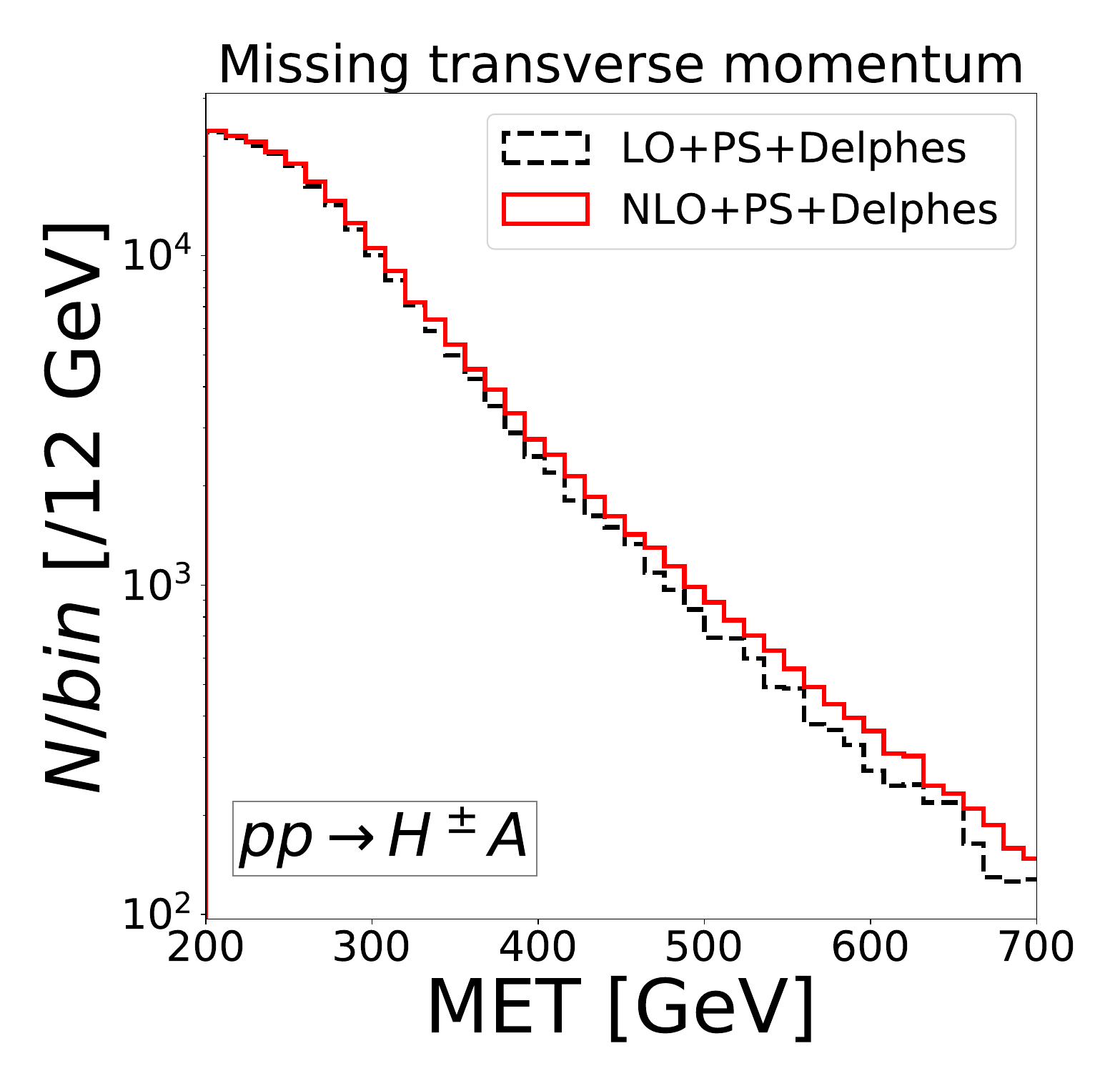}}\\
  \caption{Panels are the same as in \autoref{fig:4_1}, but for the pair production of the heavy scalars channel, such as $pp\rightarrow H^\pm A$, where both $A$, $H^\pm$ decay hadronically.}
\label{fig:4_3}
\end{figure*}

\begin{figure*}[tb]
\centering
  \subfloat[]{\label{fig:m_jo_a0a0}       \includegraphics[scale=0.215]{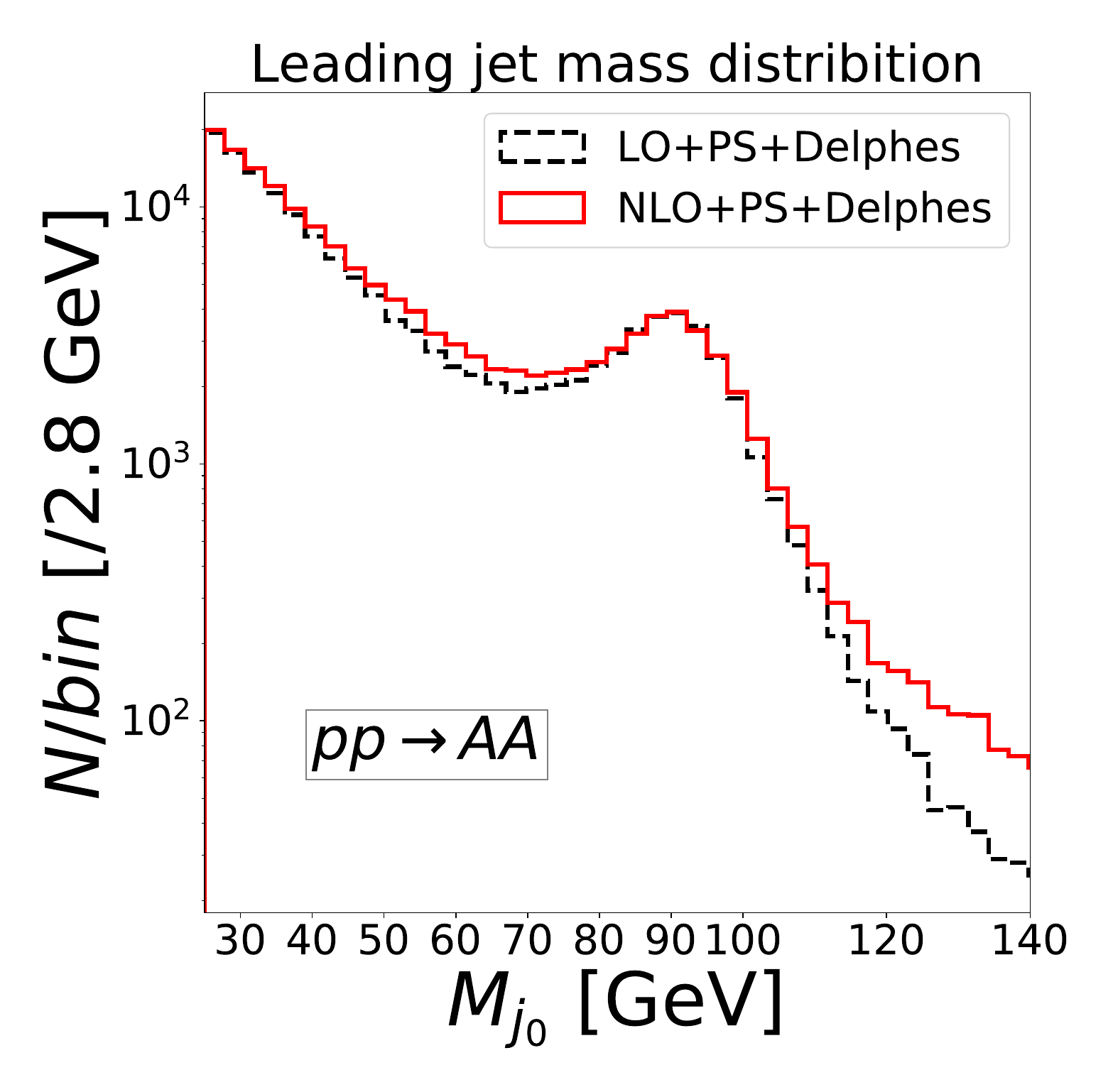}} \hspace{0.1cm}
  \subfloat[]{\label{fig:m_j1_a0a0}       \includegraphics[scale=0.215]{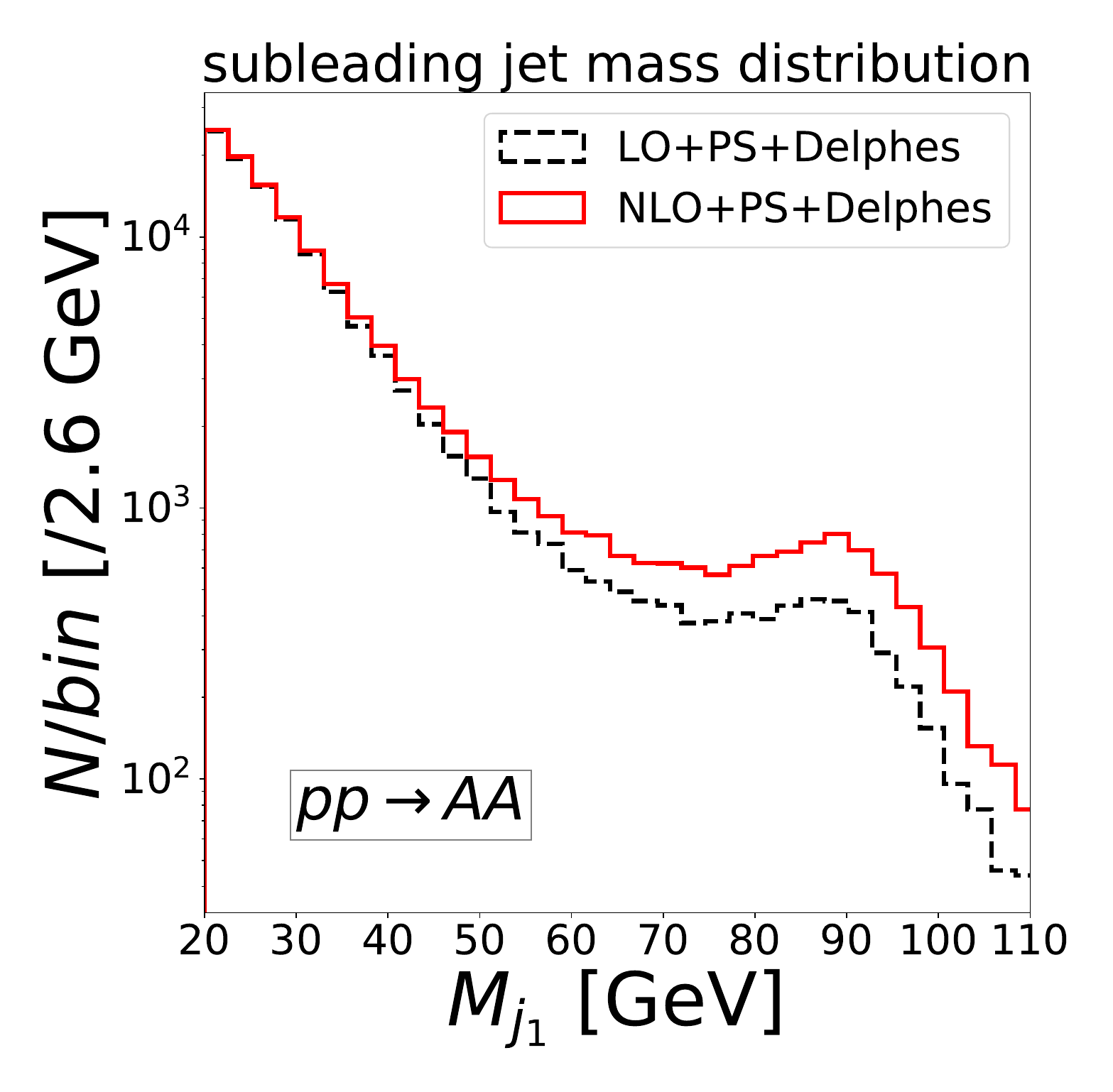}}  \hspace{0.1cm}
  \subfloat[]{\label{fig:delR_joj1_a0a0}\includegraphics[scale=0.215]{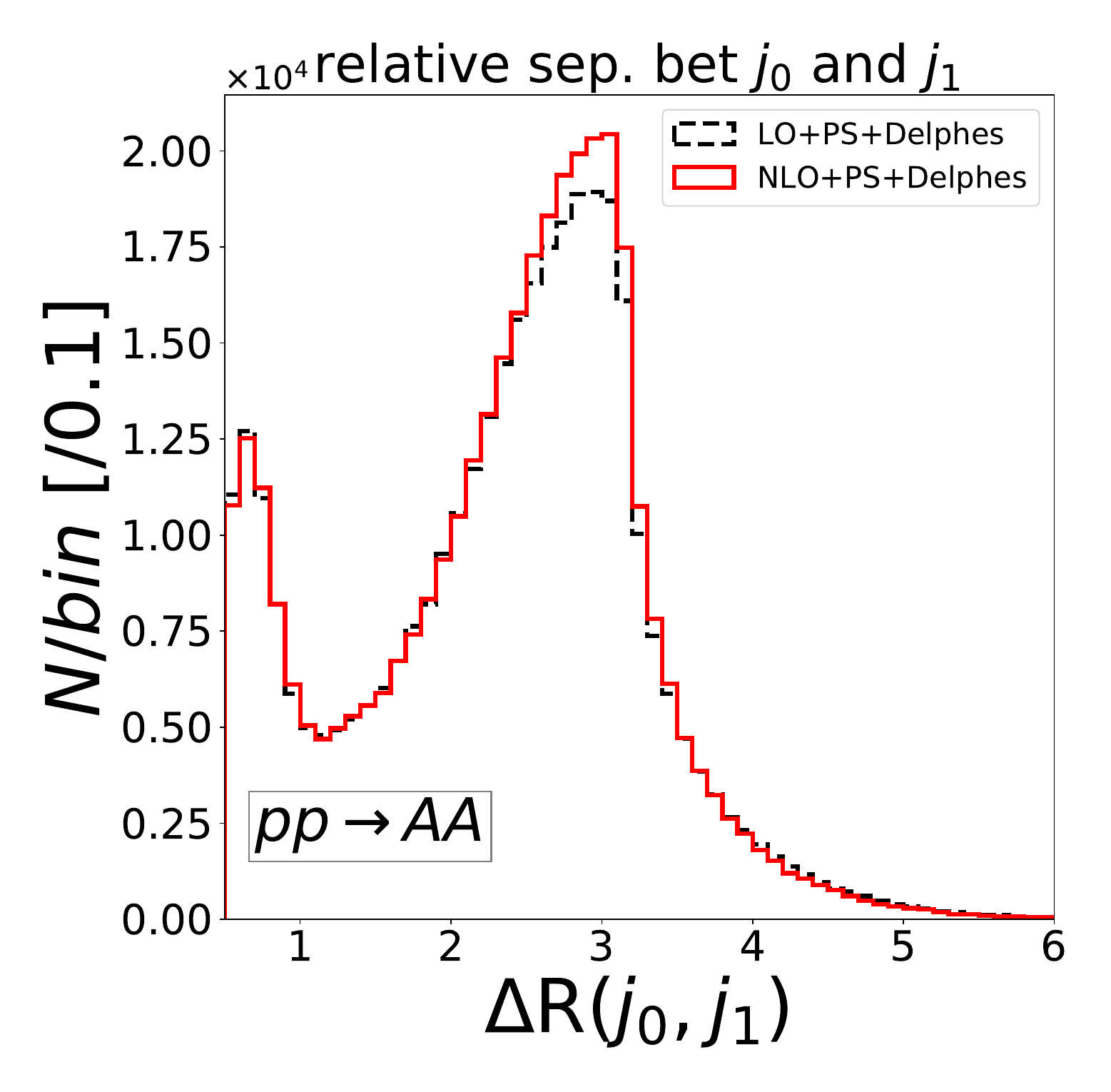}} \\
  \subfloat[]{\label{fig:pt_j0_a0a0}       \includegraphics[scale=0.215]{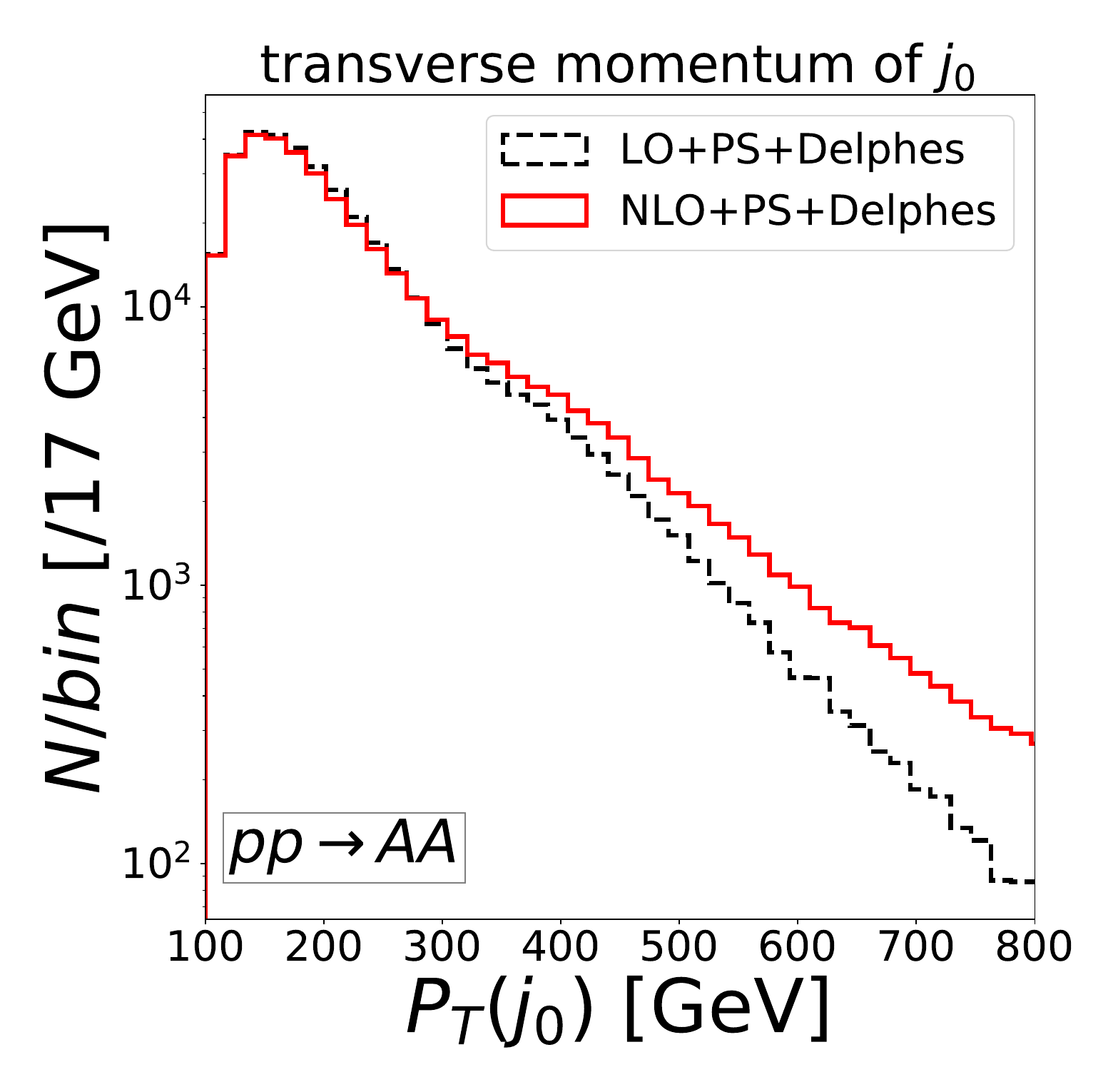}}  \hspace{0.1cm}
   \subfloat[]{\label{fig:pt_j1_a0a0}      \includegraphics[scale=0.215]{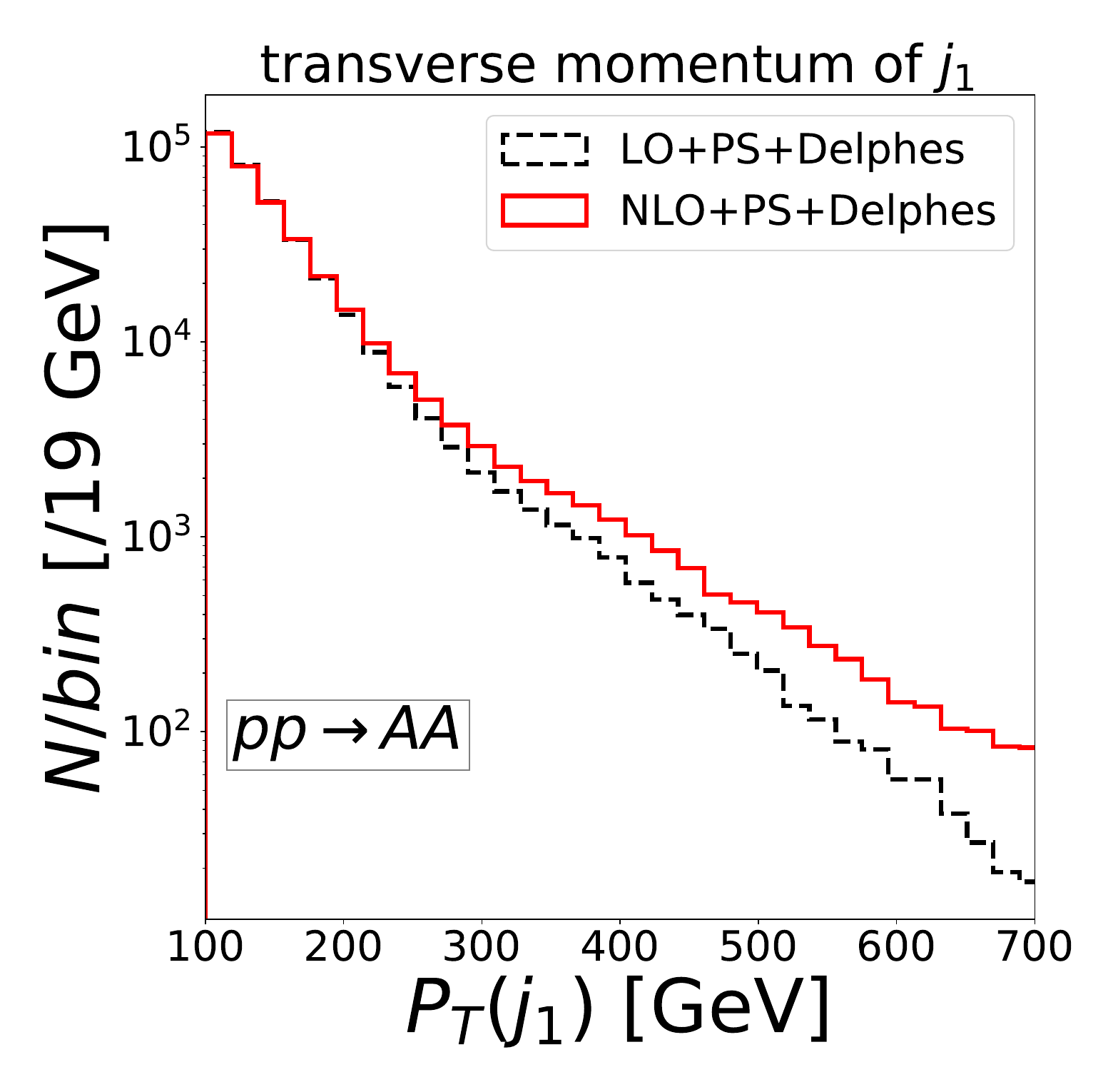}}  \hspace{0.1cm}
   \subfloat[]{\label{fig:MET_a0a0}      \includegraphics[scale=0.215]{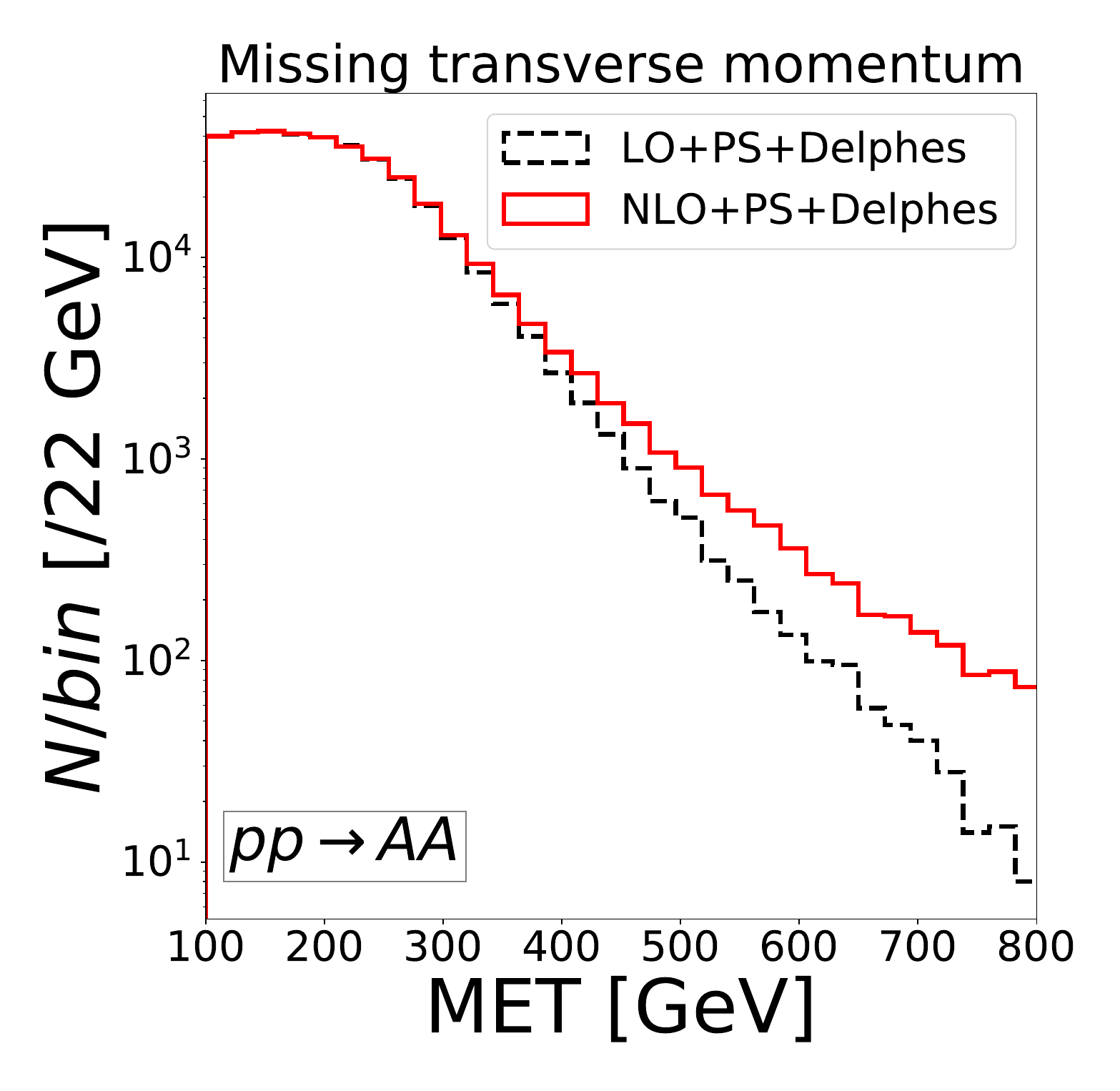}}\\  
  \caption{Panels are the same as in \autoref{fig:4_1}, but for the pair production of the heavy scalars channel, such as $pp\rightarrow A A$, where both $A$ decay hadronically.}
\label{fig:4_5}
\end{figure*}

Heavy scalars, after their creation through the associated channel along with DM candidate $H$, or from a pair production, primarily decay into $H$ and a gauge boson, which is further decayed hadronically. It is imperative to look into their dominant hadronic decay channels as a possible probe for IDM. We select the simulated events including the parton shower and detector effect with a minimum missing transverse energy, $\slashed{E}_T>100$ GeV, and the minimum transverse momentum of the two leading jets $P_T (j_i) > 100$ GeV (for $i=0, 1$). Particle-flow towers and particle-flow tracks are used as input to cluster the jets of radius parameter 0.5, where we use the $\text{anti-K}_T$ algorithm for clustering. The jet mass is defined by $M_j=(\sum_{i\in j} P_i)^2$, where $P_i$ is the four-momentum of the i-th constituent within the jet. The missing transverse energy is defined as the negative sum of the transverse momentum of all the reconstructed constituents, $\slashed{E}_T=-\sum_{i} \vec{P}_{T,i}$. The angular distance between two jets in the transverse plane is denoted as $\Delta R(j_i,j_j)$. This section aims to examine relevant distributions of the jets from the signal to motivate the significance of NLO QCD calculation over the LO. In addition to upward shift, NLO corrections can change the shape of the distribution for a variety of kinematical variables. This has a profound effect in constructing the phenomenological study.
These distributions also make a case for large-radius jets (fatjets) originated from boosted $Z/W^\pm$ boson decay which comes naturally in probing the hierarchical mass region of the IDM.

The distribution of the different high-level observables for one of the associated production channels of heavy scalar
\footnote{Both $pp\rightarrow AH$ and $pp\rightarrow H^\pm H$ channels follow similar distributions, as both $A$ and $H^\pm$ masses are nearly degenerate and produced through vector mediator.}, $pp\rightarrow AH$, and vector boson mediated pair production of the heavy scalars, $pp\rightarrow H^\pm A$ and scalar mediated pair production, $pp\rightarrow A A$ are shown in \autoref{fig:4_1}, \autoref{fig:4_3}, and \autoref{fig:4_5} respectively displaying the LO (dashed black) and NLO (solid red) contributions considering a sample benchmark point BP2. In each figure, the first two plots (a) and (b) present distributions of the leading ($j_0$) and subleading ($j_1$) jet mass, respectively. Plot (c) presents the distribution of the relative separation between these leading and subleading jets, whereas plots (d) and (e) exhibit their transverse momentum distributions, respectively.  Finally, plot (f) shows the distribution of the total missing transverse energy from such production.

The channel $pp\rightarrow AH$ at the partonic level produces three hard jets, two from Z boson decay, and the other is the NLO radiation, while at LO, it has only two hard jets from Z boson decay. The first peak in the leading jet mass distribution (\autoref{fig:m_jo_a0h2}) is generated when a QCD hard parton forms a jet after PS and detector simulation. Interestingly, this same distribution points to a second peak both for LO and NLO results. This occurs when the Z boson is produced with enough boost to form a merged jet out of its full decay products, resulting into a peak at Z boson mass. The second hard parton from the Z boson forms the subleading jet, causing a peak near $M_{j_1}=10$ GeV (\autoref{fig:m_j1_a0h2}) but no other peak in the LO $j_1$ mass distribution. However, the NLO distribution can have extra hard radiation. Occasionally when that carries enough transverse momentum to form a leading jet, Z boson decay still forms a merged subleading jet resulting in a second peak near Z boson mass (\autoref{fig:m_j1_a0h2}) deviating from a leading order estimate. Hence NLO estimate predicts an upward trend in the number of boosted di-jet events even from such associated production channels. One can also expect such abundance in boosted jets for other benchmark points with heavier scalars.
Our previous argument is even more evident in the next distribution plot of the relative separation between two leading jets (\autoref{fig:delR_joj1_a0h2}) for the same channel $pp\rightarrow AH$.
The number of events with smaller jet separation $\Delta R_{j_0,j_1}  < 1.0 $ is one order larger than in the other region. For a significant event sample, both leading and subleading jets come from the Z boson's decay and are closely separated. Naturally, the construction of large-radius jets embeds them together to form a single fatjet carrying properties of originating gauge boson. It is even more pronounced in larger masses of scalar.
The distribution of the transverse momentum of the leading (\autoref{fig:pt_j0_a0h2}) and subleading (\autoref{fig:pt_j1_a0h2}) jets and the total missing transverse energy (\autoref{fig:MET_a0h2}) shows an upswing in NLO at larger PT. This is significant in view of the final selection of events (or, during multivariate analysis at boosted decision tree) comes with higher weightage from these distribution tails to deal with a tiny signal over an overwhelmingly large background.

Similarly, one requires to follow distributions from pair production channels of the heavy scalars. 
The leading and subleading jet mass distributions for vector boson mediated (\autoref{fig:m_jo_hpma0}, \autoref{fig:m_j1_hpma0}) and scalar boson mediated (\autoref{fig:m_jo_a0a0}, \autoref{fig:m_j1_a0a0}) channels in pair production of heavy scalars demonstrates two clear mass  peaks both at LO and NLO. In this case, pairs of heavy scalars produce two boosted vector bosons, and as they have enough boost, it results into the second peak in both cases. Again, with the increase of scalar mass, the second peak rises, ensuring enhancement of di-fatjet events. The distributions of the relative separation for pair production of heavy scalars shown in  \autoref{fig:delR_joj1_hpma0} and \autoref{fig:delR_joj1_a0a0} contain two peaks. The second peak at $\Delta R \sim \pi$ appears when two jets originate from two different vector bosons. The first peak is when both the jets arrive from the same vector boson, which gradually diminishes for heavier mass.
Pair production channel $pp\rightarrow AA$ has a significant shift between NLO and LO distributions in comparison to the $pp\rightarrow H^\pm A$ channel, as the former is Higgs mediated and has a larger K-factor. 
It is evident from this discussion that the tagging of large-radius jets originating from boosted vector bosons can significantly improve the efficiency of probing the hierarchical mass spectrum of the IDM. In the next section, we will describe the selection and properties of such boosted fatjets.

\section{Boosted fatjet as a proxy for heavy scalar production} 
\label{sec:collider}

Our discussion in the previous section demonstrates that the multi-jet + $\slashed{E}_T$ search is not sufficient to explore the hierarchical mass region of IDM. Jet pair originated from the vector bosons, which comes out as boosted decay product of heavy scalar, is already collimated as a merged hadronic object. This process of getting a fatjet becomes more and more evident while probing a heavier scalar mass. A large radius fatjet can effectively identify this combined hadronic yield from the boosted vector boson. Moreover, it can carry a significant amount of information hidden inside the internal structure of jet formation through the orientation of fragmented hadrons and their energy deposits, revealing the properties and identity of the originating particle. 

\subsection{Signal and background processes}
\label{sec:S_B}

Representative LO Feynman diagrams both for associate production and pair production of heavy scalars are already depicted in \autoref{fig:Feyn1}. Our primary focus is to analyze the NLO accurate di-fatjet signal arising from heavy IDM scalar decay using jet substructure variables. We do not discriminate W-jet or Z-jet and dub them as V-fatjet ($J_V$) since we consider a suitable mass window to accommodate both in our analysis. We will  discuss the usefulness of the sophisticated multivariate analysis that can make the signature of $2J_V+\slashed{E}_T$ into the better discriminator in order to separate out tiny signal from an overwhelmingly large SM background. However, multivariate description creates a highly performant nonlinear cut at the cost of blurring the exact physical description of how different high-level variables affect our analysis. Hence, to better understand the kinematic variables that may affect LO and NLO computations, we would analyze them first with usual cut-based method before moving on to the MVA analysis. In passing, it is to be noted that the cross-section of the di-Higgs production while one Higgs boson decay into a pair of bottom quarks ($h \rightarrow b \bar{b}$) and the other decays into pair of dark matter ($h \rightarrow H H$) is 1.05 $fb$. Although this channel has a sizable effect on the di-fatjet final state, we do not include this in our analysis since this process drops sharply after applying b-veto.

All the significant backgrounds that contribute to the $2J_V+\slashed{E}_T$ signal are included in our analysis. 
We do two to four additional jets merging using the MLM matching \cite{Mangano:2006rw, Hoeche:2005vzu} scheme for different background processes, and normalize the cross-section according to the available higher-order QCD corrections.
Inclusive Z boson production is the principal background where Z boson decays invisibly ($p p \rightarrow Z + \text{jets} \rightarrow \nu \nu +\text{jets} $) and gives a large $\slashed{E}_T$ together with fatjets arising from QCD jets. This process is matched to four extra partons using the MLM scheme.
Second, inclusive $W^\pm$ boson production has a significant contribution when the lepton from the leptonic decay of the W boson remains undetected ($p p \rightarrow W + \text{jets} \rightarrow l_{e,\mu} \nu +\text{jets}$). The neutrino from W-decay gives a substantial amount of $\slashed{E}_T$ and fatjets arise from QCD jets. This process is generated up to four extra partons with MLM matching. 
Note that the contribution from the above two background processes counts only when the missing transverse momentum is sufficiently large. 
We apply the generation level hard cut $\slashed{E}_T > 100$ GeV, as the region with lower missing transverse energy is of no interest for this present analysis. 
Additionally, di-boson production can offer a considerable amount of contribution in the background. The three different di-boson processes $pp\rightarrow WZ,\hspace{1mm} WW,\hspace{1mm} \text{and} \hspace{1mm} ZZ,$ are possible, where the $WZ$ process gives the most significant contribution among these three. All three processes are generated and merged up to two extra partons. One of the vector bosons in all these processes decays hadronically, giving rise to a $J_V$. Other vector boson decaying invisibly ($Z\rightarrow \nu \nu$) or leptonically ($W\rightarrow l_{e,\mu} \nu$) with lepton being undetected, gives a large $\slashed{E}_T$. Another fatjet in all these di-boson processes arises from the QCD jets.
Single top production is possible in SM through three different types of process, S-channel ($p \hspace{1mm} p \rightarrow t \hspace{1mm} b$), t-channel ($p \hspace{1mm} p \rightarrow t \hspace{1mm} j$) and associated production ($p \hspace{1mm} p \rightarrow t \hspace{1mm} W$), where associated production gives a considerable amount of contribution to the background of our signal. This process is merged up to two extra partons using the MLM scheme.
Finally, top pair production contributes to the background when one top decays leptonically and lepton is escaping the detection. Whereas the other top decays hadronically and that essentially gives rise to a vector-like fatjet $J_V$. Since such an event comes with a couple of b-jets, b-veto can effectively reduce this background. This process is generated to two extra partons with MLM matching. The other fatjet aries from the QCD jets or untagged b-jets.
We found negligible contributions to the background from the QCD multi-jet and tri-boson processes compared to the processes mentioned above. Therefore we do not include these processes into our analysis. For our simulated backgrounds at 14 TeV LHC, we normalize their cross-section according to the available higher-order QCD corrections, as tabulated in Table IV of ref. \cite{Bhardwaj:2019mts}.

The associated production of heavy scalar with two jets merging and pair production of the heavy scalars are analyzed at LO \cite{Bhardwaj:2019mts} where it was found that the former processes contribute dominantly in the di-fatjet final state than the latter. A further estimate at NLO accuracy modifies the contribution in two vital directions. First, both for the associated production and pair production of heavy scalar processes, the differential NLO K-factor plays an important role, as already described in the previous section. Second, two jets merged associated production channels can mimic the Higgs mediated pair production of heavy scalar processes, and therefore may contribute to double-counting in a particular phase space region. NLO estimate eliminates such possibility giving non-overlapping contributions from all processes. 

Now, along with both these effects, our estimate at NLO predicts reduced contribution from associated production, thereby enhancing the part from the pair production. This has a profound significance in setting up the phenomenological analysis. On contrary to a more complex mixed-signal region analysis by taking into account the admixture of $1J_V$ and $2J_V$, that has been carried out in ref. \cite{Bhardwaj:2019mts}, it is tempting to concentrate only on the $2J_V$ identification for a significant gain. Demanding that both the fatjets have V-jet like characteristics, one finds a more effective background control and, as a result, a higher statistical significance.

\subsection{Construction of high level variables}
\label{sec:variables_cuts}

The total energy of the fatjet originated from the hadronic decay of boosted W, Z is distributed around two subjet axes. N-subjettiness ratio ($\tau_{21}$) and the jet-mass ($M_J$) are two potent variables to classify such fatjets $J_V$ from those that arise from the fragmentation of QCD parton. The jet-mass is defined by $M_J=(\sum_{i\in J} P_i)^2$, where $P_i$ is the four-momentum of the i-th hit in the calorimeter. Large-radius jets are prone to attract additional soft contributions from underlying QCD radiation, which needs to be eliminated to get reliable estimates from the different high-level variables. Pruning, filtering, and trimming \cite{Krohn:2009th, Butterworth:2008iy, Ellis:2009su, Ellis:2009me} are different grooming techniques prescribed to remove those soft and wide-angle radiations. We consider pruned jet in our analysis as discussed in refs. \cite{Ellis:2009su, Ellis:2009me}.

We run the pruning algorithm repeatedly to remove the soft and wide-angle emission and veto such recombinations. One has to estimate two variables, the angular separation of the two proto-jets, $\Delta R_{ij}$ and softness parameter $Z=\text{min}(P_{Ti},P_{Tj})/P_{T(i+j)}$, at every recombination step. The recombination between i-th and j-th proto jets is not performed dropping the softer one, if $\Delta R_{ij} > R_{fact} $ and $Z<Z_{cut}$. We choose standard default values of $R_{fact}=0.5$ and $Z_{cut}=0.1$ \cite{Ellis:2009su}.
The N-subjettiness determines the jet shape of hadronically-decaying boosted V-bosons. Considering that N number of subjets exist within the jet, N-subjettiness ($\tau_N$) is defined by the angular separation between constituents of the jet from their nearest sub-jet axis as given below \citep{Thaler:2010tr, Thaler:2011gf}.
\begin{equation}
\tau_N = \dfrac{1}{\mathcal{N}_0}\, \sum_i \, P_{T,i}\, \text{min} \{ \Delta R_{i,1},\Delta R_{i,2},...,\Delta R_{i,N}\}
\end{equation}
The summation runs over all the constituents of the jet, and $P_{T, i}$ is the transverse momentum of the i-th constituent. $\mathcal{N}_0=\sum_i \, P_{T,i} \, R$ is the normalization factor, and R is the jet radius.
$\tau_{21}$ denotes the ratio of $\tau_2$ and $\tau_1$, which is an excellent variable to tag a hadronically-decaying boosted V-boson as it tends to zero (far from zero) for a correctly identified two-prong (one-prong) jet.

\begin{table*}[tb!]
\begin{center}
 \begin{tabular}{|c|c|c|c||c|c|c|}
\hline
\multirow{3}{1em}{BP} & \multicolumn{6}{| c |}{ $\text{Pre-selection cuts}+\slashed{E}_T>200\hspace{1mm} GeV,\hspace{1mm} \text{b-veto}, \hspace{1mm} 65 \hspace{1mm} GeV< M(J_0), M(J_1)<105 \hspace{1mm} GeV, \hspace{1mm} \tau_{21}(J_0),\tau_{21}(J_1)<0.35 $}   \\
\cline{2-7}
&  \multicolumn{3}{| c |}{$H^\pm A$} & \multicolumn{3}{| c |}{ $H^+ H^-$}  \\
\cline{2-7}
& \hspace{0.5cm} $N_S^{NLO}$\hspace{0.5cm} & \hspace{0.5cm} $N_S^{LO\times K}$ \hspace{0.5cm} & \hspace{0.2cm} relative change$\%$ \hspace{0.2cm} & \hspace{0.5cm} $N_S^{NLO} $ \hspace{0.5cm} & \hspace{0.5cm} $N_S^{LO\times K}$ \hspace{0.5cm} & relative change$\%$  \\
\hline\hline
BP1 & $168.2^{+2.8}_{-2.5}$  & $119.5^{+3.3}_{-3.2}$ & $40.75\%$ & $121.2^{+4.6}_{-4.6}$  & $82.7^{+5.0}_{-4.1}$ & $46.55\%$   \\
\hline
BP2 & $190.7^{+3.1}_{-4.1}$  & $155.6^{+5.7}_{-5.5}$ & $22.56\%$ & $150.4^{+9.2}_{-8.0}$  & $111.1^{+9.8}_{-7.7}$ & $35.37\%$   \\
\hline
BP3 & $202.8^{+4.2}_{-4.2}$ & $162.8^{+7.0}_{-6.5}$ & $24.57\%$ & $153.8^{+11.8}_{-9.8}$  & $122.5^{+13.8}_{-10.9}$ & $25.55\%$ \\
\hline
 \end{tabular} 
\caption{$N_S^{NLO}$ and $N_S^{LO\times K}$ are shown to represent the expected number of the $2J_V+\slashed{E}_T$ final state events for different pair production of heavy scalars channels, $pp\rightarrow H^\pm A$, and $pp\rightarrow H^+ H^-$. These two numbers coming from NLO computation and LO with integrated NLO K-factor multiplication, respectively, at 14 TeV HL-LHC. Superscripts and subscripts are the change in the corresponding number of events due to the envelope of eight different ($\mu_R, \mu_F$) scale choices. Additional cuts over existing pre-selection (see text) are $\slashed{E}_T>200\hspace{1mm} GeV,\hspace{1mm} \text{b-veto}, \hspace{1mm} 65 \hspace{1mm} GeV< M(J_0), M(J_1)<105 \hspace{1mm} GeV, and \hspace{1mm} \tau_{21}(J_0),\tau_{21}(J_1)<0.35 $. Relative change, $(N_S^{NLO} - N_S^{LO\times K})/N_S^{LO\times K}$ reflects the importance of differential K-factor in the present computation. Relative changes are given corresponding to the central scale.}
\label{tab:fatjet_LO_NLO}
\end{center}
\end{table*}

\begin{table*}[t]
\begin{center}
 \begin{tabular}{|c|c|c|c|c|c|c|c|}
\hline\hline
Topology & BP1 & BP2 & BP3 & BP4 & BP5 & BP6 & BP7   \\
\hline
Associated production & 452.29 & 377.73 & 327.56& 266.9 & 217.53 & 176.9 & 138.11 \\
\hline
Pair production & 1677.13 & 1432.67 & 1184.16 & 969.0 & 785.63 & 622.99 & 516.62 \\
\hline\hline
Z+jets  & W+jets & tW+jets & tt+jets & WZ+jets & ZZ+jets & WW+jets & Total \\
\hline
652519  & 527312 & 46011.8 & 54635& 36126.5 &3689.51 & 12002.4 & $1.3323 \times 10^6$ \\
\hline
 \end{tabular} 
\caption{Expected number of events from different signal and background processes at 14 TeV HL-LHC corresponding to the central scale after applying the Pre-selection cuts with leading and subleading fatjet mass $M_{J_0}, M_{J_1}>40$ GeV and b-veto.}
\label{tab:MVA}
\end{center}
\end{table*}

To proceed further, we define the following \emph{pre-selection criteria} based on which signal and background event samples are prepared: 
({\em i}) each event has to have at least two fatjets constructed by the Cambridge-Aachen (CA) jet clustering algorithm with radius parameter $R=0.8$, and the minimum transverse momentum of each fatjet $P_T(J_i) > 180 \, GeV$,
({\em ii}) since pair of DM particles are produced in the signal, a minimum missing transverse energy $\slashed{E}_T > 100 \, GeV$ is applied to select the events,
({\em iii}) we also impose a minimum azimuthal angle separation between the identified fatjet and missing transverse momentum direction, so that, $|\Delta \phi (J_i,\slashed{E}_T)| > 0.2$. This minimizes any jet mismeasurement effect contributing to $\slashed{E}_T$,
({\em iv}) since no leptons are expected in signal region, backgrounds can be further suppressed by vetoing a lepton tag. So, events are vetoed if they contain leptons that have pseudorapidity $|\eta(l)| < 2.4 $ and transverse momentum $P_T(l) > 10 \, GeV$.

It is clear from our previous discussion on boosted fatjet that several interesting variables can contribute to strengthen the signal efficiency. We would demonstrate the distribution of all such variables, but before that we point out some of the significant changes that appeared due to NLO computation in the signal region. 
In \autoref{tab:fatjet_LO_NLO}, we show the expected number of $2J_V+\slashed{E}_T$ final state events corresponding to the central scale, originated from different pair production of heavy scalar processes. Such numbers for $pp\rightarrow H^\pm A$, and $pp\rightarrow H^+ H^-$ at NLO ($N_S^{NLO}$) level are given for three sample benchmark points, together with LO level numbers multiplied by overall NLO K-factor ($N_S^{LO\times K}$) for 3000 $fb^{-1}$ integrated luminosity at 14 TeV LHC. Superscripts and subscripts are the change in the corresponding number of events due to the envelope of eight different ($\mu_R, \mu_F$) scale choices. In both cases, that makes the overall cross-section  normalized to the NLO value. Signal region criteria in conjunction with pre-selection cuts are described in the table \footnote{One can, in principle, use such stiff event selection criteria for a realistic cut based analysis. Our purpose is purely for demonstration, as we would finally employ multivariate analysis to construct the suitable optimization based on rather loosely set criteria.}. Relative change, defined as $(N_S^{NLO} - N_S^{LO\times K})/N_S^{LO\times K}$, is given for the central scale.  Relative change is independent of the luminosity and ascertains the necessity of considering actual NLO events instead of using LO events multiplied by a flat K-factor. It is evident that NLO and LO computations have different efficiencies for the given kinematic cuts. Relative change between these two estimations exhibits the role of the differential NLO K-factor by changing the LO estimation up to $40\%$ for the process $pp\rightarrow H^\pm A$, and $46\%$ for $pp\rightarrow H^+ H^-$ for the given kinematic cuts mentioned at the top of the \autoref{tab:fatjet_LO_NLO}.

In addition to the pre-selection cuts described above, final event selection criteria for multivariate analysis includes a very relaxed cut on pruned jet mass. All other variables are kept free to provide the multivariate analysis with enough scope to optimize the nonlinear cut based on suitable variables. 
We select the signal and background events after applying the following cuts:
({\em i}) both leading and subleading fatjets have to have a minimum pruned jet mass of 40 GeV to reduce the contribution of fatjets originated from QCD,    
({\em ii}) b-veto is applied on the jets that are formed using $\mbox{anti-k}_t$ algorithm with radius parameter $R = 0.5$ and this significantly reduces ${t\bar{t}}$ background.
%

\begin{figure*}[tb!]
\centering
    \subfloat[] {\label{fig:m_jo_sigbg}           \includegraphics[scale=0.30]{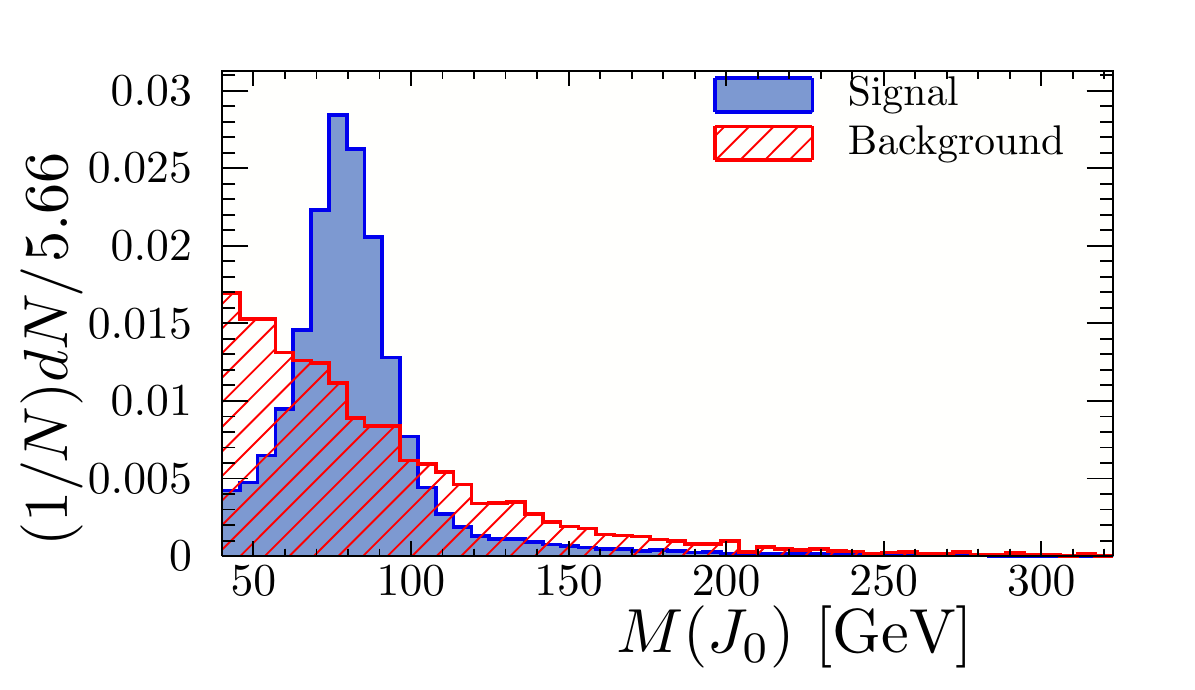}} 
    \subfloat[] {\label{fig:m_j1_sig_bg}         \includegraphics[scale=0.30]{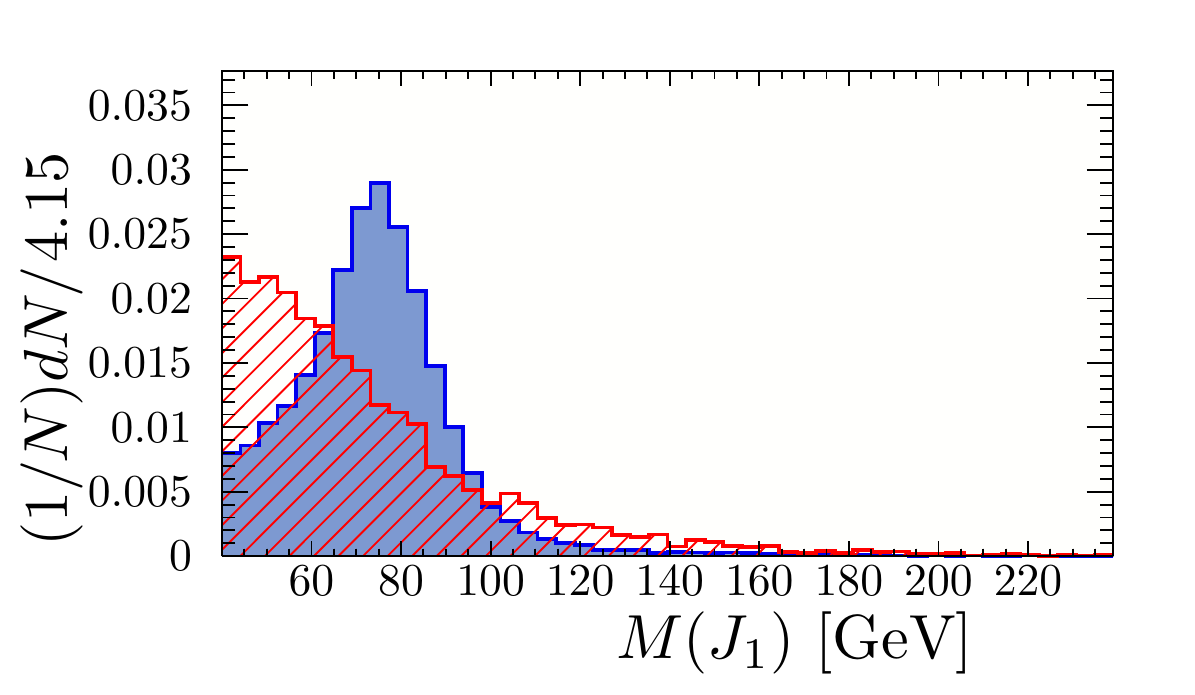}} 
    \subfloat[] {\label{fig:tau_j0_sig_bg}       \includegraphics[scale=0.30]{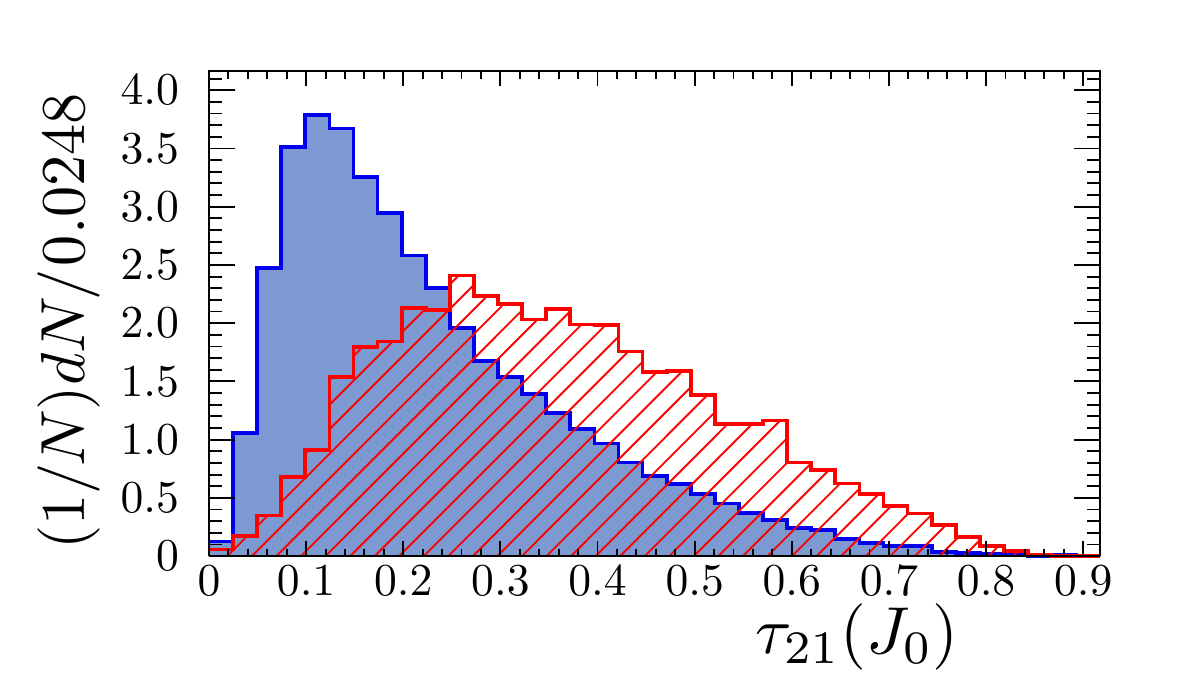}}\\ 
  \subfloat[] {\label{fig:tau_j1_sig_bg}       \includegraphics[scale=0.30]{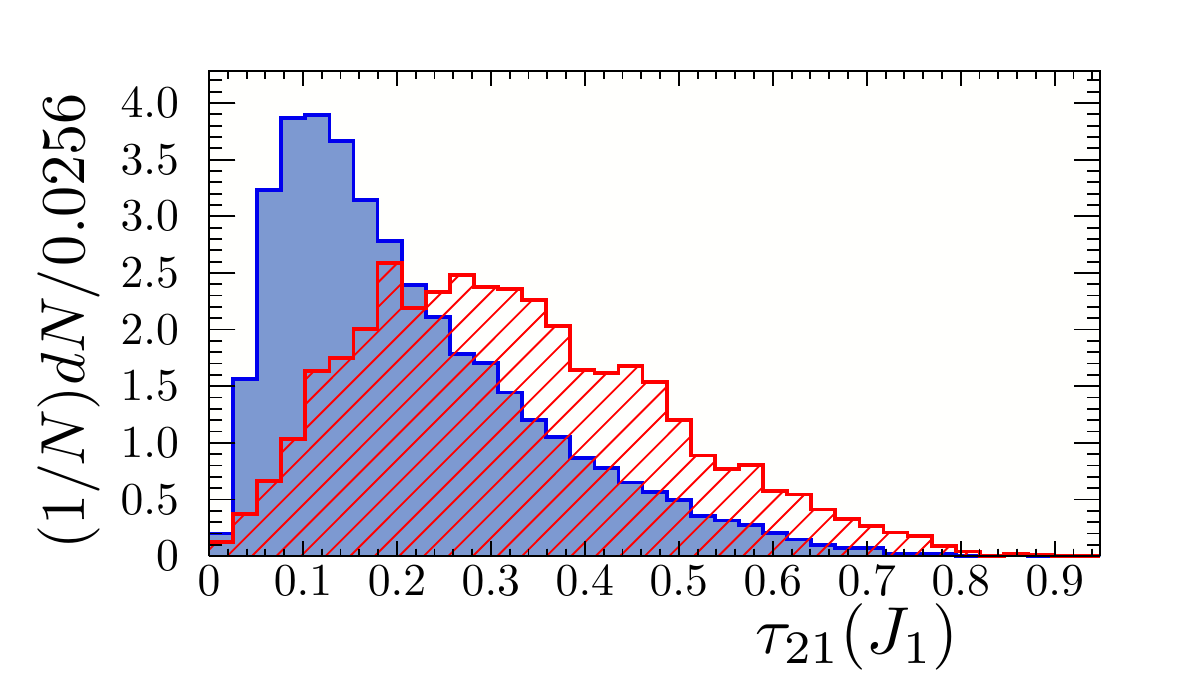}} 
  \subfloat[] {\label{DR_joj1_sig_bg}         \includegraphics[scale=0.30]{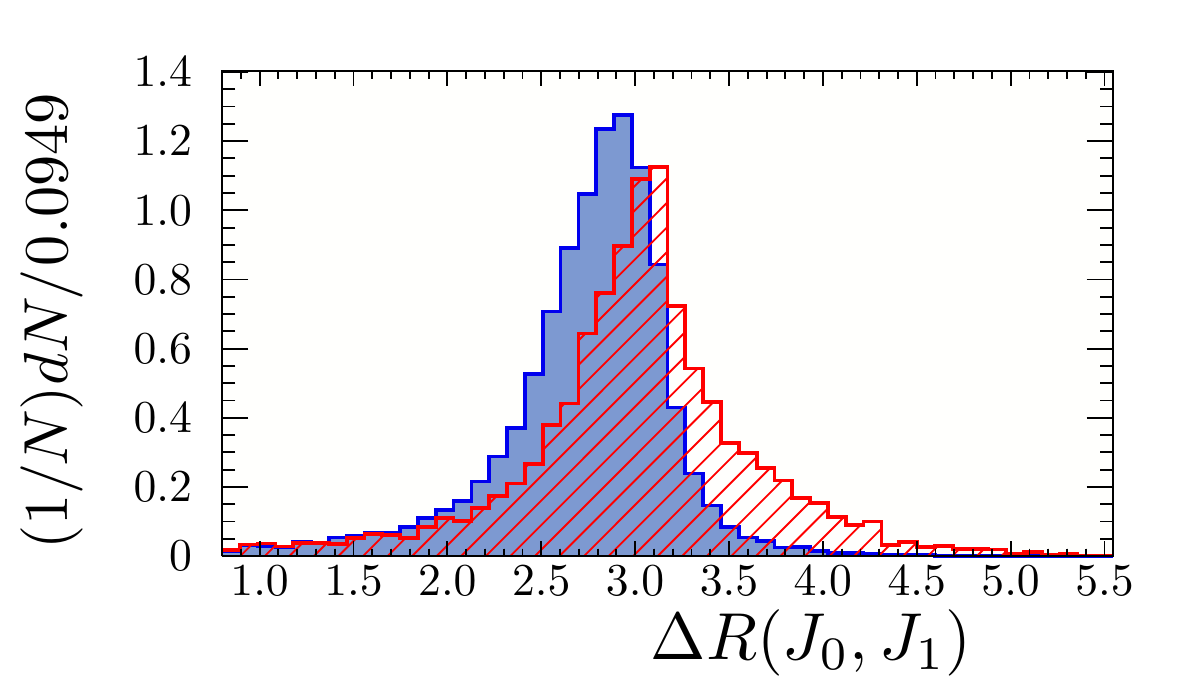}} 
  \subfloat[] {\label{fig:dphi_j1ET_sig_bg} \includegraphics[scale=0.30]{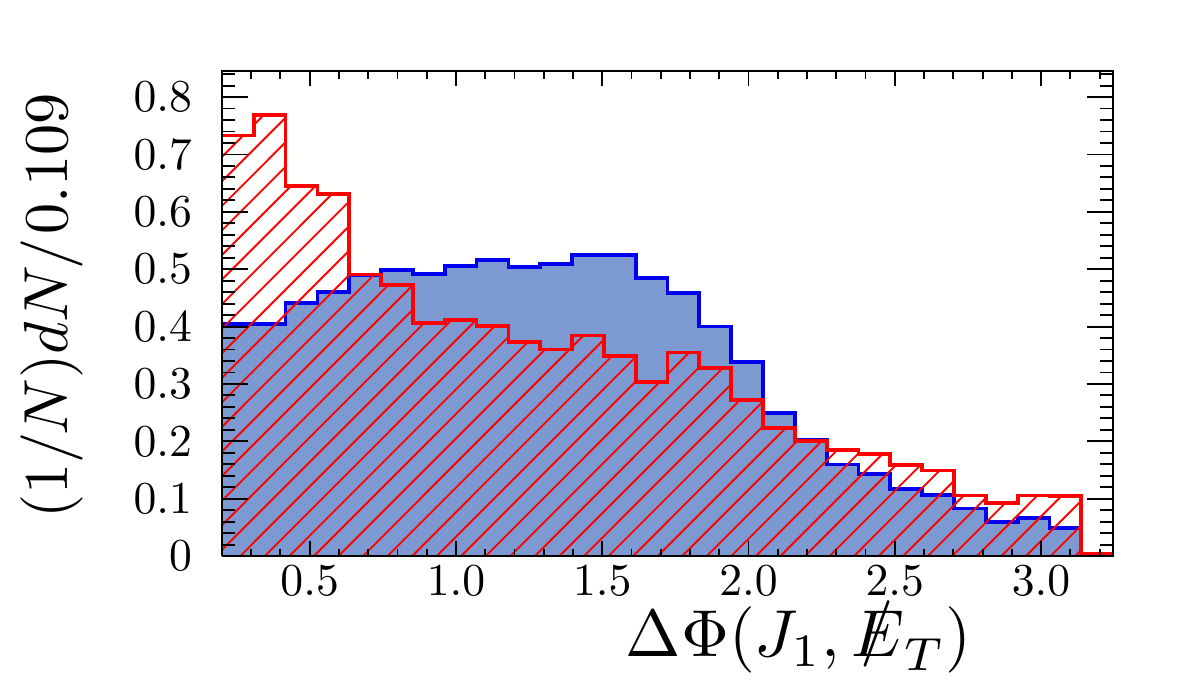}}\\ 
    \subfloat[] {\label{fig:shat_sigbg}     \includegraphics[scale=0.295]{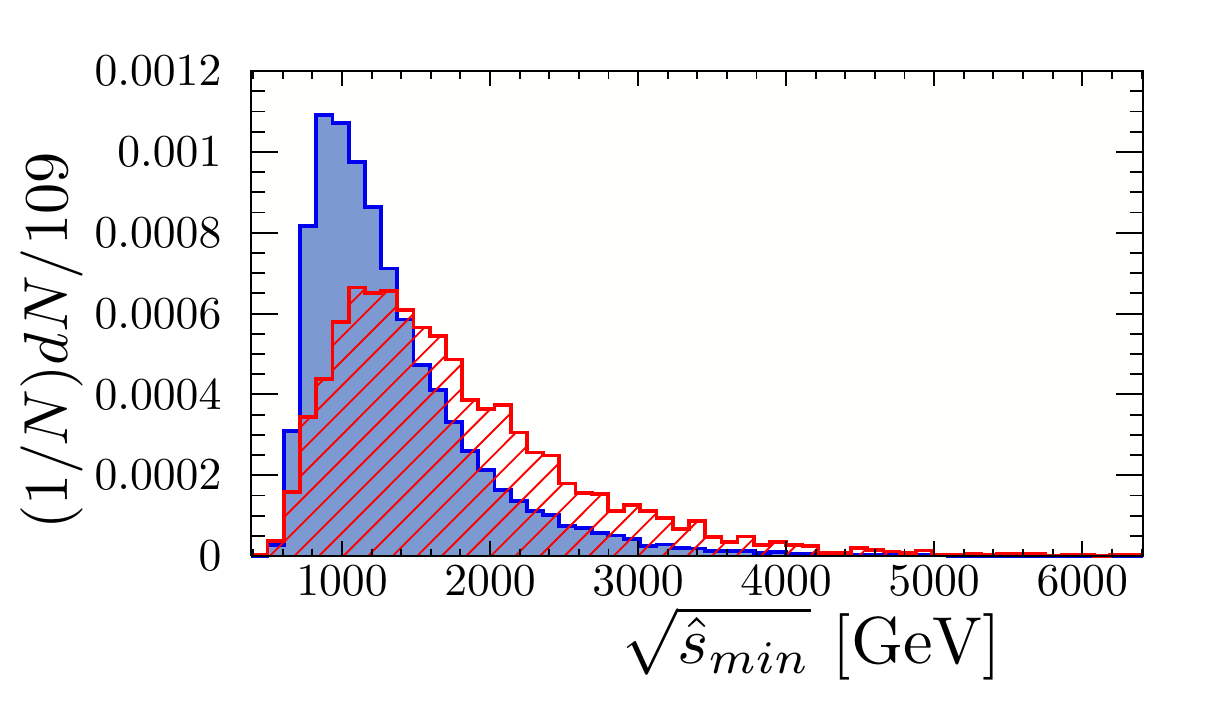}}
    \subfloat[] {\label{fig:pt_j1_sig_bg}  \includegraphics[scale=0.295]{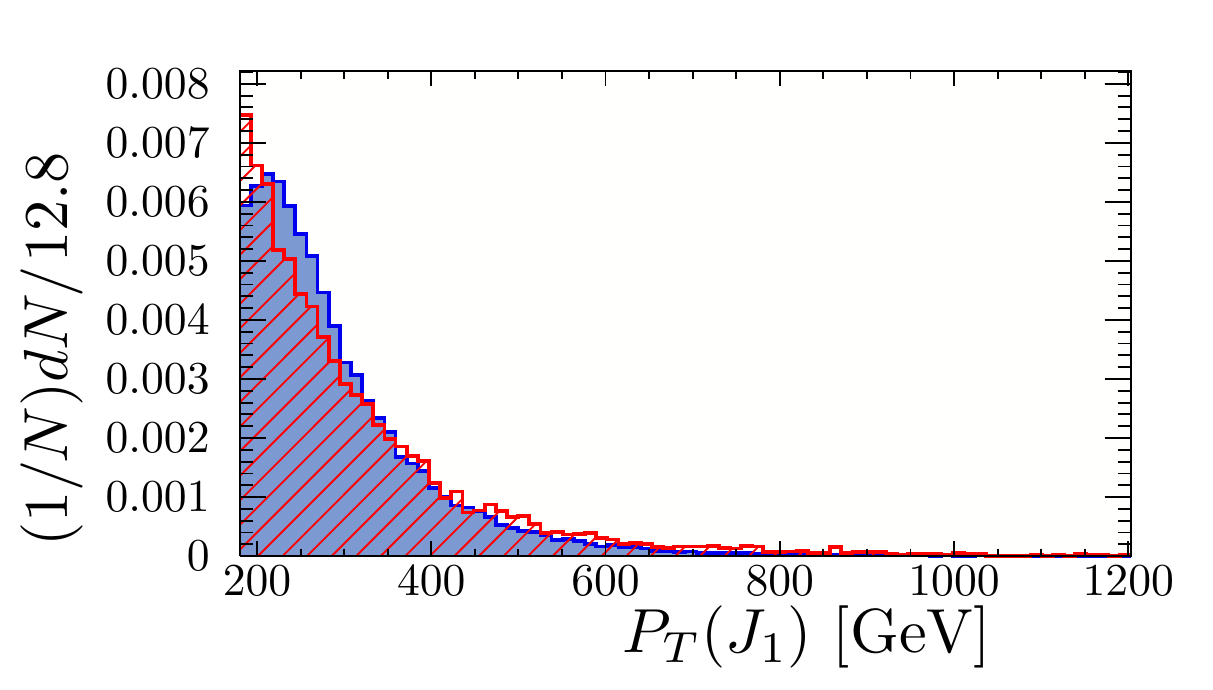}} 
  \subfloat[] {\label{fig:MET_sig_bg}     \includegraphics[scale=0.295]{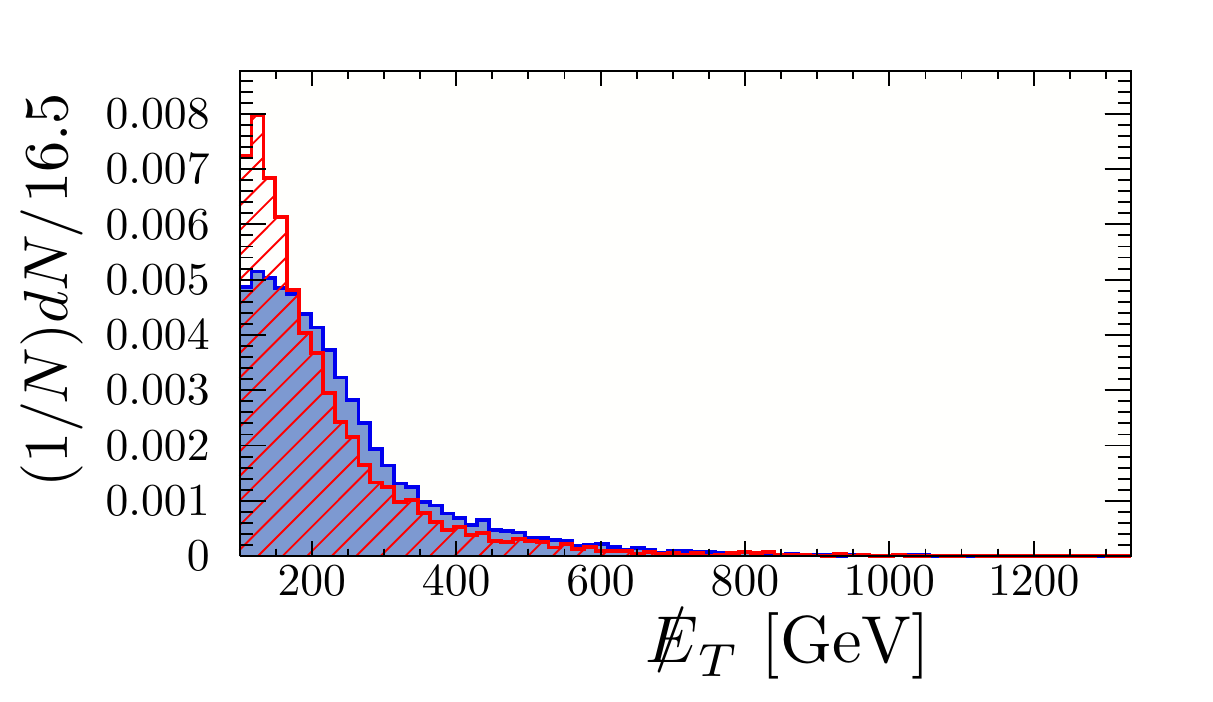}}\\
  \caption{Normalized kinematic distributions of the different input variables used in MVA for the background (red) and the signal (blue). Plot (a) and (b) represent the distribution of pruned jet mass of the leading and subleading fatjets, respectively, whereas plot (c) and (d) are the distributions of the N-subjettiness ratio of the leading and subleading fatjets, respectively. Plot (e) shows distribution of the relative separation between the two leading fatjets. Azimuthal separation distribution of the subleading fatjet from the missing energy direction is depicted in plot (f). Plot (g) shows distribution of the global inclusive variable $\sqrt{\hat{S}_{min}}$ and distribution of the transverse momentum of the subleading fatjet and total missing transverse momentum are presented in plot (h) and (i), respectively. We display the signal distributions for BP2, including all contributions from associated production of the heavy scalar and pair production of the heavy scalars at NLO. The background comprises all the processes discussed in \autoref{sec:S_B} after applying the cuts $M_{J_{0,1}}>40$
 GeV and b-veto together with the pre-selection criteria mentioned in \autoref{sec:variables_cuts}.}
\label{fig:sig_bg_1}
\end{figure*}

\subsection{Multivariate analysis (MVA)}
\label{sec:MVA}

\begin{table*}[tb]
\centering
%
 \begin{tabular}[b]{|c|c|c|c|c|c|c|c|c|c|}
\hline
Variable     & $\tau_{21}(J_0)$ & $M(J_0)$ & $\tau_{21}(J_1)$ & $M(J_1)$ & $\Delta R (J_0, J_1)$ & $\sqrt{\hat{S}_{min}}$ & $\Delta \phi(J_1,\slashed{E}_T)$ & $\slashed{E}_T$ & $P_T(J_1)$ \\ %
\hline\hline
Separation & 16.58 & 15.71 & 13.71 & 11.57 & 11.27 & 9.039 & 3.011 & 2.451 & 1.324\\
\hline
 \end{tabular} 

\caption{Method unspecific relative importance (or separation power) of the different variables according to their rank before using at MVA. We obtain the numbers for BP2 from the TMVA package during MVA. Those numbers can change modestly for different benchmark points and different algorithms. }\label{relative_imp}
\end{table*}

\begin{figure*}[tb]
\centering
  \subfloat {\label{fig:correlationS}\includegraphics[scale=0.43]{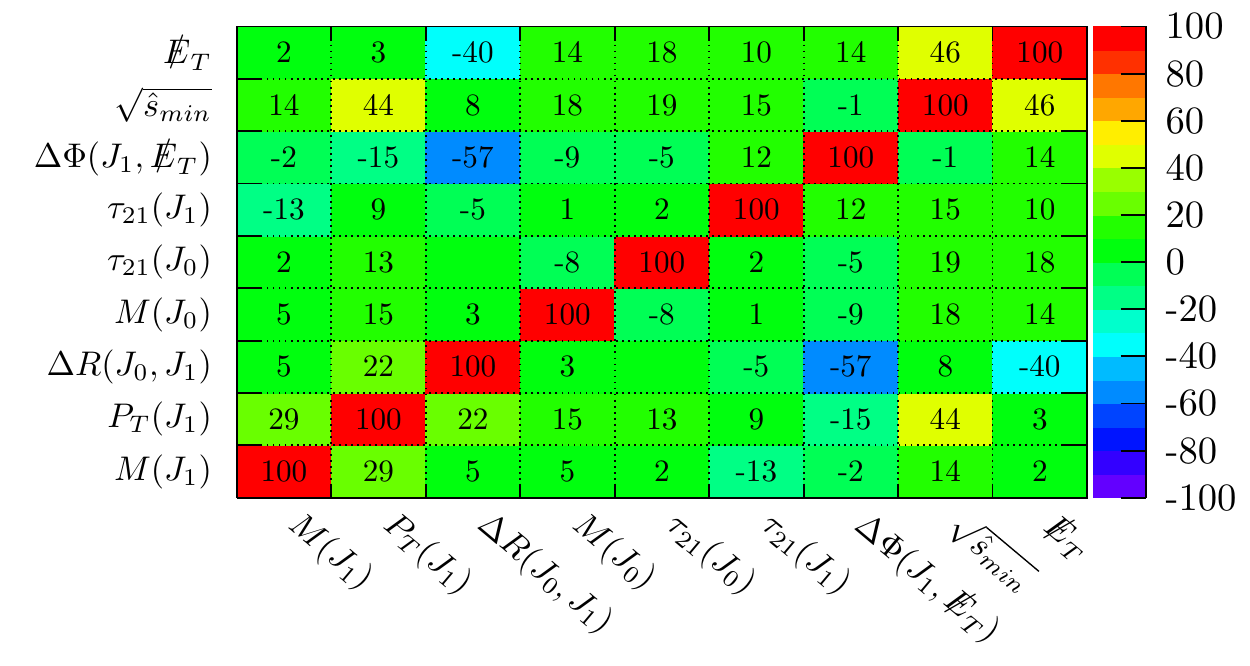}}
  \subfloat {\label{fig:correlationB}\includegraphics[scale=0.43]{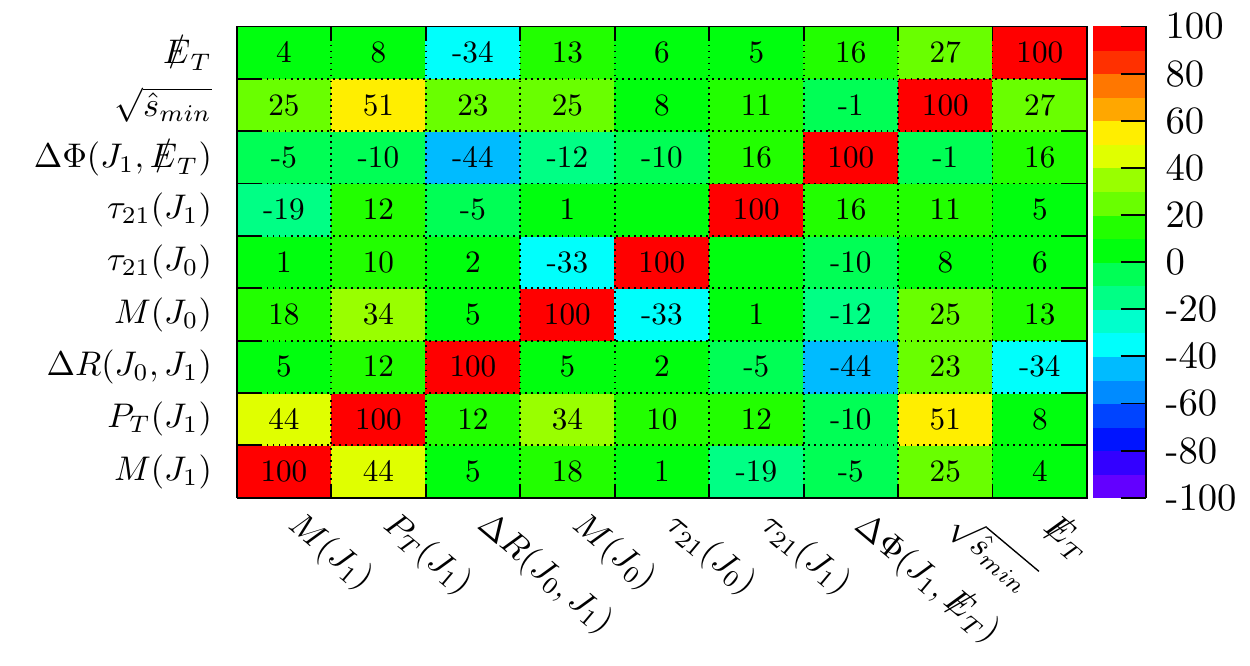}}\\
  \caption{The linear correlation coefficients among different kinematic variables used in MVA (in percentage) for the signal (left panel, BP2) and background (right panel). The positive and negative signs signify the positive and negative correlations (anti-correlated) among the two variables.}\label{correlation}
\end{figure*}

In \autoref{tab:MVA}, we present the expected number of signal events coming from the associated production and pair production of the heavy scalar channels together with all background processes at 14 TeV LHC with integrated luminosity 3000 $fb^{-1}$. From this table, we see contribution in the $2J_V+\slashed{E}_T$ final state coming from the pair production of the heavy scalars is always more prominent than the associated production after these cuts.
We construct two independent event samples for our multivariate analysis, one for the signal and another for the background. The entire dataset is splitted randomly -- $50\%$ for the training and the remaining for testing purposes for both samples. We employ an adaptive BDT algorithm for MVA. We generate different signal processes separately at NLO and combine them according to their weights to get the kinematic distributions of the combined signal. Similarly, the different background processes are generated separately at LO with two to four extra jet MLM matching and combined thereafter according to their weights to get the kinematic distributions of the combined background. A set of kinematic variables is chosen from a bigger group of variables employed in the MVA analysis depending on their relative importance while discriminating the signal class from the background class. 
We present in \autoref{fig:sig_bg_1} the normalized kinematic distributions of all nine input variables that are used in MVA. We obtain the signal distributions using sample benchmark point BP2, including all the associated production and pair production of the heavy scalars at NLO. 
We do not include the process $p p \rightarrow HH$ in our analysis although it has a larger cross-section than any other associated or pair-production channels, as b-veto and cuts on the fatjet mass and N-subjettiness ratio $\tau_{21}$ weaken its effect and the remaining events reside well away from the maximum \textit{BDT response} region.
The background comprises of all the processes discussed in \autoref{sec:S_B} after applying the cuts $M_{J_{0}}, M_{J_{1}} > 40$ GeV and b-veto along with the pre-selection cuts mentioned in \autoref{sec:variables_cuts} at 14 TeV LHC. 
The distributions of the pruned jet mass $M_{J_{0,1}}$ of the leading (\autoref{fig:m_jo_sigbg}) and subleading (\autoref{fig:m_j1_sig_bg}) fatjets, have a peak near 80-90 GeV for the signal close to the vector boson mass, however no such peak for the background reflects that fatjets are predominantly formed from QCD jets. 
The distributions of the N-subjettiness ratio, $\tau_{21}(J_{0,1})$ of the leading (\autoref{fig:tau_j0_sig_bg}) and subleading (\autoref{fig:tau_j1_sig_bg}) fatjets establish that both the fatjets of the signal have a two-prong structure as they peak at a smaller value of $\tau_{21}$. In contrast, both the fatjets in the background has a characteristic one-prong structure producing a larger value for this variable. Hence these four jet substructure variables are crucial in discriminating the signal from the background. 
The relative separation between the leading ($J_0$) and subleading ($J_1$) fatjets $\Delta R(J_0,J_1)$ (\autoref{DR_joj1_sig_bg}), azimuthal separation between  $J_1$ and $\slashed{E}_T$ is represented as $\Delta \phi(J_1,\slashed{E}_T)$ (\autoref{fig:dphi_j1ET_sig_bg}), and the inclusive global variable $\sqrt{\hat{S}_{min}}$ (\autoref{fig:shat_sigbg}) are effective observables to separate the signal from the background. The inclusive variable $\sqrt{\hat{S}_{min}}$, defined as the minimum CM energy required to satisfy all observed objects and $\slashed{E}_T$ was proposed in \cite{Konar:2008ei, Konar:2010ma, Barr:2011xt} to find the new physics mass scale for the signals containing invisible particles like ours. All the reconstructed objects of the detectors are used to construct the reconstructed object-level $\sqrt{\hat{S}_{min}}$ that demonstrate better efficiency than the other inclusive variables $H_T$, $\slashed{H}_T$ etc.

A variable is considered to be more powerful discriminator, if it possesses a larger separation between the signal and background. For different kinematic variables, the method unspecific relative importance is shown in \autoref{relative_imp}, where we principally keep the variables that have less (anti-)correlation among themselves both for the signal and background. We notice four jet substructure variables $M_{J_{0,1}}$ and $\tau_{21}(J_{0,1})$ are very good discriminators. The relative importance of the different kinematic variables can change modestly for different benchmark points. Although very high $P_T$ for both fatjets and large $\slashed{E}_T$ are considered during event selection, transverse momentum of the subleading fatjet, $P_T(J_1)$ (\autoref{fig:pt_j1_sig_bg}) and $\slashed{E}_T$ (\autoref{fig:MET_sig_bg}) still can take a role in discriminating the signal from the background in MVA. Note that $P_T(J_0)$ and $P_T(J_1)$ are highly correlated (positively) both in signal and background classes, so we keep only $P_T(J_1)$ in the analysis as it has more relative importance than $P_T(J_0)$. Similarly, $\Delta \phi(J_0,\slashed{E}_T)$ and $\Delta \phi(J_1,\slashed{E}_T)$ are highly anti-correlated, but we keep $\Delta \phi(J_1, \slashed{E}_T)$ because of its larger relative importance. The linear correlation coefficients among different kinematic variables used in MVA (in$\%$) for the signal and background are shown in \autoref{correlation}. The positive (negative) signs signify the positive (negative) correlation (anti-correlation) among the two variables. Modestly large anti-correlation between $\Delta \phi(J_1,\slashed{E}_T)$ and $\Delta R(J_0, J_1)$ is present, although we kept them both as they have large relative importance.

\begin{figure*}[tb]
\centering
  \subfloat {\label{fig:sig_bg}\includegraphics[scale=0.44]{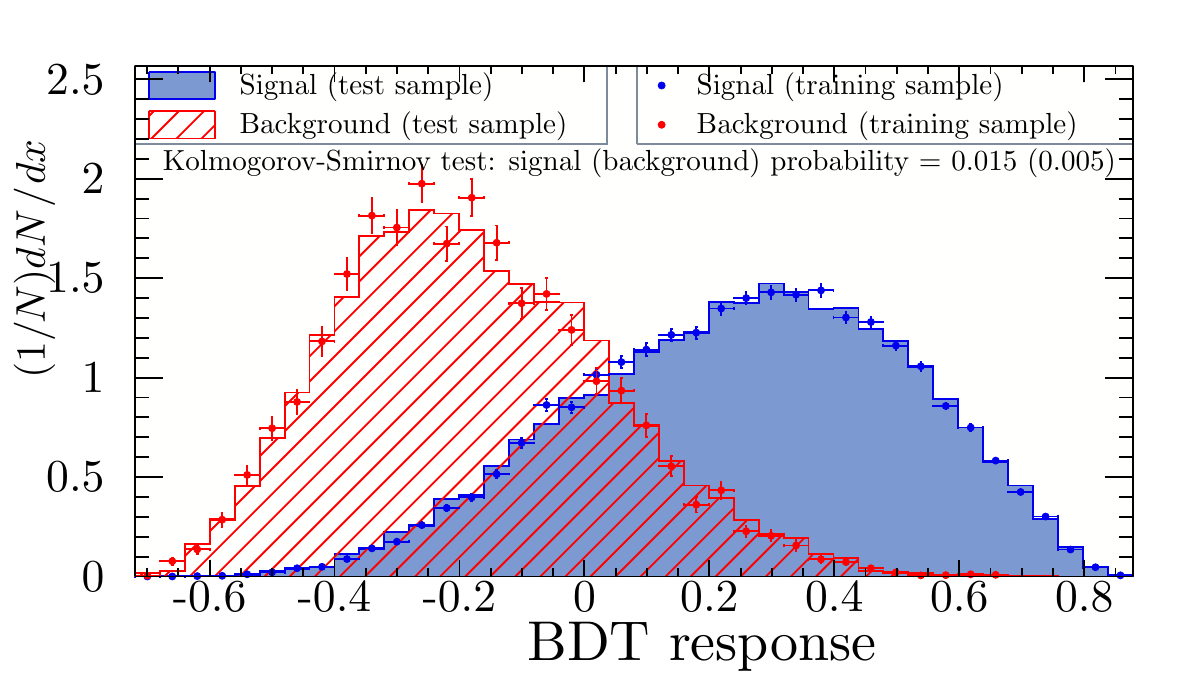}} 
  \subfloat {\label{fig:significance}\includegraphics[scale=0.44]{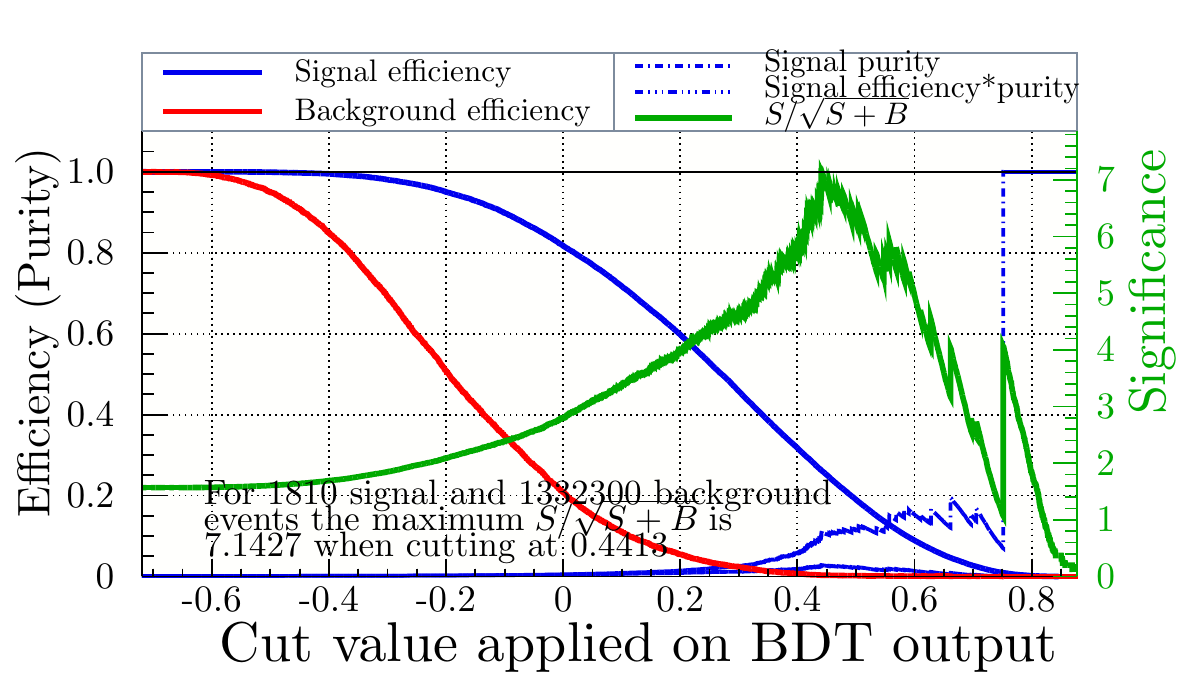}}\\
  \caption{The left panel shows the normalized BDT response for the training and testing samples for both signal (BP2) and background classes. The right panel contains the cut efficiencies for the background (red) and the signal (blue) and the statistical significance of the signal over the background (green) as a function of the cut value applied on the BDT response.}\label{sig_bg_significance}
\end{figure*}

\begin{figure}[!tb]
\centering
\includegraphics[width=0.42\textwidth]{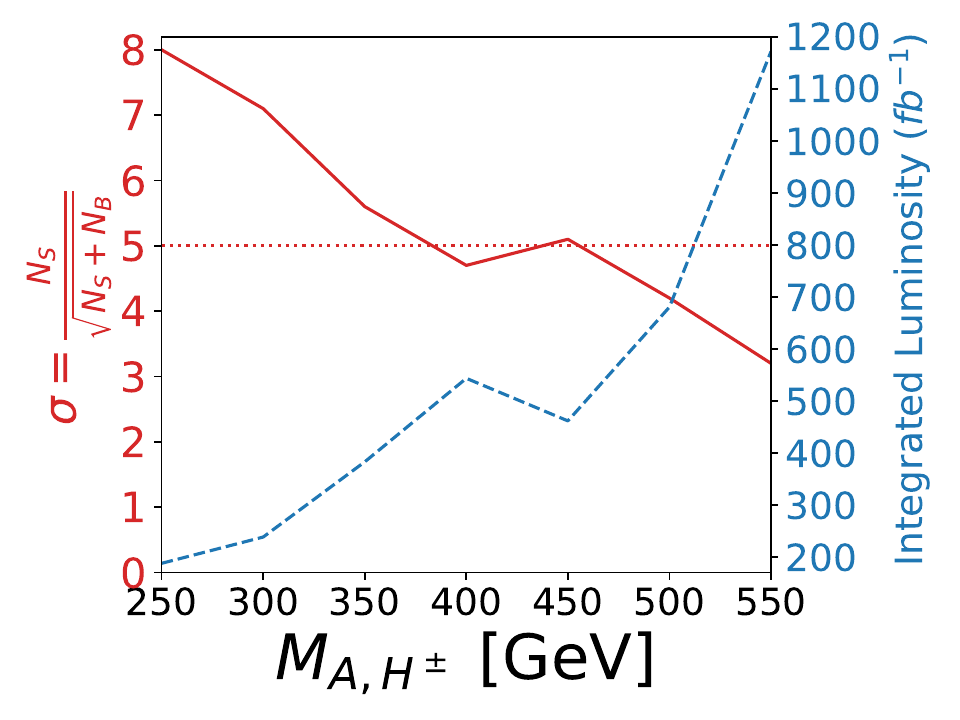} \\ \vspace{0.3cm}
\qquad 
 \begin{tabular}[b]{|c|c|c|c|c|c|}
\hline\hline
Model& $N_S^{bc}$ & BD$T_{opt}$ & $N_S$ & $N_B$ & $\sigma$ \\ %
\hline\hline
BP1 & 2129 & 0.6927 & 357 & 1650 & 8.0\\
\hline
BP2 & 1810 & 0.4413 & 474 & 3930 & 7.1\\
\hline
BP3 & 1511 & 0.6003 & 390 & 4516 & 5.6\\
\hline
BP4 & 1236 & 0.5270 & 207 & 1738 & 4.7 \\
\hline
BP5 & 1003 & 0.6499 & 178 & 1041 & 5.1 \\
\hline
BP6 & 800 & 0.6098 & 102 & 485 & 4.2 \\
\hline
BP7 & 654 & 0.6462 & 97 & 820 & 3.2 \\
\hline\hline
$N_{SM}$ & 1332300 & \multicolumn{4}{| c |}{} \\
\hline
 \end{tabular} 
\caption{The upper panel shows the statistical significance of the signal over the background as a function of masses of the heavy BSM scalars (solid red line) at 14 TeV HL-LHC. The dashed blue curve on the same plot exhibits the required luminosity for two sigmas ($2\sigma$) exclusion for different benchmark points--the horizontal dotted red line to mark 
$5\sigma$ discovery potential. 
The lower panel demonstrates the corresponding expected number of signal events ($N_S^{bc}$) at NLO and background events ($N_{SM}$) before applying the BDT cut, where $N_S$ and $N_B$ are the expected number of signal and background events that survive after applying the optimum $\text{BDT}_{opt}$ cut, respectively.}
\label{tab:BDT}
\label{Fig:discovery_pot}
  \end{figure}

Finally, we present the normalized BDT response for the training and testing samples for both signal and background classes in the left panel of \autoref{sig_bg_significance}. The signal distribution is presented for BP2. The distributions of the BDT response get well separated for the signal and background. Cut efficiencies can be estimated by applying a cut $\text{BDT}_{res}>\text{BDT}_{cut}$  on the BDT response. In the right panel of \autoref{sig_bg_significance}, such cut efficiencies are demonstrated for the background (red) and signal (blue), along with the statistical significance of the signal over the background (green) as a function of the cut value applied on the BDT response. We use the prescription $\sigma=\frac{N_S}{\sqrt{N_S+N_B}}$ for computing the statistical significance. $N_S$ and $N_B$ are respectively the expected number of signal and background events after using the optimal cut $\text{BDT}_{opt}$  at 3000 $fb^{-1}$ luminosity at 14 TeV LHC. $N_S$, $N_B$, and $\sigma$ are shown in the right panel of \autoref{tab:BDT} for different benchmark points. We find more than $5\sigma$ discovery potential for four different benchmark points. In the left panel of \autoref{Fig:discovery_pot} we summarize the result in terms of statistical significance of the signal as a function of the masses of the heavy BSM scalars (solid red) at 14 Tev LHC with integrated luminosity 3000 $fb^{-1}$. At the same time, the dashed blue line exhibits the required luminosity for $2\sigma$ exclusion for different benchmark points.

\section{Conclusions}
\label{sec:conclude}

IDM is a simple extension of the SM where a new $SU(2)_L$ scalar doublet owning a discrete $\mathbb{Z}_2$ symmetry provides a viable DM candidate together with additional heavy BSM scalars. This model offers two distinct parameter spaces, consisting of hierarchical mass spectrum and degenerate mass spectrum of these scalars, that satisfy the observed relic density of the dark matter and other theoretical and experimental constraints.

Despite of several studies being performed in exploring this viable dark matter model at the LHC, in this paper we initiate the effort of looking into a promising channel with NLO QCD precision. This study focuses on the hierarchical mass region and considers NLO QCD corrections on the associated and pair production channels of heavy scalars.  We find that the effect of QCD correction is significant for encrypting the correct search strategy at the LHC. \autoref{tab:crosssection_2}, and \autoref{tab:crosssection_3} encapsulate the correction factors for different benchmark points. We get an overall correction of about $33\%$-$39\%$ for the associated production processes and for a gauge boson mediated pair production channel, $p p \rightarrow H^\pm A$. Similarly, the $p p \rightarrow H^+ H^-$ process, which encompasses both gauge boson and Higgs mediator, has the correction factor in between $38\%$ and $56\%$. In contrast, $p p \rightarrow A A$ being scalar mediated, receives a correction factor in the range of $70\%$-$92\%$. Nevertheless, notable improvement on scale uncertainties is achieved due to the inclusion of NLO corrections. We also take into account the parton shower effect and demonstrate its practicality at the low transverse momentum region.

After jet clustering and detector simulation, we compare distributions of various crucial kinematic observables at LO and NLO.  Noted shifts in the shape of these distributions over the LO computation can significantly influence the construction of phenomenological analysis. 
We notice a substantial relative change in the number of survived signal events as an effect of the differential NLO K-factor. For example, this change is up to $46\%$ for the gauge mediated pair production of heavy scalar processes. 
We also emphasize that gauge boson mediated decay products of hadronically decayed heavy scalars are highly collimated in this signal region and therefore large-radius fatjets come naturally in probing the hierarchical mass region. 
The internal structure and properties of the fatjet are key ingredients to know about their genesis. Fatjets originated from the QCD radiation of partons pose different characteristics compared to the fatjets generated from boosted vector boson. The jet substructure is a powerful tool to get control over the colossal SM background and identify the signal correctly. We find jet-substructure observables $M_{J_{0,1}}$ and $\tau_{21}(J_{0,1})$ are excellent discriminators in discriminating fatjets originated from the boosted vector boson and the QCD jets. We work with parton shower matched NLO QCD corrected signal and employ sophisticated multivariate analysis to distinguish the signal using these powerful jet-substructure variables. We discuss the set of nine variables that are used in the MVA analysis and their linear correlation coefficients are presented for the signal at a sample benchmark point and for the background.

We observe that the discovery potential for different benchmark points nearly up to 350 GeV of heavy scalar mass in the hierarchical mass region has a statistical significance above $5\sigma$ at the HL-LHC. Hence this parameter space of the hierarchical mass spectrum which is well motivated having a dark matter candidate of mass $m_{\rm DM} \sim m_h/2$, would be quite interesting to look into. We also notice through this study that the heavy BSM scalar mass falling in the range of 250-550 GeV can be excluded with 1200 $fb^{-1}$ integrated luminosity at the 14 TeV LHC. 

\acknowledgments
We thank Akanksha Bhardwaj for her continuous help and fruitful discussions regarding phenomenological part of this analysis. This work is supported by the Physical Research Laboratory (PRL), Department of Space, Government of India. Computational work was performed using the HPC resources (Vikram-100 HPC) and TDP project at PRL.

\appendix
\section{Feynman Rules}
\label{appendixFR}

\begin{equation}
\vcenter{\hbox{\includegraphics[width=0.2\textwidth]{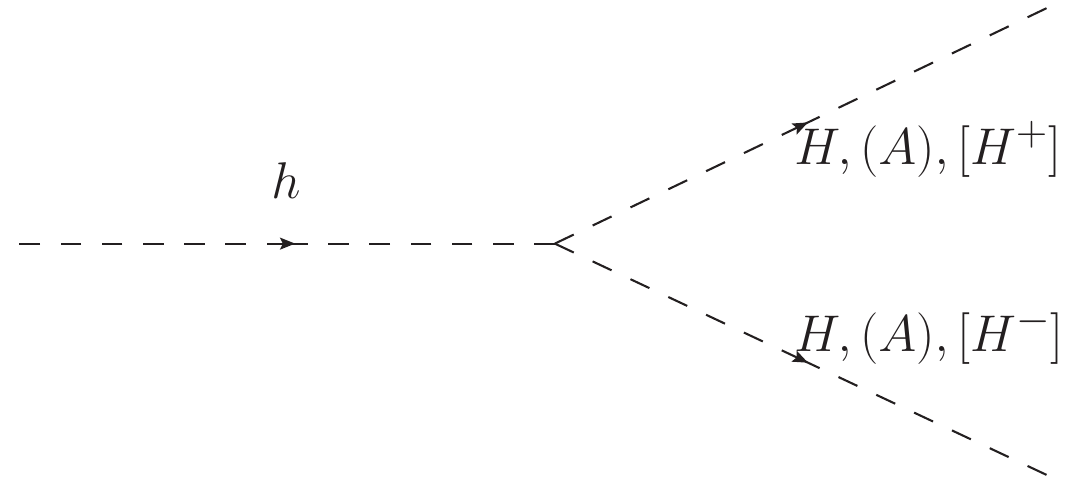}}}
= -i \Lambda v =
\begin{cases}
 -i\lambda_c v \hspace{2mm}\text{for}\hspace{1mm} A A\\
 -i\lambda_L v \hspace{2mm}\text{for}\hspace{1mm} H H\\
 -i\lambda_3 v \hspace{2mm}\text{for}\hspace{1mm} H^+ H^-
\end{cases}
\end{equation}
where $\lambda_{L/c}=(\lambda_3+\lambda_4\pm \lambda_5)$.

\begin{equation}
\vcenter{\hbox{\includegraphics[width=0.2\textwidth]{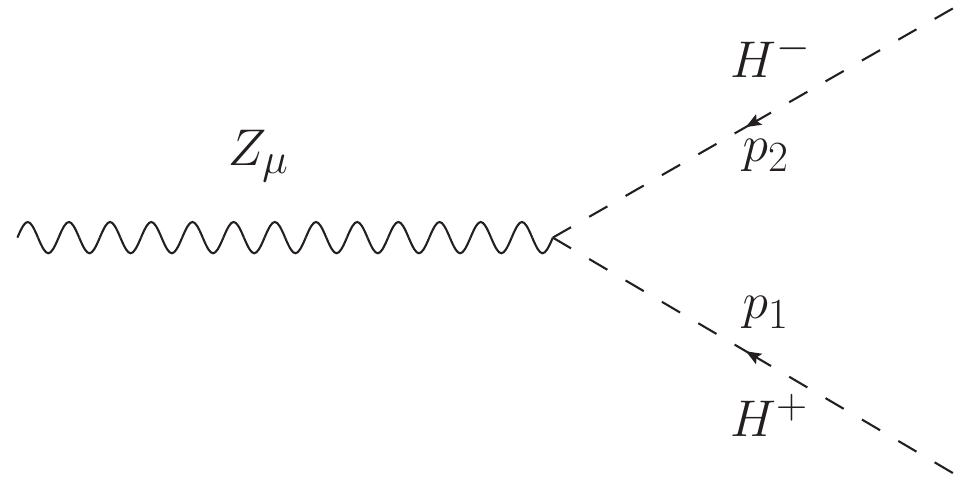}}}
= i\hspace{1mm} e \hspace{1mm} (p_1-p_2)_\mu \frac{\cos(2\theta_W)}{2\hspace{1mm} C_W \hspace{1mm} S_W}
\end{equation}
where $\theta_W$ is the Weinberg angle. $S_W$ and $C_W$ correspond to the Sine and Cosine of the Weinberg angle respectively: 
\begin{equation}
S_W =\sqrt{1-(M_W/M_Z)^2} \hspace{3mm} \text{and} \hspace{3mm} C_W=\sqrt{1-S_W^2} \, .
\end{equation}

\begin{equation}
\vcenter{\hbox{\includegraphics[width=0.2\textwidth]{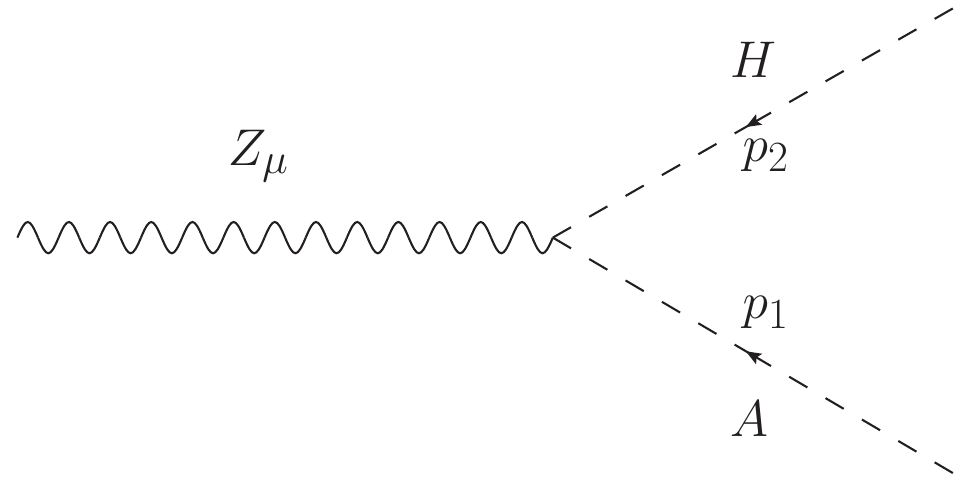}}}
= e \hspace{1mm} \frac{(p_1-p_2)_\mu}{2\hspace{1mm} C_W \hspace{1mm} S_W}
\end{equation}

\begin{equation}
\vcenter{\hbox{\includegraphics[width=0.2\textwidth]{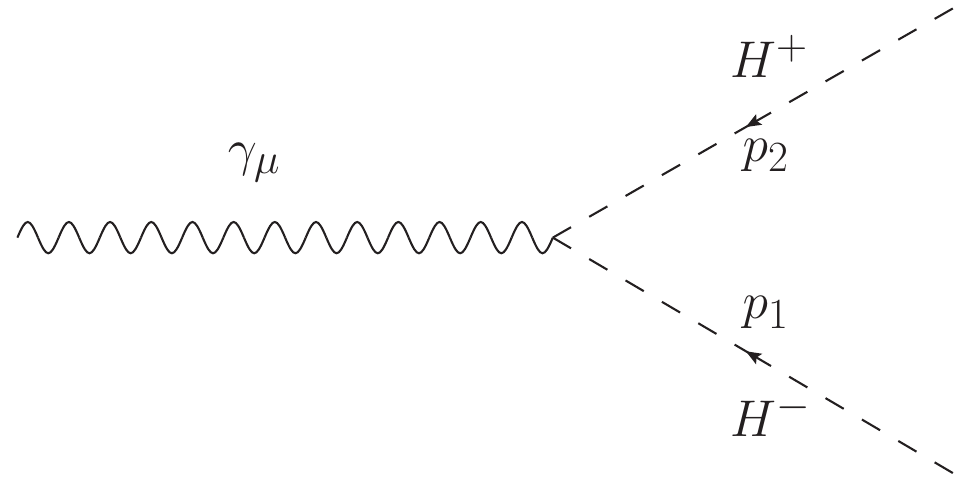}}}
= i\hspace{1mm} e \hspace{1mm} (p_2-p_1)_\mu
\end{equation}

\begin{equation}
\vcenter{\hbox{\includegraphics[width=0.2\textwidth]{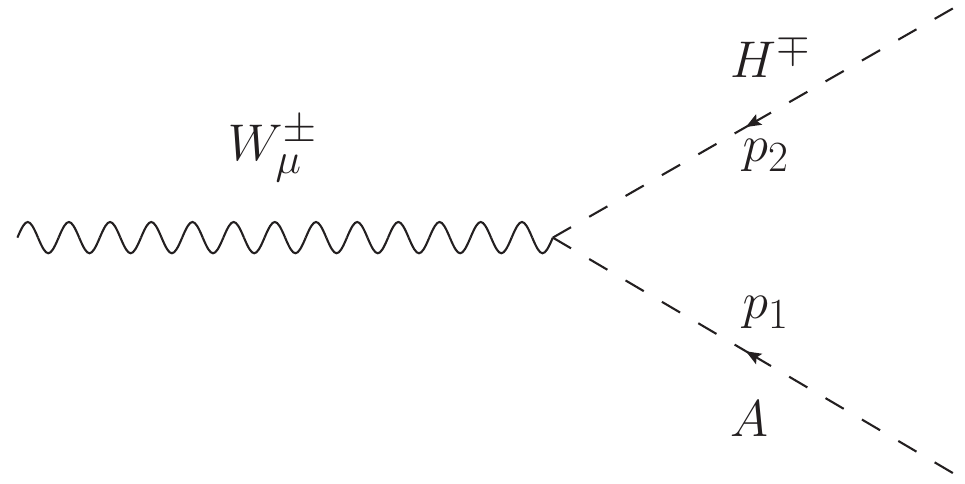}}}
= -e \hspace{1mm} \frac{(p_1-p_2)_\mu}{2\hspace{1mm} S_W}
\end{equation}

\begin{equation}
\vcenter{\hbox{\includegraphics[width=0.2\textwidth]{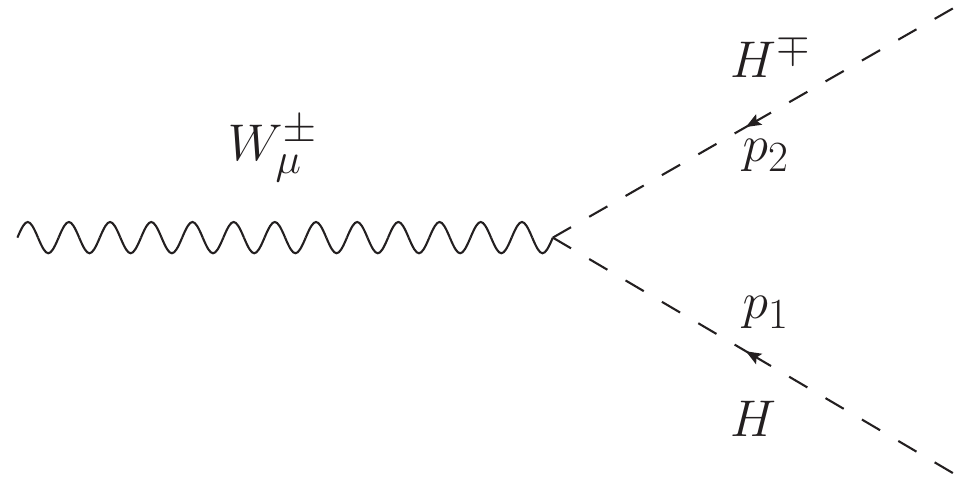}}}
= \pm \hspace{1mm} i \hspace{1mm} e \hspace{1mm} \frac{(p_1-p_2)_\mu}{2\hspace{1mm} S_W}
\end{equation}

\begin{equation}
\vcenter{\hbox{\includegraphics[width=0.2\textwidth]{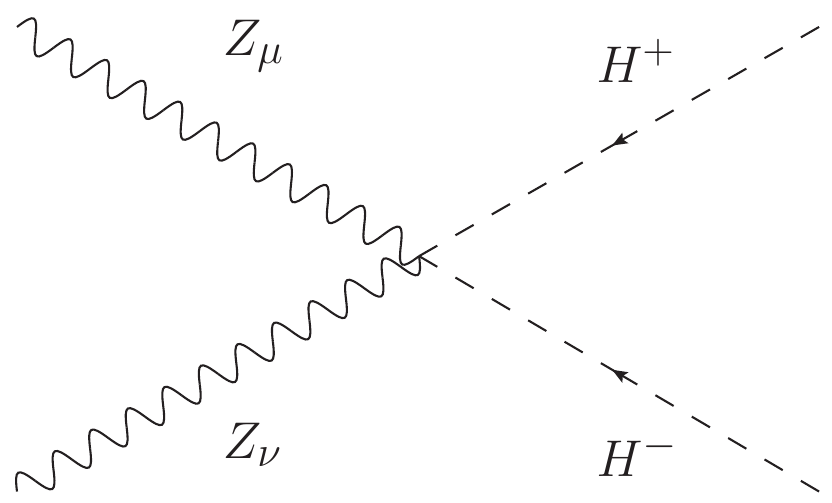}}}
= i \hspace{1mm} e^2 \hspace{1mm} \frac{\cos^2(2\theta_W)\hspace{1mm} g_{\mu\nu}}{2\hspace{1mm} C_W^2 \hspace{1mm} S_W^2}
\end{equation}

\begin{equation}
\vcenter{\hbox{\includegraphics[width=0.2\textwidth]{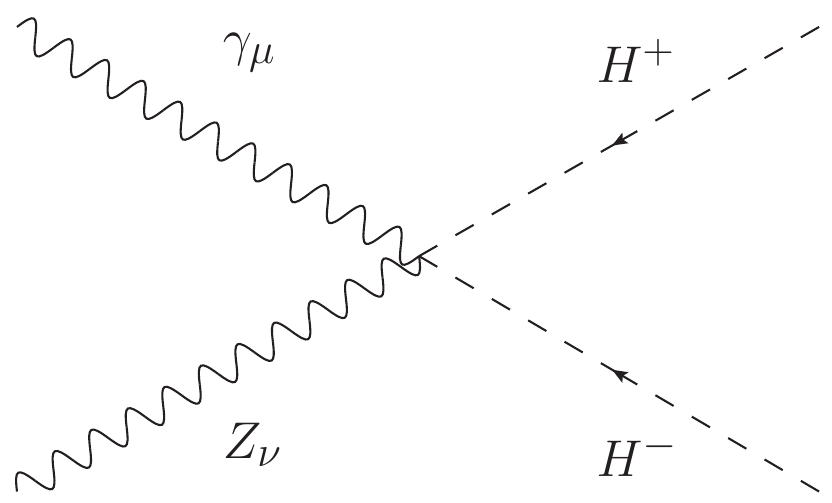}}}
= i \hspace{1mm} e^2 \hspace{1mm} \frac{\cos(2\theta_W)\hspace{1mm} g_{\mu\nu}}{C_W \hspace{1mm} S_W}
\end{equation}

\begin{equation}
\vcenter{\hbox{\includegraphics[width=0.2\textwidth]{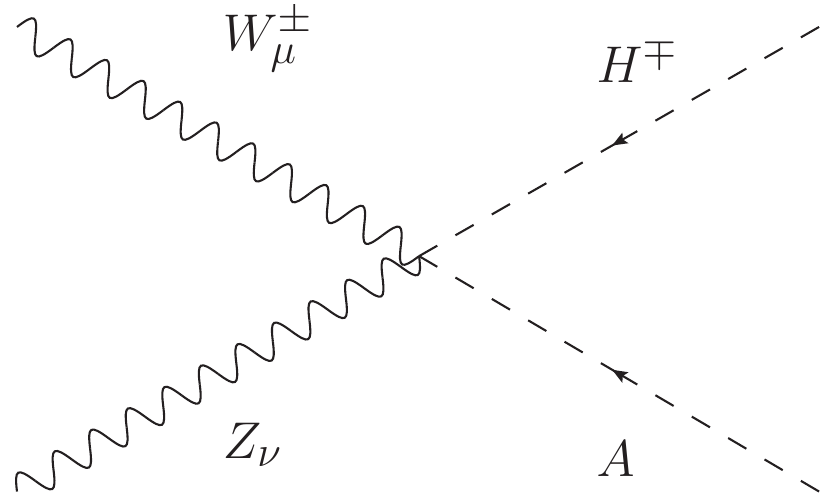}}}
= \pm \hspace{1mm} e^2 \hspace{1mm} \frac{g_{\mu\nu}}{2 \hspace{1mm} C_W}
\end{equation}

\begin{equation}
\vcenter{\hbox{\includegraphics[width=0.2\textwidth]{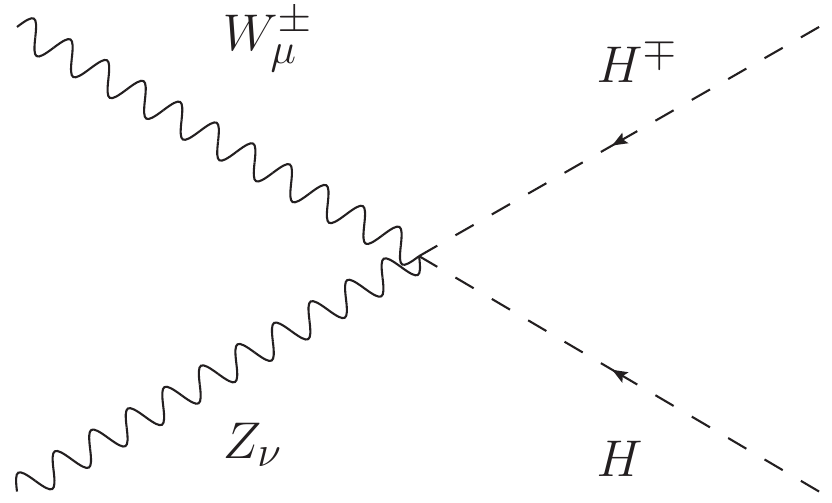}}}
= -i \hspace{1mm} e^2  \hspace{1mm} \frac{g_{\mu\nu}}{2 \hspace{1mm} C_W}
\end{equation}

\begin{equation}
\vcenter{\hbox{\includegraphics[width=0.2\textwidth]{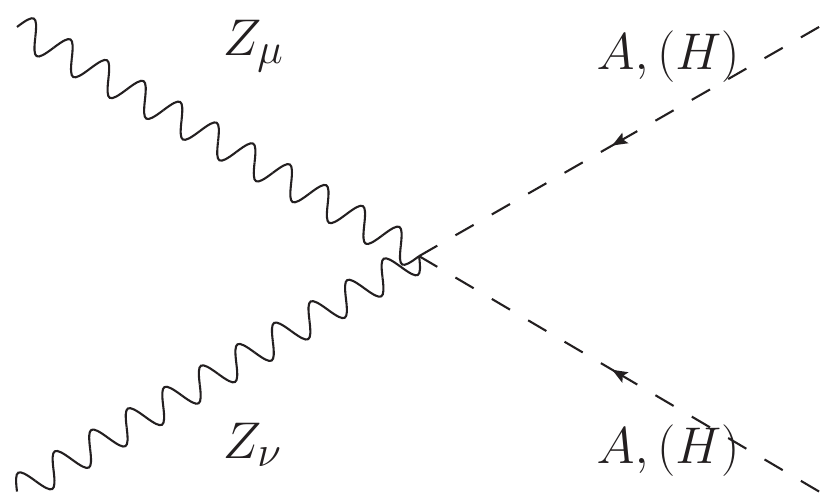}}}
= i \hspace{1mm} e^2  \hspace{1mm} \frac{g_{\mu\nu}}{2 \hspace{1mm} C_W^2 S_W^2}
\end{equation}

\begin{equation}
\vcenter{\hbox{\includegraphics[width=0.2\textwidth]{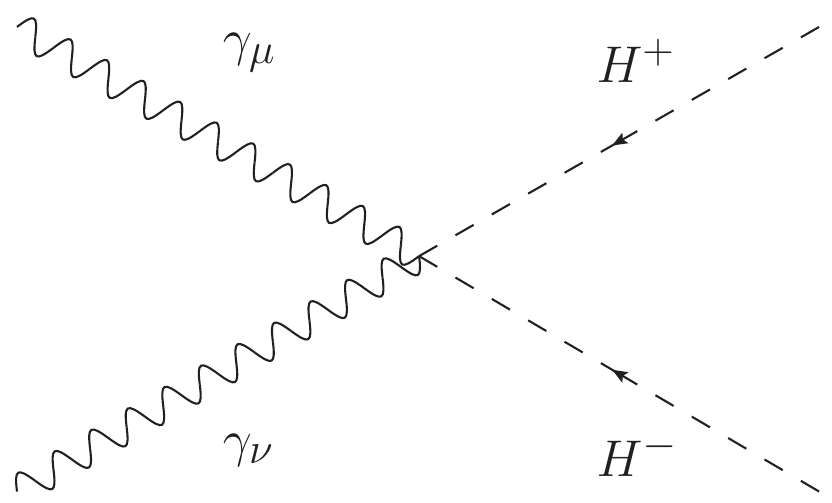}}}
= i \hspace{1mm} 2 e^2  \hspace{1mm} g_{\mu\nu}
\end{equation}

\begin{equation}
\vcenter{\hbox{\includegraphics[width=0.2\textwidth]{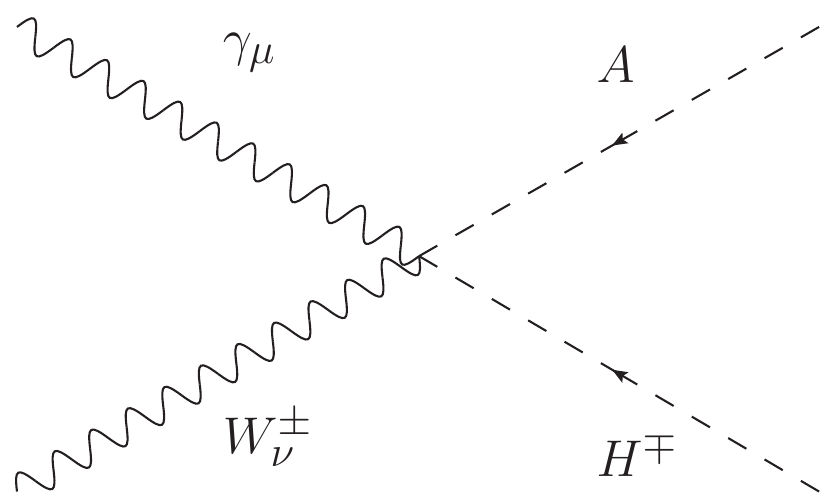}}}
= \mp \hspace{1mm} e^2  \hspace{1mm} \frac{g_{\mu\nu}}{2\hspace{1mm}S_W}
\end{equation}

\begin{equation}
\vcenter{\hbox{\includegraphics[width=0.2\textwidth]{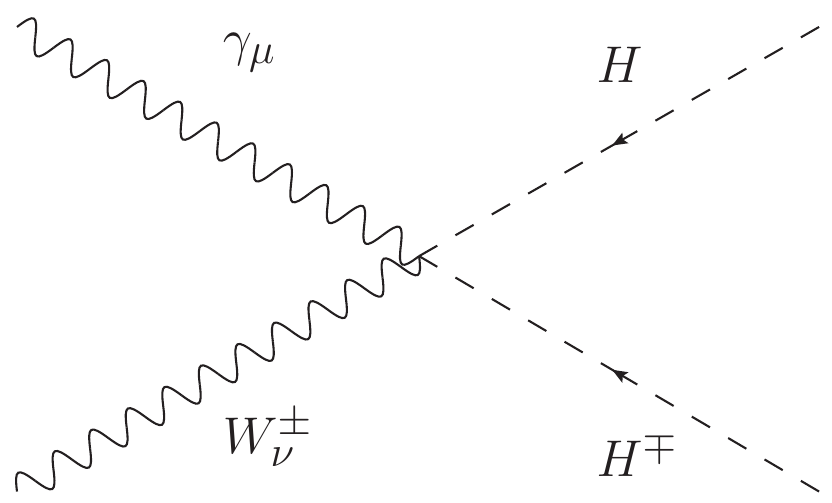}}}
= i \hspace{1mm} e^2  \hspace{1mm} \frac{g_{\mu\nu}}{2\hspace{1mm}S_W}
\end{equation}

\begin{equation}
\vcenter{\hbox{\includegraphics[width=0.2\textwidth]{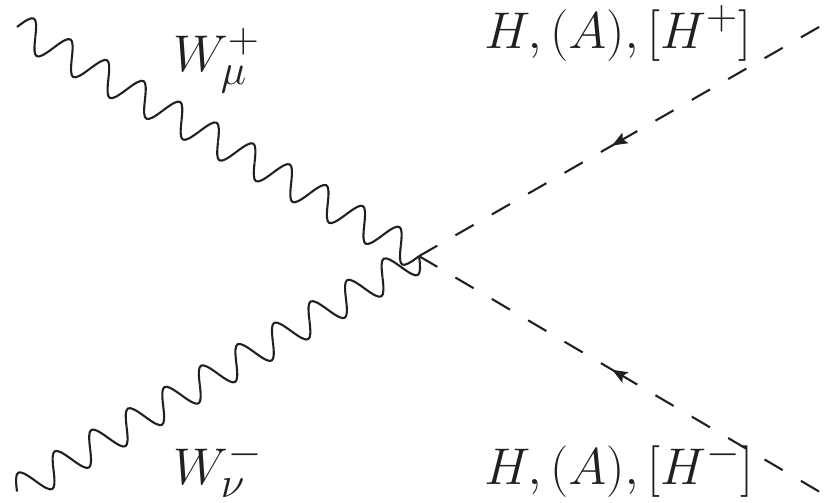}}}
= i \hspace{1mm} e^2  \hspace{1mm} \frac{g_{\mu\nu}}{2\hspace{1mm}S_W^2}
\end{equation}

\begin{equation}
\vcenter{\hbox{\includegraphics[width=0.2\textwidth]{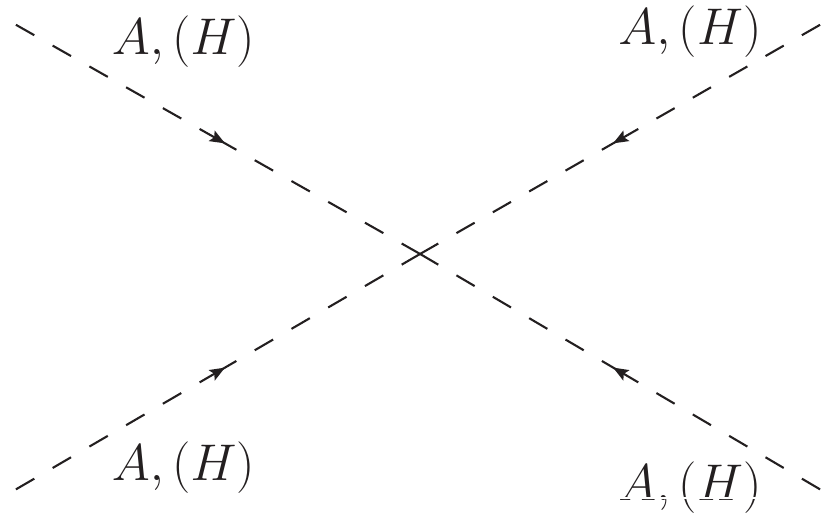}}}
= -i\hspace{1mm} 6 \lambda_2
\end{equation}

\begin{equation}
\vcenter{\hbox{\includegraphics[width=0.2\textwidth]{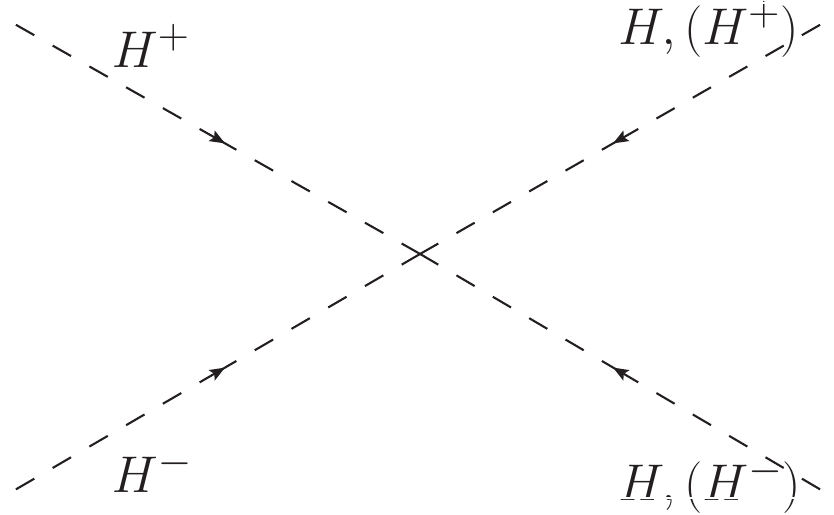}}}
= -i\hspace{1mm} 2 \lambda_2 \hspace{1mm} (-i\hspace{1mm} 4 \lambda_2)
\end{equation}

\begin{equation}
\vcenter{\hbox{\includegraphics[width=0.2\textwidth]{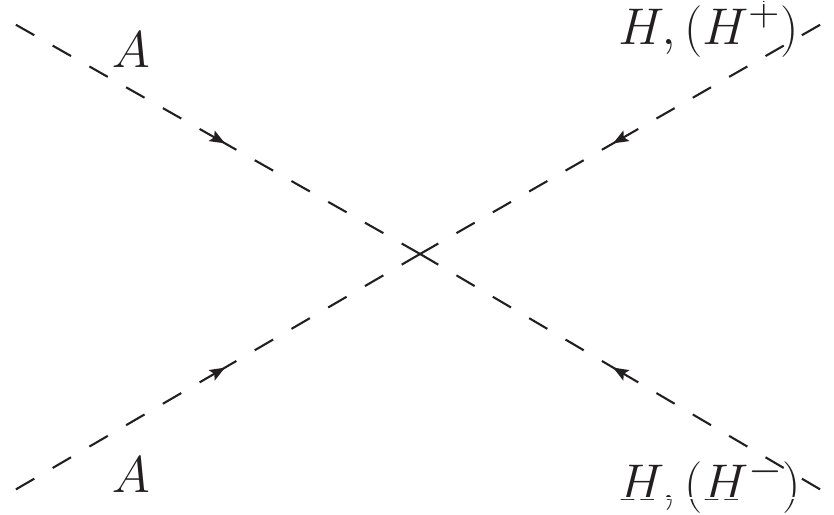}}}
= -i\hspace{1mm} 2 \lambda_2
\end{equation}

\begin{equation}
\vcenter{\hbox{\includegraphics[width=0.2\textwidth]{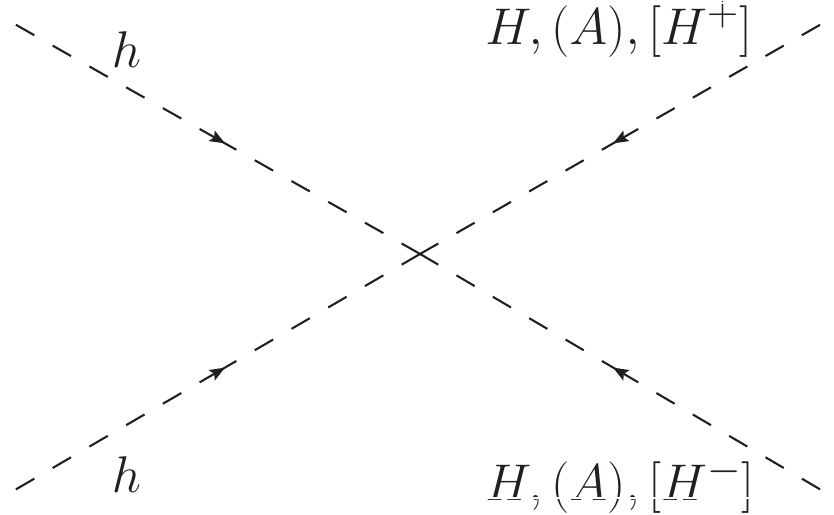}}}
= -i\hspace{1mm} \lambda_L,(-i\hspace{1mm} \lambda_c),[-i\hspace{1mm} \lambda_3]
\end{equation}

\begin{equation}
\vcenter{\hbox{\includegraphics[width=0.2\textwidth]{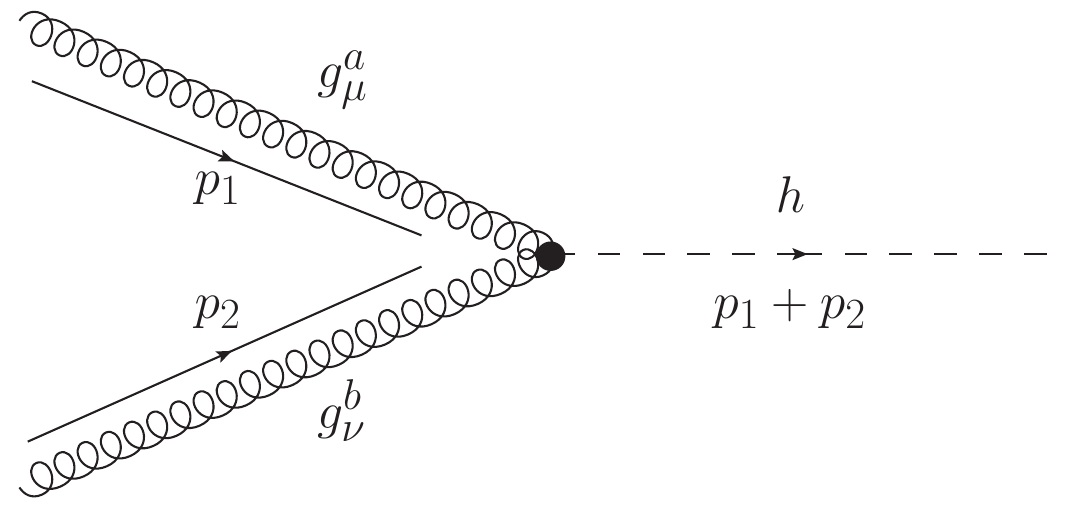}}}
=-i C_{eff}\delta^{ab}\hspace{1mm}(p_1^\nu p_2^\mu -g^{\mu\nu}\dfrac{\hat{s}_{12}}{2})
\end{equation}
where $\hat{s}_{12}  = (p_1+p_2)^2$ and 
the black blob indicates to the effective vertex arising from the Lagrangian $\mathcal{L}_{HEFT}$, given in \autoref{eftL}.

\begin{equation}
\vcenter{\hbox{\includegraphics[width=0.2\textwidth]{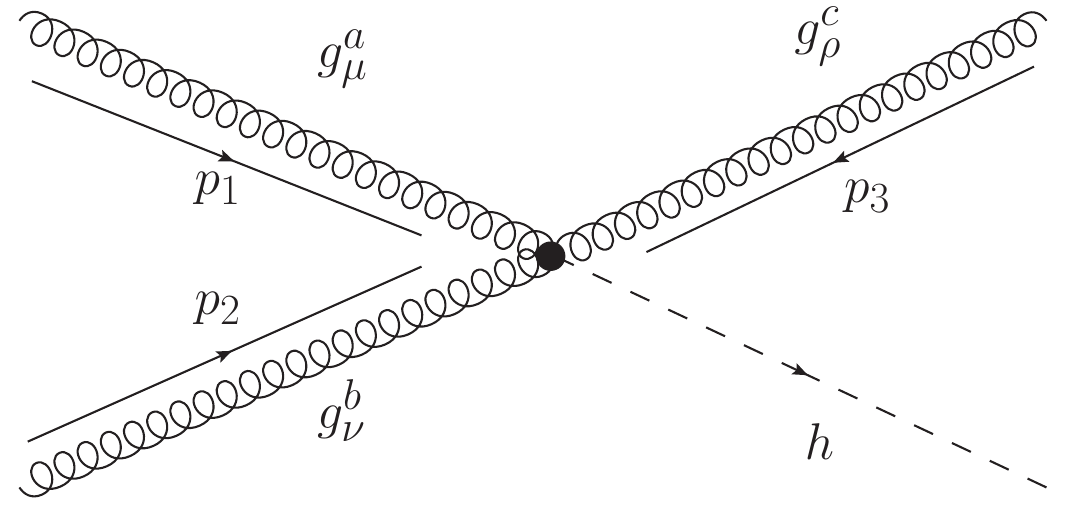}}}
\begin{aligned}
= & - g_s \hspace{1mm} C_{eff} f^{abc} [ (p_1-p_2)^\rho \hspace{1mm} g^{\mu\nu} \\& 
+ (p_2-p_3)^\mu \hspace{1mm} g^{\nu\rho}  + (p_3-p_1)^\nu \hspace{1mm} g^{\rho\mu} ]
\end{aligned}
\end{equation}

\begin{equation}
\vcenter{\hbox{\includegraphics[width=0.2\textwidth]{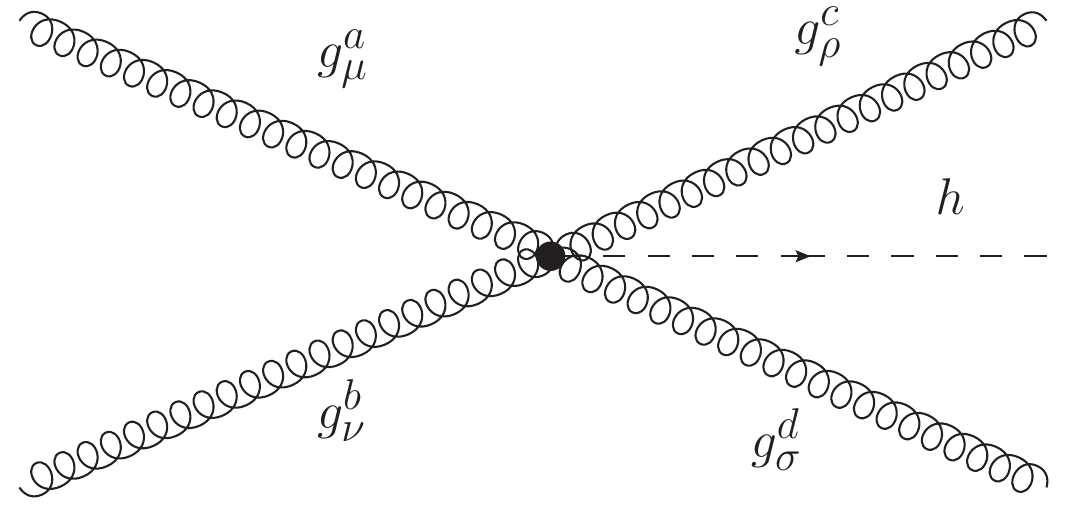}}}
\begin{aligned}
=  i g_s^2 \hspace{1mm} & C_{eff}  [ f^{abe} f^{cde} ( g^{\mu\sigma} g^{\nu\rho} - g^{\mu\rho} g^{\nu\sigma}) \\ & + f^{ace} f^{bde} ( g^{\mu\sigma} g^{\nu\rho} - g^{\mu\nu} g^{\rho\sigma})\\ & + f^{ade} f^{bce} ( g^{\mu\rho} g^{\nu\sigma} - g^{\mu\nu} g^{\rho\sigma}) ]
\end{aligned}
\end{equation}

%

%
\bibliographystyle{unsrt} 
\bibliography{references.bib}

\end{document}